\documentclass{elsartL}
\usepackage{graphicx}
\usepackage{amssymb}
\newsavebox{\myhbar}
\savebox{\myhbar}{$\hbar$}
\usepackage{amsmath}
\usepackage{mathrsfs}
\usepackage{mathtools}
\newcommand{\avg}[1]{\left< #1 \right>} % For a nice display of the average value
\begin{document}

\begin{frontmatter}
\title{Update of the phase-shift analysis of the low-energy $\pi N$ data}
%\author{Evangelos Matsinos{$^*$}},
\author{Evangelos Matsinos},
\author{G{\"u}nther Rasche}
%\address[GR]{Physik-Institut der Universit\"at Z{\"u}rich, Winterthurerstrasse 190, CH-8057 Z{\"u}rich, Switzerland}

\begin{abstract}
This paper presents the results of an updated phase-shift analysis (PSA) of the low-energy (pion laboratory kinetic energy $T \leq 100$ MeV) pion-nucleon ($\pi N$) data; this solution will be referred to as `ZRH20'. As in our 
earlier PSAs, the modelling of the $s$- and $p$-wave $K$-matrix elements will be achieved by following either of two methods: a) suitable low-energy parameterisations or b) the analytical expressions stemming from a hadronic 
model (`ETH model'), which is mainly based on $f_0(500)$- and $\rho(770)$-exchange $t$-channel Feynman graphs, as well as on $s$- and $u$-channel graphs with $N$ and $\Delta(1232)$ intermediate states. During the last two 
decades, used in our PSAs is the Arndt-Roper formula, which enables the controlled (i.e., regulated by the reported - or assigned - normalisation uncertainties) rescaling of the datasets.\\
Regarding the $s$- and $p$-wave $K$-matrix parameterisations, analytically included are the important direct ($s$-channel) contributions from nearby higher baryon resonances (HBRs) in the partial waves $P_{33}$ and $P_{11}$. 
As it does not impose any theoretical constraints onto the data, the analysis with the $K$-matrix parameterisations is suitable for the reliable identification of the outliers in the database (DB) and for the preparation of 
consistent input for further analysis. In contrast, the ETH model obeys the theoretical constraints of crossing symmetry and isospin invariance. Recent enhancement of the analysis software has enabled the inclusion in the ETH 
model not only of the $s$- and $u$-channel graphs with all well-established HBRs with masses below $2$ GeV as intermediate states, but also of the $t$-channel graphs with all scalar-isoscalar and vector-isovector mesons (in 
the same mass range) with known branching fractions to $\pi \pi$ decay modes.\\
After the removal of the outliers using the former method ($K$-matrix parameterisations), the two elastic-scattering (ES) DBs are jointly submitted to an analysis with the ETH model. The optimal values of the model parameters 
and the corresponding Hessian matrices are obtained from fits to the input data, and yield Monte-Carlo predictions for the low-energy constants of the $\pi N$ system, for the $\pi N$ phase shifts, and for the standard low-energy 
$\pi N$ observables. From these fits, we obtained for the charged-pion coupling constant the result: $f_c^2=0.0753(15)$. The combined $\pi^+ p$ and $\pi^- p$ charge-exchange (CX) DBs are also analysed following the same procedure.\\
The essential difference to our PSAs, which had been carried out before 2019, relates to the inclusion in the DB of the two available estimates for the $\pi^- p$ $s$-wave ES length, extracted from measurements of the 
strong-interaction shift of the ground state in pionic hydrogen. Although (in comparison with our former analyses) the addition of these highly precise measurements to the DB does not give rise to pronounced effects (e.g., to 
the identification as outliers of data which had not posed problems in our former PSAs), this modification has nevertheless one important consequence: the long-standing discrepancy, between the value obtained (via the 
extrapolation to $T=0$ of the $\pi^- p$ $s$-wave scattering amplitude of the ETH model) from the PSA of the ES DBs and the results extracted from pionic hydrogen, has been resolved. In spite of this development, other 
discrepancies persist; they comprise: differences in the fitted values of the model parameters between the two categories of fits, namely to the $\pi^\pm p$ ES DBs and to the combined $\pi^+ p$ and $\pi^- p$ CX DBs; a bias in 
the scale factors of the fit to the combined $\pi^+ p$ and $\pi^- p$ CX DBs; and significant energy-dependent effects in the reproduction of the absolute normalisation of the experimental data of the $\pi^- p$ CX reaction on 
the basis of the fitted amplitudes obtained from the $\pi^\pm p$ ES DBs. Assuming the absence of significant systematic effects (e.g., of a systematic underestimation) in the determination of the absolute normalisation of the 
low-energy DB and the negligibility of any residual electromagnetic contributions, one may interpret these discrepancies as strong indication that the isospin invariance is violated in the hadronic part of the $\pi N$ interaction 
at low energy.
\end{abstract}
%13.75.Gx: neutron-pion interactions, nucleon-meson interactions, nucleon-pion interactions, pion-baryon reactions, proton-pion interactions
%25.80.Dj: elastic scattering (meson-induced reactions), elastic scattering (pion-nucleus)
%25.80.Gn: charge-exchange reactions (pion), pion absorption and capture
%11.30.-j: symmetry in theory of fields and particles, conservation laws fields and particles
%{$^*$}{Corresponding author. Electronic address: evangelos (dot) Matsinos (at) sunrise (dot) ch}
\end{frontmatter}

\section{\label{sec:Introduction}Introduction}

The papers under this link represent updates of the analysis of the low-energy pion-nucleon ($\pi N$) measurements, resting upon the use of the Arndt-Roper minimisation function \cite{Arndt1972}, which allows for the controlled 
(regulated by the reported or assigned normalisation uncertainties) rescaling of the datasets. To enable easy access to the history of the development of this research programme and provide updates whenever new results become 
available, the creation of a dedicated web site was at first believed to be an apposite solution; such a web site, hosted by the Paul Scherrer Institut (PSI), was available in the remote past. In fact, there is no better place 
for uploading such material than the preprint archive: the user may easily access not only the latest update, but also the history of the development (both of the theoretical framework used in the description of the experimental 
data, as well as of the techniques employed in the statistical analysis of the data), contained in the earlier versions of the paper. Furthermore, the notification system of arXiv\textregistered~is unbeatable.

The low-energy $\pi N$ database (DB) comprises measurements of the differential cross section (DCS), analysing power (AP), partial-total cross section (PTCS), and total cross section (TCS) for the two elastic-scattering (ES) 
processes ($\pi^\pm p \to \pi^\pm p$) and for the $\pi^- p$ charge-exchange (CX) reaction ($\pi^- p \to \pi^0 n$). In two of the experiments contained in the $\pi^- p$ CX DB, only the first three coefficients in the Legendre 
expansion of the DCS have been reported. Unlike all other analyses of the $\pi N$ measurements, the studies of this programme have (after the mid 1990s) made use of the $\pi^+ p$ PTCSs and TCSs, as well as of the $\pi^- p$ CX TCSs.

Also included in the DB during the last two years are the two $\pi^- p$ $s$-wave scattering lengths $a_{cc}$ and $a_{c0}$ (pertaining to $\pi^- p$ ES and CX, respectively), obtained via the Deser formulae \cite{Deser1954,Trueman1961} 
from measurements of the strong-interaction shift (henceforth, strong shift) $\epsilon_{1 s}$ \cite{Schroeder2001,Hennebach2014} and of the total decay width $\Gamma_{1s}$ \cite{Schroeder2001} of the ground state in pionic 
hydrogen. Prior to the ZRH19 phase-shift analysis (PSA), only the $a_{c0}$ result had been used in the optimisation; given their unprecedented precision (for Pion-Physics standards), the $a_{cc}$ results had been used at former 
times as a means for assessing the consistency of the analysis in terms of the accuracy of the absolute normalisation of the datasets, of the negligibility of residual electromagnetic (EM) contributions in the corrections 
applied to the experimental data, and of the fulfilment of the theoretical constraints of crossing symmetry~\footnote{The scattering amplitudes of the two ES processes are linked via the interchange $s \leftrightarrow u$ in the 
two invariant amplitudes $A_{\pm}(s,t,u)$ and $B_{\pm}(s,t,u)$, where $s$, $t$, and $u$ are the standard Mandelstam variables.} and isospin invariance~\footnote{Assuming that the isospin invariance holds in the $\pi N$ interaction, 
only two (complex) scattering amplitudes enter the description of the three low-energy $\pi N$ reactions: the isospin $I=3/2$ amplitude ($f_3$) and the $I=1/2$ amplitude ($f_1$). Fulfilment of the isospin invariance in the 
hadronic part of the $\pi N$ interaction implies that the $\pi^+ p$ reaction is described by $f_3$, the $\pi^- p$ ES reaction by the linear combination $(2 f_1 + f_3)/3$, and the $\pi^- p$ CX reaction by $\sqrt{2} (f_3-f_1)/3$. 
From these relations, the following expression (known as `triangle identity') links together the three corresponding amplitudes $f_{\pi^+ p}$, $f_{\pi^- p}$, and $f_{\rm CX}$:
\begin{equation*}
f_{\pi^+ p} - f_{\pi^- p} = \sqrt{2} f_{\rm CX} \, \, \, .
\end{equation*}
}.

The backbone of our PSAs of the low-energy $\pi N$ data is a hadronic model which was developed at the ETH (Zurich) during the early 1990s. This model (conveniently named `ETH model' henceforth) is based on four main Feynman 
graphs (henceforth, graphs). Two of these graphs relate to the $t$-channel exchanges of the lowest $I^G(J^{PC})=0^+(0^{++})$ (scalar-isoscalar) and $1^+(1^{--})$ (vector-isovector) mesons, namely of the $f_0(500)$ and of the 
$\rho(770)$. The $s$- and $u$-channel contributions are modelled via $N$ and $\Delta(1232)$ graphs. All remaining (analytical) contributions are small: at former times, they comprised only the $s$- and $u$-channel amplitudes of 
the well-established $s$ and $p$ higher baryon resonances (HBRs) with masses below $2$ GeV. For the sake of consistency, the $t$-channel contributions from all scalar-isoscalar mesons with masses below $2$ GeV and known branching 
fractions to $\pi \pi$ decay modes were recently included: the contributions from the $f_0(980)$ and $f_0(1500)$ were added in the ZRH19 PSA, whereas those from the $f_0(1710)$ are added in this work. Similarly, the $t$-channel 
contributions from the $\rho(1700)$ (i.e., from the only vector-isovector meson with mass between the $\rho(770)$ mass and $2$ GeV, and known branching fraction to the $\pi^+ \pi^-$ decay mode) are added herein. In this form, 
the ETH model may be considered to be complete, as it contains the effects of the exchange of all well-established light unflavoured mesons, as well as $N$ and $\Delta$ baryons below $2$ GeV as intermediate states. Hadronic 
states with higher masses are too distant to have discernible impact on the results of our PSAs of the low-energy $\pi N$ data.

The term `low-energy' implies the restriction of the pion laboratory kinetic energy $T$ below $100$ MeV. After 1994, all PSAs of this programme have been performed exclusively in the energy range from $T=0$ (known as $\pi N$ 
threshold) to $100$ MeV. There are four reasons why our PSAs are restricted to this energy domain.
\begin{itemize}
\item The low-energy DB is extensive enough to enable an exclusive analysis.
\item In its current form, the ETH model is suitable for PSAs at small values of the $4$-momentum transfer $Q^2$. In all probability, the use of the model above the $\Delta(1232)$ resonance would entail the introduction of 
hadronic form factors.
\item It is debatable whether theoretical constraints, which are valid in the region of asymptotic freedom, also hold at low energy. One such constraint is the isospin invariance in the hadronic part of the $\pi N$ interaction~\footnote{In 
the following, `isospin invariance in the $\pi N$ interaction' will be used as the short form of `isospin invariance in the hadronic part of the $\pi N$ interaction'. It is known that the isospin invariance is broken in the EM 
part of the interaction.}. To estimate the dispersion integrals, dispersion-relation analyses rely (by and large) on high-energy data. The analysis of the low-energy measurements in an unbiased way (i.e., without any influence 
from high energy) is not possible in such schemes. It was recently demonstrated that the product of one such partial-wave analysis (PWA) \cite{Arndt2006}, namely the popular WI08 solution (which is still considered to be the 
`current solution' of the SAID Analysis Program), is biased in the low-energy region, in that it does not reflect the behaviour of the bulk of the experimental data \cite{Matsinos2017}.
\item Interest in the low-energy $\pi N$ interaction was maintained for almost three decades by the possibility of extracting predictions for the $\pi N$ $\Sigma$ term using the low-energy phase shifts as input. The extrapolation 
of the $\pi N$ scattering amplitude into the unphysical region is expected to be more reliable when exclusively based on `close-by' measurements, thus avoiding high-energy influences. This possibility was explored and realised 
in Ref.~\cite{Alarcon2012}.
\end{itemize}

The following notation facilitates the repetitive referencing to the DBs of this work.
\begin{itemize}
\item DB$_+$ for the $\pi^+ p$ DB,
\item DB$_-$ for the $\pi^- p$ ES DB,
\item DB$_0$ for the $\pi^- p$ CX DB,
\item DB$_{+/-}$ for the combined DB$_+$ and DB$_-$, and
\item DB$_{+/0}$ for the combined DB$_+$ and DB$_0$.
\end{itemize}
In addition, the prefix `t' (as, for instance, in tDB$_+$) will denote a `truncated' DB, i.e., a DB after the removal of the outliers (i.e., of the measurements in the DB which do not tally with the general behaviour of the 
bulk of the data). Finally, DoF will stand for `degree of freedom' and NDF for `number of DoFs'.

The structure of this paper is as follows.
\begin{itemize}
\item In Section \ref{sec:Modelling}, the two ways of modelling the $s$- and $p$-wave $K$-matrix elements are described. a) The hadronic model (i.e., the ETH model) is isospin-invariant and also incorporates the constraint of 
crossing symmetry. The data description is currently achieved on the basis of seven real parameters. b) The second model employs simple low-energy parameterisations of the $s$- and $p$-wave $K$-matrix elements; to distinguish 
it from the ETH model, it will be referred to as `phenomenological'. As the forms, used in the low-energy parameterisations of the $s$- and $p$-wave $K$-matrix elements, do not impose theoretical constraints onto the data, the 
phenomenological model is suitable for the identification of the outliers in the DB and for assessing the consistency of the data prior to their submission to further analysis (with the ETH model). The data description with the 
phenomenological model involves seven real parameters for each value of the total isospin $I$ (i.e., for $I=1/2, 3/2$). At the end of Section \ref{sec:Modelling}, a few important details are given regarding the minimisation 
function and the quantities which are necessary for quantifying the quality of the reproduction of the datasets.
\item Section \ref{sec:PSA} presents an updated PSA of the low-energy $\pi N$ data. Results for the parameters of the ETH model are given, as well as the corresponding predictions for the low-energy constants of the $\pi N$ 
system and for the phase shifts. A thorough analysis of the scale factors of the experiments contained in the tDBs follows, along with the details of the reproduction of datasets which have not been included in our analysis DBs. 
Finally, the reproduction of the DB$_0$ is investigated on the basis of predictions extracted from the fits of the ETH model to the tDB$_{+/-}$.
\item Section \ref{sec:Causes} speculates on the origin of the discrepancies observed in the analysis of the low-energy tDBs. Three possibilities are addressed. The first is a trivial one, laying the blame for the discrepancies 
on experimental mismatches. The second attributes the effects (at least, in part) to the incompleteness of the EM corrections, applied to the hadronic part of the $\pi N$ scattering amplitude on the way to the evaluation of the 
observables. In Physics terms, the third possibility is the most compelling one. It posits the thesis that the discrepancies point to a departure from the triangle identity, thus attributing the effects to the violation of the 
isospin invariance in the $\pi N$ interaction.
\item The results are discussed in the last section, and our understanding of the dynamics of the $\pi N$ system at low energy is summarised.
\item Our predictions for the ES DCS, obtained from one of the PSAs of this paper, at three pion laboratory momenta, will be given in Appendix \ref{App:AppA} (as well as in the form of an Excel file, uploaded as ancillary 
material). These predictions contain meaningful (i.e., reflecting the statistical and systematic effects of the fitted data) uncertainties, and may be of interest to the members of the MUSE Collaboration, who plan (along with 
their scheduled measurements of the $\mu p$ and $e p$ DCS) to acquire also measurements of the $\pi^\pm p$ ES DCS at PSI.
\end{itemize}

It will be assumed that the physical quantities appearing in this paper, be they model parameters, scattering lengths/volumes, phase shifts, etc., are not purely hadronic; they possibly contain residual EM contributions. Such 
effects are predominantly associated with the use of the physical (instead of the unknown hadronic) masses for the proton, for the neutron, and for the charged and neutral pions (in the hadronic part of the interaction) in the 
determination of the EM corrections. Although part of these effects might have already been captured by the procedure put forward in the determination of the EM corrections in Refs.~\cite{Gashi2001a,Gashi2001b}, it remains 
unknown how large any residual contributions might be. Evidently, the importance of the residual EM contributions needs to be assessed prior to advancing to definite conclusions regarding the level of the isospin breaking in 
the $\pi N$ interaction. At present, one cannot but retain the cautious attitude of considering all hadronic quantities in all analyses of the $\pi N$ data as `EM-modified'~\footnote{In some former works, all such quantities 
were marked by the symbol `\~{}'. In view of the fact that all hadronic quantities obtained from all PSAs/PWAs of the data (be they parameters used in the modelling or predictions derived thereof) are unavoidably affected, there 
is no such need. Owing to the presence of these residual EM effects, there is no purely hadronic quantity in any of the PSAs/PWAs of the $\pi N$ data, not only in those conducted within this programme.}. However, as the repetitive 
use of this term is tedious, it will be omitted.

The proper references to the low-energy $\pi N$ measurements, on which our PSAs are based, may be found in former papers. Only those of the experimental reports, which attract particular attention in parts of this work, will 
be explicitly cited. The masses and the momenta of the particles will be expressed in energy units, mostly in MeV.

\section{\label{sec:Modelling}Modelling of the $s$- and $p$-wave $K$-matrix elements}

\subsection{\label{sec:ETHHistory}The history of the ETH model}

The ETH model is the product of the study of the properties of pionic-atom data of isoscalar nuclei \cite{Goudsmit1991,Goudsmit1992a,Goudsmit1992b} within the framework of the relativistic mean-field theory of the 1980s; the 
principal objective in that investigation was an explanation for the phenomenon of the $s$-wave repulsion in the $\pi$-nucleus interaction. In its original form, the model did not contain any off-shell contributions in the 
$\Delta(1232)$ propagator. Owing to the interest of the ETH(Zurich)-Neuch{\^a}tel-PSI Collaboration to measure $\epsilon_{1 s}$ and $\Gamma_{1s}$ in pionic hydrogen and deuterium, emphasis in the early 1990s was predominantly 
placed on the extraction of predictions for the scattering lengths. The first attempts to account for the energy dependence of the then available phase shifts (up to the $\Delta(1232)$-resonance pole) turned out to be successful 
after the inclusion of the spin-$1/2$ contributions in the $\Delta(1232)$ propagator \cite{Goudsmit1992c,Goudsmit1993a,Goudsmit1993b}.

Of importance in the foundation of the model was Ref.~\cite{Goudsmit1994a}. The analytical contributions from the main graphs of the model to the partial-wave amplitudes are detailed in Appendix A of that paper. Also extracted 
in Ref.~\cite{Goudsmit1994a} were estimates for the model parameters, obtained from fits to three popular (at that time) phase-shift solutions, for the scattering lengths and volumes, and for the $\pi N$ $\Sigma$ term. A 
subsequent paper \cite{Goudsmit1994b} examined the reproduction of the modern (meson-factory) low-energy $\pi N$ measurements: established therein were sizeable discrepancies between the model predictions (derived on the basis 
of the results of Ref.~\cite{Goudsmit1994a}) and most of the modern data.

In order to validate the results obtained with the model, a novel parameterisation of the $s$- and $p$-wave $K$-matrix elements was developed and applied to the $\pi^+ p$ DCSs in Ref.~\cite{Fettes1997}. The method, put forward 
in that paper, became indispensable in subsequent works. On the one hand, the modelling of the hadronic part of the $\pi N$ interaction in terms of such parameterisations is devoid of theoretical constraints, other than the 
expected low-energy behaviour of the $K$-matrix elements. As a result, this approach is suitable for assessing the consistency of the DB and for reliably identifying the outliers. On the other hand, deployed for the first time 
in Ref.~\cite{Fettes1997} was an alternative (in comparison with the earlier works) plan of action, one which all subsequent works adhered to, namely the extraction of the important information from direct fits to the low-energy 
$\pi N$ measurements, rather than to the `fitted' phase-shift results of other works.

After the compatibility of the results obtained with the phenomenological model and those extracted from the fits of the ETH model to the (same) data was confirmed, the investigation turned upon the development of the methodology 
aiming at the extraction of reliable hadronic information from the low-energy $\pi N$ measurements. The first attempts towards this objective were made after the implementation of a robust optimisation scheme 
\cite{Matsinos1997a,Matsinos1997b}; in those works, the isospin invariance in the $\pi N$ interaction was also addressed. However, experience showed that the general lack of familiarity with robust regression techniques in the 
$\pi N$ domain presented an impediment to the dissemination of the important results. Consequently, it was decided (in 1998) that more conventional statistical approaches be followed in subsequent works.

An important step was next taken. The EM corrections (which must be applied to the phase shifts and to the partial-wave amplitudes on the way to the evaluation of the observables) were obtained in an iterative procedure comprising 
two stages: a) fits of the ETH model to the modern data and b) the numerical solution of relativised Schr{\"o}dinger equations containing EM and effective hadronic potentials. A few iteration steps sufficed to achieve convergence 
of the EM corrections. The newly obtained EM corrections \cite{Gashi2001a,Gashi2001b} replaced (in our subsequent PSAs) those of the NORDITA group \cite{Tromborg1976,Tromborg1977,Tromborg1978}. An upgraded PSA of the DB$_{+/-}$ 
with the EM corrections of Refs.~\cite{Gashi2001a,Gashi2001b}, also featuring the use of the Arndt-Roper minimisation function \cite{Arndt1972}, was presented in Ref.~\cite{Matsinos2006}.

A number of subjects were addressed in more recent papers. A new PSA was performed in Ref.~\cite{Matsinos2012}, improving on Ref.~\cite{Matsinos2006} in regard to two issues: a) only one test for the acceptance of each dataset 
was performed in Ref.~\cite{Matsinos2012} (on the basis of the contribution of that dataset to the overall $\chi^2_{\rm min}$) and b) a more stringent acceptance criterion was adopted in the statistical tests, namely the 
p-value~\footnote{The definition of the p-value will be given later on, see Eq.~(\ref{eq:EQ007}).} threshold $\mathrm{p}_{\rm min}$ which is associated with $2.5 \sigma$ effects in the normal distribution. That $\mathrm{p}_{\rm min}$ 
value is approximately equal to $1.24 \cdot 10^{-2}$, i.e., slightly exceeding $1.00 \cdot 10^{-2}$, the threshold regarded by most statisticians as the outset of statistical significance. The isospin invariance in the $\pi N$ 
interaction was revisited in Ref.~\cite{Matsinos2013a}; the results were found compatible with those reported in Refs.~\cite{Matsinos1997a,Matsinos2006}.

In two subsequent papers \cite{Matsinos2013b,Matsinos2015}, exclusive (i.e., without any involvement of the other $\pi N$ measurements) analyses of the extensive~\footnote{The modifier `extensive' applies both to the size of the 
acquired data as well as to the angular domain which they cover.} low-energy $\pi^\pm p$ DCSs of the CHAOS Collaboration \cite{Denz2006} (referred to as `DENZ04' in our DBs) were conducted. The important conclusion in both works 
was that the angular distribution of their $\pi^+ p$ datasets clashes with the one emerging from the rest of the modern DB$_+$ (which had been established as consistent in Refs.~\cite{Matsinos1997a,Matsinos2006,Matsinos2012}). 
Unable to advance an explanation for this mismatch, we decided to refrain from using any part of the CHAOS DCSs in the PSAs of this programme.

Finally, the derivation of the contributions from all graphs of the ETH model to the partial-wave amplitudes and a new evaluation of the $\Sigma$ term appeared in Ref.~\cite{Matsinos2014}, a paper which is as important as 
Ref.~\cite{Goudsmit1994a} in providing insight into the various contributions to the model amplitudes and in enabling the straightforward implementation of the $K$-matrix elements of the ETH model in other works.

\subsection{\label{sec:ETHModel}The physical content of the ETH model}

\begin{figure}
\begin{center}
\includegraphics [width=15.5cm] {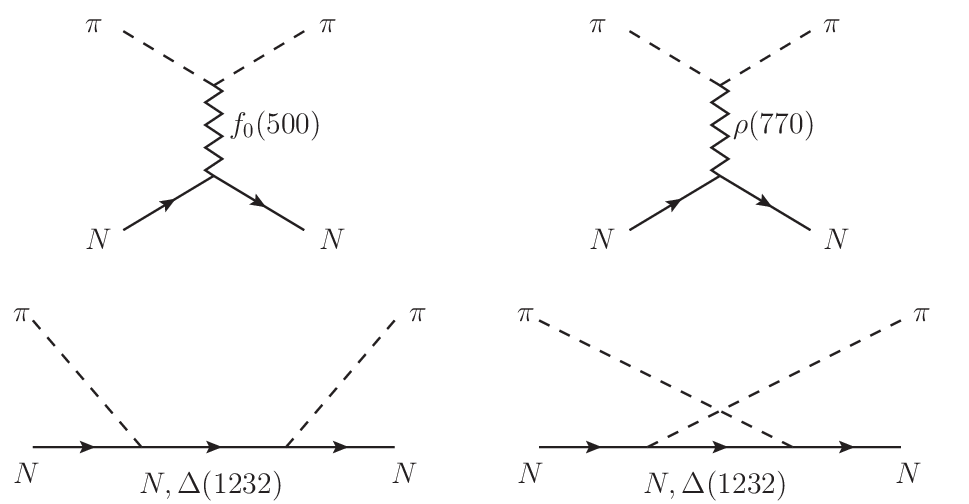}
\caption{\label{fig:FeynmanGraphsETHZ}The main Feynman graphs of the ETH model: scalar-isoscalar ($I=J=0$) and vector-isovector ($I=J=1$) $t$-channel graphs (upper part), and $N$ and $\Delta(1232)$ $s$- and $u$-channel graphs 
(lower part). Not shown in this figure, but also analytically included in the model, are the small contributions from the well-established $s$ and $p$ HBRs with masses below $2$ GeV, as well as those from the $t$-channel 
exchanges of four (three scalar-isoscalar and one vector-isovector) mesons with masses below $2$ GeV and known branching fractions to $\pi \pi$ decay modes, see Section \ref{sec:PSA} for details.}
\vspace{0.35cm}
\end{center}
\end{figure}

As already mentioned in the introduction, the ETH model is mainly based on $f_0(500)$ and $\rho(770)$ $t$-channel exchanges, as well as on $N$ and $\Delta(1232)$ $s$- and $u$-channel contributions (see Fig.~\ref{fig:FeynmanGraphsETHZ}). 
By 1994, also the small contributions to the $s$ and $p$ partial waves from the well-established (four-star) $s$ and $p$ HBRs with masses below $2$ GeV had been analytically included \cite{Goudsmit1994a}. The derivative coupling 
in the $I=J=0$ $t$-channel graph was added in Ref.~\cite{Matsinos1997a} for the sake of completeness. 

Before 2019, the $t$-channel contributions to the partial-wave amplitudes of the ETH model were accounted for by the exchange of one scalar-isoscalar ($I^G(J^{PC})=0^+(0^{++})$ or $I=J=0$) meson (i.e., of the $f_0(500)$, simply 
named $\sigma$-meson in earlier works) and of one vector-isovector ($I^G(J^{PC})=1^+(1^{--})$ or $I=J=1$) meson (i.e., of the $\rho(770)$). On account of consistency, there is no reason to refrain from including in the model 
the $t$-channel exchanges of all scalar-isoscalar and vector-isovector mesons with masses below $2$ GeV and known branching fractions to $\pi \pi$ decay modes, given that the corresponding contributions from the HBRs (in that 
mass range) to the $s$ and $u$ channels have been part of the ETH model for over $25$ years. The current version of the model includes four such graphs, three of which relate to the exchange of scalar-isoscalar mesons and one 
to the exchange of the only vector-isovector meson above the $\rho(770)$ (and below $2$ GeV) with known branching fraction to the $\pi^+ \pi^-$ decay mode (see Section \ref{sec:PSA}). The effort notwithstanding, the impact of 
these modifications on the important results of the analysis is insignificant.

Information on the model parameters may be obtained from several earlier papers, e.g., from Refs.~\cite{Goudsmit1994a,Matsinos2014}. Regarding the $f_0(500)$, the recommendation by the Particle-Data Group (PDG) is to make use 
of a Breit-Wigner mass between $400$ and $550$ MeV, see the properties of the $f_0$($500$) in the recent PDG compilation \cite{PDG2020}. To account for this broad domain, the approach in our recent PSAs was to perform the fits 
of the ETH model at seven (evenly spaced and equally weighted) $m_\sigma$ values between $400$ and $550$ MeV. A new approach will be followed in this work: the fits will be performed at one hundred $m_\sigma$ values, randomly 
selected according to the recent result: $m_\sigma=497^{+28}_{-33}$ MeV \cite{Matsinos2020a}. All uncertainties herein (in the values of the model parameters, in the predictions for the low-energy constants of the $\pi N$ system, 
in the phase shifts, etc.) contain the effects of the $m_\sigma$ variation, as well as (if exceeding $1$) the Birge factor $\sqrt{\chi^2_{\rm min}/{\rm NDF}}$ which takes account of the quality of each fit \cite{Birge1932}.

When a fit to the data is made using all eight parameters of the ETH model, it turns out that there are strong correlations among $G_\sigma$, $G_\rho$, and $x$. To suppress these correlations, one of these three parameters needs 
to be fixed. To this end, the fits of the ETH model have been performed for a long time using $x=0$ (pure pseudovector $\pi N$ coupling). Therefore, each fit of the ETH model (at fixed $m_\sigma$) is achieved by variation of the 
following seven parameters.
\begin{itemize}
\item Scalar-isoscalar $t$-channel graph ($f_0(500)$ exchange): $G_\sigma$ and $\kappa_\sigma$;
\item Vector-isovector $t$-channel graph ($\rho(770)$ exchange): $G_\rho$ and $\kappa_\rho$;
\item $N$ $s$- and $u$-channel graphs: $g_{\pi N N}$; and
\item $\Delta(1232)$ $s$- and $u$-channel graphs: $g_{\pi N \Delta}$ and $Z$.
\end{itemize}
The $s$ and $p$ HBRs do not introduce any additional free parameters \cite{Matsinos2014}; the same applies to the $t$-channel contributions of the $f_0(980)$, $f_0(1500)$, $f_0(1710)$, and $\rho(1700)$, see Section \ref{sec:PSA}.

It needs to be mentioned that the low-energy $\pi N$ data may be fitted to with fewer model parameters. For instance, the coupling constant $g_{\pi N \Delta}$ could be fixed from the decay width of the $\Delta(1232)$ resonance, 
see comments in Section \ref{sec:PSAETHPar}. In addition, the derivative-to-normal coupling (of the scalar-isoscalar mesons $f_0$ to the pion) $\kappa_\sigma$ could be fixed to $0$; since the introduction of this coupling (in 
the mid 1990s), the fitted values of $\kappa_\sigma$ have always come up compatible with $0$. Therefore, the low-energy $\pi N$ data could be fitted to with just five model parameters. Although this possibility might be enquired 
into in the future, we would rather retain the freedom and flexibility of a seven-parameter optimisation at this time.

A number of physical constants need to be fixed in our PSAs. Those of the values, which have been taken from the recent PDG compilation \cite{PDG2020}, are detailed in Table \ref{tab:Constants}; the remaining constants have been 
fixed from Refs.~\cite{Matsinos2020a,Matsinos2020b}. Henceforth, the small $d$ and $f$ waves will be taken from the SAID solution XP15 \cite{XP15}, which is more recent than the WI08 solution \cite{Arndt2006} by the same group; 
the differences between these two solutions are small.

\begin{table}%[h!]
{\bf \caption{\label{tab:Constants}}}The values of the physical constants which have been imported into this work from the recent PDG compilation \cite{PDG2020}.
\vspace{0.2cm}
\begin{center}
\begin{tabular}{|l|c|}
\hline
Physical quantity (unit) & Value\\
\hline
\hline
Inverse of the fine-structure constant $\alpha^{-1}$ & $137.035999084$\\
$\hbar c$ (MeV fm) & $197.3269804$\\
\hline
Electron mass $m_e$ (MeV) & $0.5109989461$\\
Charged-pion mass $m_c$ (MeV) & $139.57039$\\
Neutral-pion mass $m_0$ (MeV) & $134.9768$\\
Proton mass $m_p$ (MeV) & $938.2720813$\\
Neutron mass $m_n$ (MeV) & $939.5654133$\\
\hline
Pion charge radius (fm) & $0.659$\\
Proton magnetic moment $\kappa_p+1$ ($\mu_N$) & $2.79284734462$\\
Pion-decay constant $F_\pi$ (MeV) & $130.20$\\
\hline
\end{tabular}
\end{center}
\vspace{0.5cm}
\end{table}

\subsection{\label{sec:KM}The phenomenological model}

The assumptions in the analyses employing the $K$-matrix parameterisations of this section relate to just two issues: a) the number of terms which one retains from the infinite power series (expansion of the hadronic $K$-matrix 
elements in terms of a suitable variable, e.g., of the pion kinetic energy $\epsilon$ in the center-of-momentum (CM) frame of reference), and b) the forms used in the modelling of the resonant contributions. Experience shows 
that our low-energy parameterisations of the $s$- and $p$-wave $K$-matrix elements successfully capture the dynamics of the $\pi N$ system. A number of tests have been carried out, demonstrating beyond doubt that the outliers 
are flanked by measurements which can successfully be accounted for by the phenomenological model. Therefore, the identification of datapoints as outliers in the fits cannot be attributed to the inadequacy of the parametric 
forms of this section to account for the energy dependence of the phase shifts; consequently, the outliers are suggestive of experimental discrepancies.

In the analysis of the measurements with the phenomenological model, terms up to $\mathcal{O}(\epsilon^2)$ are retained. Owing to the current uncertainties in the measurements, the coefficients of higher orders in the expansion 
of the $K$-matrix elements cannot be determined reliably from the available low-energy $\pi N$ data.

\subsubsection{\label{sec:KMPIP}Fits of the phenomenological model to the low-energy DB$_+$}

For the $\pi^+ p$ reaction, the $s$-wave phase shift is parameterised as
\begin{equation} \label{eq:EQ001}
k \cot \delta_{0+}^{3/2}=(a_{0+}^{3/2})^{-1} + b_3 \epsilon + c_3 \epsilon^2 \, \, \, ,
\end{equation}
where $k$ denotes the magnitude of the CM $3$-momentum. The $p_{1/2}$-wave phase shift is parameterised according to the form
\begin{equation} \label{eq:EQ002}
\tan \delta_{1-}^{3/2}/k = d_{31} \epsilon + e_{31} \epsilon^2 \, \, \, .
\end{equation}

As the $p_{3/2}$ wave contains the $\Delta(1232)$ resonance, a singular (at $W=M_\Delta$) term must be added to the background term, leading to the expression
\begin{align} \label{eq:EQ003}
\tan \delta_{1+}^{3/2}/k = d_{33} \epsilon + e_{33} \epsilon^2 & + \frac{\Gamma_\Delta M_\Delta}{2 k_\Delta^3 (p_{0 \Delta} + m_p)} \frac{(p_0 + m_p) k^2}{W (M_\Delta-W)}\nonumber\\
& + \text{Corresponding contribution from $\Delta(1600)$} \, \, \, ,
\end{align}
where $p_0$ is the proton CM energy and $W=\sqrt{s}$ is the total energy. The quantities $M_\Delta$ and $\Gamma_\Delta$ denote the Breit-Wigner mass and the partial decay width of the $\Delta(1232)$ resonance to $\pi N$ decay 
modes. The quantities $k_\Delta$ and $p_{0 \Delta}$ denote the values of the variables $k$ and $p_0$, respectively, at $W=M_\Delta$. The singular term in Eq.~(\ref{eq:EQ003}) was obtained from Ref.~\cite{Matsinos2014}, see 
$K_{1+}$ in Eqs.~(39) and the corresponding $K^{3/2}_{1+}$ element (after the isospin decomposition of $K_{1+}$ is taken into account), as well as footnote 10 therein. Also added is the contribution of the $\Delta(1600)$, which 
is easy to obtain from Section 3.4 of Ref.~\cite{Matsinos2014} by simply replacing $M_\Delta$ and the partial decay width of the $\Delta(1232)$ resonance with the corresponding quantities of the $\Delta(1600)$ resonance.

\subsubsection{\label{sec:KMPIM}Fits of the phenomenological model to the low-energy DB$_-$ and DB$_0$}

The isospin $I=3/2$ amplitudes, obtained from the final fit of the phenomenological model to the tDB$_+$, are imported into the analysis of the low-energy DB$_-$ and DB$_0$. In this part, another seven parameters (different for 
these two DBs) are introduced, to parameterise the $I=1/2$ amplitudes. The necessary parametric forms are similar to those given by Eqs.~(\ref{eq:EQ001})-(\ref{eq:EQ003}), with parameters: $a_{0+}^{1/2}$, $b_1$, $c_1$, $d_{13}$, 
$e_{13}$, $d_{11}$, and $e_{11}$. It is necessary to explicitly include in the phase shift $\delta_{1-}^{1/2}$ ($P_{11}$) the effects of a $P_{11}$ HBR (known as `Roper resonance'), which may be of importance due to the proximity 
of this resonance's pole to the low-energy region. Given that this contribution is taken into account, the second $P_{11}$ resonance of Table 1 of Ref.~\cite{Matsinos2020b} has also been included in $\delta_{1-}^{1/2}$, 
though its pole is distant for any sizeable impact at low energy.
\begin{align} \label{eq:EQ004}
\tan \delta_{1-}^{1/2}/k &= d_{11} \epsilon + e_{11} \epsilon^2\nonumber\\
 &+ \sum_{i=1}^2 \frac{(\Gamma_R)_i (M_R)_i \left( (p_{0 R})_i + m_p \right)}{2 (k_R^3)_i \left( (M_R)_i+m_p \right)^2} \frac{(W + m_p)^2 k^2}{W \left( (M_R)_i - W \right) (p_0 + m_p)} \, \, \, ,
\end{align}
where $(\Gamma_R)_i$ is the partial decay width of each contributing $P_{11}$ resonance to $\pi N$ decay modes and $(M_R)_i$ is the Breit-Wigner mass of that state. The quantities $k_R$ and $p_{0 R}$ denote the $k$ and $p_0$ 
values at the pole of each resonance ($W=(M_R)_i$). The singular terms in Eq.~(\ref{eq:EQ004}) were obtained from Ref.~\cite{Matsinos2014}, see Section 3.5.1 therein, in particular, Eq.~(54) for $K_{1-}$.

\subsection{\label{sec:MF}Minimisation function}

In our PSAs of the $\pi N$ data, we use the minimisation function which Arndt and Roper introduced in 1972 \cite{Arndt1972}, i.e., the function which the SAID group also employ in their PWAs. The contribution of the $j$-th 
dataset to the overall $\chi^2$ reads as:
\begin{equation} \label{eq:EQ004_5}
\chi^2_j=\sum_{i=1}^{N_j} \left( \frac{y_{ij}^{\rm exp} - z_j y_{ij}^{\rm th}}{\delta y_{ij}^{\rm exp} } \right)^2 + \left( \frac{z_j-1}{\delta z_j} \right)^2 \, \, \, ,
\end{equation}
where $y_{ij}^{\rm exp}$ denotes the $i$-th datapoint of the $j$-th dataset, $y_{ij}^{\rm th}$ the corresponding fitted (`theoretical') value, $\delta y_{ij}^{\rm exp}$ the statistical uncertainty of $y_{ij}^{\rm exp}$, $z_j$ 
a scale factor (applied to the entire dataset), $\delta z_j$ the normalisation uncertainty (reported or assigned), and $N_j$ the number of the accepted (i.e., not identified as outliers) datapoints in the dataset. The fitted 
values $y_{ij}^{\rm th}$ are obtained by means of the parameterised forms of the $s$- and $p$-wave scattering amplitudes of Sections \ref{sec:ETHModel} and \ref{sec:KM}. As the scale factor $z_j$ appears only in $\chi^2_j$, the 
minimisation of the overall $\chi^2 \coloneqq \sum_{j=1}^{N} \chi^2_j$ (where $N$ denotes the number of the accepted datasets in the fit) implies the fixation of $z_j$ from the condition $\partial \chi^2_j / \partial z_j = 0$. 
The unique solution
\begin{equation} \label{eq:EQ005}
z_j = \frac{\sum_{i=1}^{N_j} y_{ij}^{\rm exp} y_{ij}^{\rm th} / (\delta y_{ij}^{\rm exp} )^2 + (\delta z_j)^{-2}} {\sum_{i=1}^{N_j} (y_{ij}^{\rm th} / \delta y_{ij}^{\rm exp})^2 + (\delta z_j)^{-2}}
\end{equation}
leads to
\begin{equation} \label{eq:EQ006}
(\chi^2_j)_{\rm min} = \sum_{i=1}^{N_j} \frac{ (y_{ij}^{\rm exp} - y_{ij}^{\rm th})^2}{(\delta y_{ij}^{\rm exp})^2 } - \frac {\left( \sum_{i=1}^{N_j} (y_{ij}^{\rm exp} - y_{ij}^{\rm th}) y_{ij}^{\rm th} / (\delta y_{ij}^{\rm exp} )^2 \right)^2} 
{ \sum_{i=1}^{N_j}(y_{ij}^{\rm th}/\delta y_{ij}^{\rm exp})^2 + (\delta z_j)^{-2} } \, \, \, .
\end{equation}
The sum of the contributions $\sum_{j=1}^{N} (\chi^2_j)_{\rm min}$ is a function of the parameters entering the modelling of the $s$- and $p$-wave scattering amplitudes. By variation of these parameters, the overall $\chi^2$ 
is minimised, yielding $\chi^2_{\rm min}$.

For the optimisation, the MINUIT package \cite{James} of the CERN library (FORTRAN version) has exclusively been used throughout this programme. Each optimisation is achieved using the sequence: SIMPLEX, MINIMIZE, MIGRAD, and 
MINOS. The calls to the last two methods involve the high-level strategy in the function minimisation.
\begin{itemize}
\item SIMPLEX uses the simplex method of Nelder and Mead \cite{Nelder1965}.
\item MINIMIZE minimises the user-defined function by calling MIGRAD, but reverts to SIMPLEX in case that MIGRAD fails to converge.
\item MIGRAD, undoubtedly the workhorse of the MINUIT software package, is a variable-metric method, based on the Davidon-Fletcher-Powell algorithm. The method checks for the positive-definiteness of the Hessian matrix.
\item MINOS performs a detailed error analysis, separately for each free parameter. Given that it takes into account the non-linearities in the problem, as well as the correlations among the model parameters, MINOS yields 
reliable estimates for the fitted uncertainties.
\end{itemize}
All aforementioned methods admit an optional argument, fixing the maximal number of calls of each particular method; if this limit is reached, the corresponding method is terminated (by MINUIT, internally) regardless of whether 
or not that method converged. To ensure the successful termination of the MINUIT application and the convergence of its methods, the MINUIT output is routinely inspected.

\subsection{\label{sec:Reproduction}Assessment of the quality of the reproduction of a dataset by a given solution}

To quantify the quality of the reproduction of datasets by a reference or baseline solution (BLS), one may follow the methodology put forward in Ref.~\cite{Matsinos2015}; the details are repeated here for the sake of 
self-sufficiency. A few definitions will be given first.
\begin{itemize}
\item A BLS is a set of values and associated uncertainties ($y_{ij}^{\rm th}$, $\delta y_{ij}^{\rm th}$, $i \in \left[ 1,N_j \right]$), corresponding to the values of the kinematical variables, e.g., of $T$ and of the CM 
scattering angle $\theta$ for the DCS and the AP, at which the measurements ($y_{ij}^{\rm exp}$, $\delta y_{ij}^{\rm exp}$, $i \in \left[ 1,N_j \right]$) have been acquired.
\item A BLS comprises predictions obtained by means of a Monte-Carlo (MC) simulation, taking into account the results of the optimisation (the fitted values and the uncertainties of the model parameters, as well as the Hessian 
matrix of each fit) of a PWA of the $\pi N$ data.
\end{itemize}

Being a sum of independent normalised residuals, each following the normal distribution, the test-statistic - to be introduced by Eq.~(\ref{eq:EQ010}) - is expected to follow the $\chi^2$ distribution. As the objective is the 
identification of datasets which are poorly reproduced, the expressions of this section are tailored to \emph{one-sided} tests (right-tail events).

Let the background process, underlying the phenomenon under investigation, be a stochastic one, described by the probability density function $f(x) \geq 0$, where $x \in [0,\infty)$ is a numerical result obtained from a 
measurement conducted on the system. Kolmogorov's second axiom dictates that
\begin{equation*}
\int_{0}^\infty f(x) dx = 1 \, \, \, .
\end{equation*}
The p-value is defined as the upper tail of the corresponding cumulative distribution function
\begin{equation} \label{eq:EQ007}
{\rm p} (x_0) = \int_{x_0}^\infty f(x) dx \, \, \, ;
\end{equation}
therefore, ${\rm p} (x_0)$ represents the probability that a new measurement on the same system (under identical conditions) yield a result $x$ which is more statistically significant than $x_0$ (in this case, the result $x>x_0$). 
Assuming the validity of the null hypothesis, the p-value may therefore be interpreted as the measure of the rarity of the result $x_0$: `small' p-values attest to the statistical significance of the measurements which yielded 
that result.

Evidently, prior to assessing the statistical significance, one must define what `small' means. The fixation of the $\mathrm{p}_{\rm min}$ value, signifying the outset of statistical significance~\footnote{The threshold of 
statistical significance is usually denoted as $\alpha$.}, may be thought of as involving the only subjective decision in Statistics. In practice, the fixation of $\mathrm{p}_{\rm min}$ rests upon a delicate trade-off between 
two risks: a) of accepting the alternative hypothesis (of an effect not being due to statistical contrivance) when it is false and b) of rejecting the alternative hypothesis when it is true. Of relevance in the fixation of the 
$\mathrm{p}_{\rm min}$ threshold is a decision as to which of these two risks is being assigned greater importance. For instance, if the implications of risk (b) are deemed to be more serious compared with those of risk (a), 
an increase in the $\mathrm{p}_{\rm min}$ value is tenable.

Most statisticians accept $\mathrm{p}_{\rm min}=1.00 \cdot 10^{-2}$ as the outset of statistical significance (and $\mathrm{p}_{\rm min}=5.00 \cdot 10^{-2}$ as the threshold signifying probable statistical significance). An 
interesting 2013 paper \cite{Johnson2013} interprets the lack of reproducibility of the results in various scientific disciplines as evidence that the currently accepted $\mathrm{p}_{\rm min}$ values are rather `optimistic'. 
To compensate, the author recommends the reduction of the established thresholds by one order of magnitude~\footnote{Although Ref.~\cite{Johnson2013} states that ``nonreproducibility in scientific studies can be attributed to 
a number of factors, including poor research designs, flawed statistical analyses, and scientific misconduct,'' it is more likely that, at least regarding the $\pi N$ domain, the main reason is `excessive optimism' when 
assessing the systematic effects in the experiments. In short, it is likely that these uncertainties are underestimated (on average).}. Although that paper stimulated a lively debate, in particular in 2014, it is unlikely that 
the currently accepted thresholds of statistical significance will be revised any time soon. For the time being, the $2.5 \sigma$ threshold (our default $\mathrm{p}_{\rm min}$ threshold since 2012) will continue to mark the 
outset of statistical significance in this programme.

The probability density function of the $\chi^2$ distribution with $\nu>0$ DoFs reads as:
\begin{equation} \label{eq:EQ008}
f(x,\nu) = \begin{cases} \frac{1}{2^{\nu/2} \Gamma(\nu/2)} x^{\nu/2-1} \exp(-x/2), & \mbox{for } x > 0\\ 0 , & \mbox{otherwise} \end{cases}
\end{equation}
where $\Gamma(y)$ is the gamma function
\begin{equation*}
\Gamma (y) = \int_{0}^\infty t^{y-1} \exp(-t) dt \, \, \, .
\end{equation*}
For a quantity $x$ following the $\chi^2$ distribution, the expectation value $E[x]$ is simply equal to $\nu$ and the variance $E[x^2]-(E[x])^2$ is equal to $2 \nu$. The relation $E[x]=\nu$ has led most physicists to the use 
of the reduced $\chi^2$ value (i.e., of the ratio $\chi^2/\nu$) as a measure of the quality of the data description in the modelling or in the reproduction of measurements; provided that $\chi^2/\nu \approx 1$, the outcomes of 
tests are claimed to be satisfactory. At this point, the following two remarks need to be made.
\begin{itemize}
\item The statistical hypothesis testing formally relies on the p-value. The use of the reduced $\chi^2$ to quantify the statistical significance is an approximate `rule of thumb', an informal one which, moreover, is frequently 
misleading (e.g., for small $\nu$ values).
\item The interesting issue in the statistical hypothesis testing relates to the value of the $\chi^2/\nu$ at which the results start to become \emph{un}satisfactory. Evidently, a threshold value for $\chi^2/\nu$ may be extracted 
from $\mathrm{p}_{\rm min}$, yet it is $\nu$-dependent, hence cumbersome to use.
\end{itemize}
Evidently, such a departure from simplicity is counterproductive. To assess the statistical significance of a result, one simply needs to compare the corresponding p-value, associated with the estimated $\chi^2$ for $\nu$ DoFs, 
with $\mathrm{p}_{\rm min}$. This is achieved by simply inserting $f(x,\nu)$ of Eq.~(\ref{eq:EQ008}) into Eq.~(\ref{eq:EQ007}), along with $x_0=\chi^2$, and evaluating the integral. Several software implementations of dedicated 
algorithms are available, e.g., see Refs.~\cite{Abramowitz1972} (Chapter on `Gamma Function and Related Functions') and \cite{Press2007}, the routine PROB of the FORTRAN implementation of the CERN software library (which, unlike 
most other CERNLIB routines, is available only in single-precision floating-point format), the functions CHIDIST/CHISQ.DIST.RT of Microsoft Excel, etc.

Various definitions of the `dataset' have been in use, involving different choices of the experimental conditions which must remain stable/constant during the data-acquisition session. The properties of the incident beam and 
the (physical, geometrical) properties of the target were used in the past in order to distinguish the datasets of experiments performed at one place over a short period of time. However, datasets have appeared in experimental 
reports relevant to the $\pi N$ interaction, which not only involved different beam energies, but also contained measurements of different reactions (mixing $\pi^+ p$ and $\pi^- p$ ES measurements). As a result, the prerequisite 
for accepting datapoints as comprising one dataset is that they share the same measurement of the absolute normalisation (and, consequently, normalisation uncertainty $\delta z_j$). Of course, this is a necessary, not a 
sufficient, condition. Regarding the acceptance of a set of measurements as comprising one dataset, the decision cannot be made without an investigation of the stability of the experimental conditions at which the measurements 
had been acquired (which `outsiders' can hardly assess), as well as of their (on/off-line) processing on the way to the extraction of the final experimental results.

It is time we entered the details of the reproduction of the datasets. Described in the remaining part of this section are tests of a dataset's overall reproduction by the BLS, as well as of the reproductions of its shape and 
absolute normalisation. It is assumed that the absolute normalisation of the dataset is known to a relative uncertainty $\delta z_j$ and that none of the important quantities, appearing in the denominators of the expressions 
below, vanishes.

As mentioned in the beginning of the section, the methodology for assessing the quality of the reproduction of the dataset was put forward in Ref.~\cite{Matsinos2015}; it involves the evaluation of the amount of (the controlled) 
rescaling, to be applied to the BLS ($y_{ij}^{\rm th}$, $\delta y_{ij}^{\rm th}$, $i \in \left[ 1,N_j \right]$) in order that it `best' accounts for the dataset ($y_{ij}^{\rm exp}$, $\delta y_{ij}^{\rm exp}$, $i \in \left[ 1,N_j \right]$).

First, the ratios $r_{ij}=y_{ij}^{\rm exp}/y_{ij}^{\rm th}$ are evaluated. If the quantities $y_{ij}^{\rm exp}$ and $y_{ij}^{\rm th}$ are independent (which is certainly true herein, given that the tested datasets have not been 
used in the determination of the BLS), the uncertainties $\delta r_{ij}$ are obtained via the application of Gauss' error-propagation formula:
\begin{equation} \label{eq:EQ009}
\delta r_{ij}=r_{ij} \sqrt{\left( \frac{\delta y_{ij}^{\rm exp}}{y_{ij}^{\rm exp}} \right)^2 + \left( \frac{\delta y_{ij}^{\rm th}}{y_{ij}^{\rm th}} \right)^2} \, \, \, .
\end{equation}

The quality of the reproduction is assessed on the basis of the function $\chi^2_j (z_j)$ defined as
\begin{equation} \label{eq:EQ010}
\chi^2_j (z_j) = \sum_{i=1}^{N_j} \left( \frac{r_{ij} - z_j}{\delta r_{ij}} \right)^2 + \left( \frac{z_j - 1}{\delta z_j} \right)^2 \, \, \, .
\end{equation}
It will be convenient to introduce the weights $w_{ij}$ via the relation $w_{ij} = (\delta r_{ij})^{-2}$.

The second term on the right-hand side of Eq.~(\ref{eq:EQ010}) takes account of the rescaling of the BLS. This contribution depends on how well the absolute normalisation of the dataset is known: if it is poorly known, $\delta z_j$ 
is large and the resulting contribution from the rescaling of the dataset is small; the opposite is true for a well-known absolute normalisation. Evidently, the `best' reproduction of the dataset is achieved when, by varying 
$z_j$, the function $\chi^2_j (z_j)$ is minimised, resulting in the condition
\begin{equation*}
\frac{\partial \chi^2_j (z_j)}{\partial z_j} = 0 \, \, \, .
\end{equation*}
The solution of this equation is
\begin{equation} \label{eq:EQ011}
z_j = \frac{\sum_{i=1}^{N_j} w_{ij} r_{ij} + (\delta z_j)^{-2}}{\sum_{i=1}^{N_j} w_{ij} + (\delta z_j)^{-2}} \, \, \, .
\end{equation}
Inserting this expression for $z_j$ into Eq.~(\ref{eq:EQ010}), one obtains
\begin{align} \label{eq:EQ012}
(\chi^2_j)_{\rm min} = \left( \sum_{i=1}^{N_j} w_{ij} + (\delta z_j)^{-2} \right)^{-1} \Bigg( & \sum_{i=1}^{N_j} w_{ij} \sum_{i=1}^{N_j} w_{ij} r_{ij}^2 - \big( \sum_{i=1}^{N_j} w_{ij} r_{ij} \big)^2 \nonumber\\
& + (\delta z_j)^{-2} \sum_{i=1}^{N_j} w_{ij} (r_{ij} - 1)^2 \Bigg) \, \, \, .
\end{align}

Expression (\ref{eq:EQ012}) yields the minimal $\chi^2$ value for the reproduction of the dataset, containing $N_j$ datapoints. In fact, one additional measurement had been made on that dataset, namely the one fixing its absolute 
normalisation, which is known with relative uncertainty $\delta z_j$. Therefore, the NDF for this dataset is equal to $N_j + 1 - 1 = N_j$; the subtraction of one unit is due to the use of Eq.~(\ref{eq:EQ011}) as a constraint, 
fixing the value of the scale factor $z_j$. Therefore, the quantity $(\chi^2_j)_{\rm min}$ of Eq.~(\ref{eq:EQ012}) is expected to follow the $\chi^2$ distribution with $\nu = N_j$ DoFs. To obtain the p-value of the overall 
reproduction of the dataset, one uses Eq.~(\ref{eq:EQ007}) with $f(x)=f(x,\nu)$ of Eq.~(\ref{eq:EQ008}), along with $x_0=(\chi^2_j)_{\rm min}$ and $\nu=N_j$.

Two additional tests on each dataset are possible. These tests are helpful when the overall reproduction of a dataset is poor; they may determine whether the deficient reproduction is to be blamed on the shape or on the absolute 
normalisation of the dataset.
\begin{itemize}
\item To examine the shape of the dataset (with respect to that of the BLS), one needs to allow the BLS to reproduce the dataset regardless of the rescaling contribution in Eq.~(\ref{eq:EQ010}). This is equivalent to setting 
$\delta z_j \to \infty$ or $(\delta z_j)^{-2}=0$ in Eqs.~(\ref{eq:EQ011}) and (\ref{eq:EQ012}). The corresponding quantities will be denoted as $\hat{z}_j$ (optimal scale factor) and $(\chi^2_j)_{\rm stat}$, respectively. The 
quantity $(\chi^2_j)_{\rm stat}$ represents the fluctuation in the dataset which (assuming the correctness of the shape of the BLS) is of pure statistical nature.
\begin{equation} \label{eq:EQ013}
\hat{z}_j = \frac{\sum_{i=1}^{N_j} w_{ij} r_{ij}}{\sum_{i=1}^{N_j} w_{ij}}
\end{equation}
\begin{equation} \label{eq:EQ014}
(\chi^2_j)_{\rm stat} = \left( \sum_{i=1}^{N_j} w_{ij} \right)^{-1} \left( \sum_{i=1}^{N_j} w_{ij} \sum_{i=1}^{N_j} w_{ij} r_{ij}^2 - \left( \sum_{i=1}^{N_j} w_{ij} r_{ij} \right)^2 \right)
\end{equation}
As expected, both expressions are identical to those derived for the weighted average of a set of independent measurements and for the corresponding $\chi^2$ value for constancy. Owing to the fact that the normalisation 
uncertainty is not used in Eq.~(\ref{eq:EQ014}), the quantity $(\chi^2_j)_{\rm stat}$ is expected to follow the $\chi^2$ distribution with $\nu = N_j - 1$ DoFs. The p-value, obtained from Eq.~(\ref{eq:EQ007}) with 
$x_0=(\chi^2_j)_{\rm stat}$ and $\nu = N_j - 1$, may be used in order to assess the constancy of the values $r_{ij}$ or, equivalently in this case, to examine the shape of the dataset with respect to the BLS~\footnote{In reality, 
the test simply assesses how well the datapoints of the set are represented by their average value. A failure suggests either a bad shape (e.g., a slope being present in the data) or `scattered' input values with small 
uncertainties.}.
\item To assess the compatibility of the absolute normalisations of the dataset and of the BLS, one first estimates the rescaling contribution to $(\chi^2_j)_{\rm min}$ via the relation
\begin{equation} \label{eq:EQ015}
(\chi^2_j)_{\rm sc} \coloneqq (\chi^2_j)_{\rm min}-(\chi^2_j)_{\rm stat}=\frac{(\delta z_j)^{-2} \left( \sum_{i=1}^{N_j} w_{ij} (r_{ij}-1) \right) ^2}{\left( \sum_{i=1}^{N_j} w_{ij} + (\delta z_j)^{-2} \right) \sum_{i=1}^{N_j} w_{ij}} \, \, \, ,
\end{equation}
where use of Eqs.~(\ref{eq:EQ012}) and (\ref{eq:EQ014}) has been made. The quantity $(\chi^2_j)_{\rm sc}$ is expected to follow the $\chi^2$ distribution with $1$ DoF (which, of course, is the normal distribution).
\end{itemize}

To summarise, the following tests of the quality of the reproduction of the dataset by a BLS may be carried out.
\begin{itemize}
\item The overall reproduction is tested using $(\chi^2_j)_{\rm min}$ of Eq.~(\ref{eq:EQ012}) as $x_0$ in Eq.~(\ref{eq:EQ007}) and $\nu = N_j$ DoFs. If this test fails (i.e., if it returns a p-value below $\mathrm{p}_{\rm min}$), 
the next two tests may point to the source of the deficiency.
\item The shape (statistical fluctuation) is tested using $(\chi^2_j)_{\rm stat}$ of Eq.~(\ref{eq:EQ014}) as $x_0$ in Eq.~(\ref{eq:EQ007}) and $\nu=N_j-1$ DoFs.
\item The absolute normalisation is tested using $(\chi^2_j)_{\rm sc}$ of Eq.~(\ref{eq:EQ015}) as $x_0$ in Eq.~(\ref{eq:EQ007}) and $\nu=1$ DoF.
\end{itemize}

Evidently, the \emph{only} subjective aspect in the tests outlined in this section concerns the choice of the $\mathrm{p}_{\rm min}$ value which is taken to signify the outset of statistical significance.

\section{\label{sec:PSA}The 2020 PSAs}

The solution, presented in this paper, will be referred to in the future as `ZRH20'.

\subsection{\label{sec:MainDifference}On the inclusion in the DB$_-$ of the $a_{cc}$ values extracted from pionic hydrogen}

The essential difference of our last two PSAs (i.e., ZRH19 and ZRH20) to former works relates to the inclusion in the DB$_-$ of the two $a_{cc}$ results, extracted (via the first of the Deser formulae \cite{Deser1954,Trueman1961}) 
from PSI measurements of $\epsilon_{1 s}$ in pionic hydrogen \cite{Schroeder2001,Hennebach2014}. As mentioned in the introduction, these values had been used until 2019 for assessing the consistency of the analysis; given their 
unprecedented precision (for Pion-Physics standards), it appeared to us that such a role would maximise the potential of the measurements. However, the comparison between these results and the ones obtained via the extrapolation 
of the $\pi^- p$ scattering amplitude to threshold gave rise to a persisting discrepancy, see Refs.~\cite{Matsinos2006,Matsinos2012,Matsinos2013b,Oades2007}. Arguments were put forward (e.g., see end of Section 3.2.3 in the 
ZRH17 report) for a possible explanation for this discrepancy in terms of a mismatch in the absolute normalisation between the `experimentally obtained' $a_{cc}$ values (converted into relevant cross sections) and the five 
$\pi^- p$ ES DCS datasets \cite{Frank1983,Brack1990,Joram1995a,Joram1995b}, representing the closest (to threshold) DCS measurements in the DB$_-$.

In late 2018, we asked ourselves whether the inclusion of the two $a_{cc}$ results from pionic hydrogen in the DB$_-$ would necessitate the removal of a considerable amount of other data when formally applying our rejection 
criteria to the DB$_-$ enhanced with these two datapoints. This subject was explored in ZRH19 and will further be examined herein. It will be demonstrated (again) that the inclusion of the two $a_{cc}$ values in the input DB$_-$ 
does not lead to an appreciable deterioration of the fits; the optimisation simply moves away from the established (in our PSAs before 2019) minimum, towards one with enhanced isoscalar components in the low-energy $\pi N$ 
scattering amplitude. As a) the fits do not fail and b) there is no reason to doubt the correctness of the PSI measurements of $\epsilon_{1 s}$, we came to the decision in 2018 to retain both $a_{cc}$ results in the DB$_-$.

Corrected for the EM effects according to Ref.~\cite{Oades2007} (and using the values of the physical constants as reported herein for the main corrections of Ref.~\cite{Schlesser2011}), the $a_{cc}$ result of Ref.~\cite{Schroeder2001} 
would be equal to $0.08556(16)(70)~m_c^{-1}$ and the one of Ref.~\cite{Hennebach2014} to $0.08557(9)(57)~m_c^{-1}$, where the first uncertainty is statistical (taking account of the statistical uncertainty of the experimental 
result) and the second systematic (reflecting all `external' uncertainties, i.e., the ones contained in the systematic uncertainty of the experimental result, as well as those relating to the EM corrections of Ref.~\cite{Oades2007}), 
see Appendix A of Ref.~\cite{Matsinos2019a} for some details.

One word of caution is in order. The inevitable consequence of the inclusion of highly precise measurements in a DB, comprising data several times less precise, is that the optimisation, resting upon the use of $\chi^2$-based 
minimisation functions (as is the one of Eq.~(\ref{eq:EQ004_5})), will tend to `gravitate' towards the highly precise measurements, at the expense of a deterioration in the description of the other data: such measurements become 
`anchor points' in the optimisation. Provided that such precise results are also correct, there is no doubt that their inclusion in the DB is highly favourable, as it improves the accuracy of the fitted values of the model 
parameters, as well as of any predictions extracted thereof. On the other hand, if such measurements - whichever the reason~\footnote{Who would have suspected, slightly over one decade ago, that the proton size would come out 
significantly different from measurements of the Lamb shift in muonic hydrogen and from extrapolations to $Q^2=0$ of the $e p$ scattering amplitude (for a discussion, see Section $3$ of Ref.~\cite{Matsinos2020c} and the 
references therein)? However, this discrepancy, the so-called `proton-radius puzzle', still persists and is multifaceted. One issue is to determine whether the root-mean-square (rms) electric-charge radius of the proton is `small' 
(i.e., close to the results from muonic hydrogen and deuterium) or `large' (i.e., compatible with most of the earlier determinations using $e p$ elastic-scattering data). Another, perhaps more important, issue is to pinpoint 
the source of the discrepancy between the two sets of values. Does Perturbative QED fail in bound systems? Does the proton size depend on the generation of the lepton probing it? Is this discrepancy a consequence of the 
comparison between measurements taken at threshold (particles nearly `at rest') and results obtained on the basis of measurements taken at finite (non-zero) energy? In any case, it is not inconceivable that a similar effect 
could enter the low-energy $\pi N$ Physics, thus invalidating the comparisons between quantities obtained at the $\pi N$ threshold and extrapolations to threshold of the $\pi N$ scattering amplitudes.} - turn out to be erroneous 
or if quantities extracted at threshold should not be compared with extrapolated results to threshold, then any solutions, obtained on their basis in works which do not make use of robust optimisation techniques, are bound to 
be equally flawed. In short, the sensitivity of a non-robust analysis to the treatment of the anchor points is dangerously high.

One might argue that we simply used the results as reported by all experimental groups, accepting the uncertainties which they chose to assign to their measurements. Although this may be true, one could also argue that it is 
the responsibility of the analyst to safeguard against the possibility of faulty input, in particular in a domain which is replete with persisting discrepancies and haunting inconsistencies for decades. Although it is frequently 
assumed that the pionic-hydrogen data suffer from fewer problems (in comparison with the measurements of the DCS above threshold), this assumption might be fallacious (for instance, see footnote 6 of Ref.~\cite{Matsinos2019a}). 
It needs to be emphasised that, as in the previous version of the preprint, we have not placed any safeguards against the possibility of erroneous experimental input from pionic hydrogen. Therefore, the accuracy of the solution, 
obtained and presented in this work (as well as in ZRH19), rests upon the reliability of the experimental results for $\epsilon_{1 s}$ of Refs.~\cite{Schroeder2001,Hennebach2014}, as well as of the procedure yielding $a_{cc}$ 
from $\epsilon_{1 s}$.

\subsection{\label{sec:AdditionalDifferences}Similarities and differences to the ZRH19 PSA}

This PSA differs from the previous one (ZRH19) on account of the following points, which - not surprisingly - have a minor only impact on the results.
\begin{itemize}
\item Apart from small changes in some of the values of the other physical constants, new estimates for the charged- and neutral-pion masses appeared in the recent PDG compilation \cite{PDG2020}, see Table \ref{tab:Constants}.
\item The modelling of the $t$-channel scalar-isoscalar contributions involves now also the $f_0(1710)$-exchange graph. Therefore, four contributions in total are now used, corresponding to the exchanges of an $f_0(500)$ (main 
graph), $f_0(980)$, $f_0(1500)$, and $f_0(1710)$.
\item Included now in the modelling of the $t$-channel vector-isovector contributions is the $\rho(1700)$-exchange graph. Therefore, two contributions are currently in place: the main graph, involving the exchange of a $\rho(770)$, 
and the newly added graph, involving the exchange of a $\rho(1700)$.
\item The physical properties of the scalar-isoscalar mesons, used in the modelling of the $t$-channel contributions, are fixed to the central values of Table 1 of Ref.~\cite{Matsinos2020a}. Similarly, those of the vector-isovector 
mesons are fixed to the central values of Table 9 of the same paper. The physical properties of the $N$ HBRs are taken from Table 1 of Ref.~\cite{Matsinos2020b}, whereas those of the $\Delta$ HBRs from Table 2 of the same paper. 
Finally, new estimates for the mass and for the total decay width of the $\Delta(1232)$ resonance, obtained from all available data in Ref.~\cite{Matsinos2020b}, may be found in Appendix A therein; the branching fraction of the 
$\Delta(1232)$, also included in Table A.1 of Ref.~\cite{Matsinos2020b}, represents the recommendation by the PDG (see also the text therein).
\item In our PSAs since 2015, we had been using the proton EM form factors of Ref.~\cite{Venkat2011}, see also Ref.~\cite{Matsinos2020c} for a discussion. To reduce the dependence of this research programme on extraneous sources, 
we will replace those form factors with simple dipole forms \cite{Matsinos2020c}, featuring as parameters the rms electric-charge and magnetic radii of the proton, denoted as $\sqrt{\avg{r^2_E}_p}$ and $\sqrt{\avg{r^2_M}_p}$, 
respectively. Reference \cite{Matsinos2020c} demonstrated that the differences between the dipole forms and more elaborate form-factor schemes are small at low $Q^2$. Of course, one remaining issue, admittedly not a trivial one 
given the persisting discrepancies, is the fixation of the values of these two parameters. This subject will be addressed shortly. For the EM form factor of the pion, a monopole form is used \cite{Matsinos2020c}, after fixing 
the rms charge radius of the pion to the PDG recommendation, see Table \ref{tab:Constants}.
\item The last change relates to the fixation in our PSAs of the small $d$ and $f$ waves. Heretofore, we had made use of the $d$ and $f$ waves of the WI08 solution by the SAID group, i.e., results which were extracted more 
than ten years ago. From now on, we will make use of the more recent XP15 solution \cite{XP15} by the same group.
\end{itemize}

The only parameter in the dipole form of the electric Sachs form factor of the proton $G^p_E$ is the particle's rms electric-charge radius. The discrepancies among the currently available results are discussed in Ref.~\cite{Matsinos2020c} 
(as well as in a plethora of other works, several of which are cited in Ref.~\cite{Matsinos2020c}). To make the long story short, we used as the appropriate $\sqrt{\avg{r^2_E}_p}$ value the weighted average of the two muonic-hydrogen 
results \cite{Pohl2010,Antognini2013}, after combining the reported statistical and systematic uncertainties in quadrature: $\sqrt{\avg{r^2_E}_p} = 0.84112(42)$ fm. Regarding the rms magnetic radius of the proton (which enters 
the parameterisation of the magnetic Sachs form factor $G^p_M$), the available information is equally confusing. We fitted a constant to the values reported in 
Refs.~\cite{Belushkin2007,Borisyuk2010,Bernauer2010,Zhan2011,Lorenz2012,Epstein2014,Lorenz2015,Lee2015} (omitting one of the two $\sqrt{\avg{r^2_M}_p}$ results of Ref.~\cite{Lee2015}, namely the one extracted from an analysis 
of the Mainz data, as we preferred to use instead the original $\sqrt{\avg{r^2_M}_p}$ result of Ref.~\cite{Bernauer2010}). After imposing a restriction on the weight of each individual datapoint~\footnote{All uncertainties, 
which were short of the median uncertainty in the input dataset, were replaced by that median uncertainty.} and using a robust optimisation with Andrews weights, we obtained: $\sqrt{\avg{r^2_M}_p}=0.863^{+0.010}_{-0.011}$ fm.

To enable the investigation of a possible bias in the analysis, all datasets (even those containing one datapoint) must be accompanied by a normalisation uncertainty. As a result, realistic uncertainties must be assigned to all 
datasets with unknown normalisation uncertainty; the alternative would be to exclude all such datasets. Regarding this subject, there have been no changes to the procedure put forward in the ZRH19 PSA. Normalisation uncertainties 
were assigned to $126$ out of a total of $1133$ datapoints contained in the three initial DBs. The relevant datasets are:
\begin{itemize}
\item the $\pi^+ p$ BERTIN76 \cite{Bertin1976} and AULD79 \cite{Auld1979} DCSs: $8$ ($7+1$) datasets;
\item the $\pi^\pm p$ FRIEDMAN90 \cite{Friedman1990} and FRIEDMAN99 \cite{Friedman1999} PTCSs, as well as the CARTER71 \cite{Carter1971} and PEDRONI78 \cite{Pedroni1978} TCSs (named `total-nuclear' in the two papers): $12$ one- 
or two-point datasets;
\item the $\pi^- p$ CX DUCLOS73 \cite{Duclos1973} DCSs: $3$ one-point datasets;
\item the $\pi^- p$ CX SALOMON84 \cite{Salomon1984} DCSs: $2$ sets of the first $3$ coefficients in the Legendre expansion of the measured DCS (the DCSs were not reported); and
\item the $\pi^- p$ CX BUGG71 \cite{Bugg1971} and BREITSCHOPF06 \cite{Breitschopf2006} TCSs: $10$ ($1+9$) one-point datasets.
\end{itemize}

A robust fit to the normalisation uncertainties ($T$ being the independent variable), reported in the modern $\pi^+ p$ DCS experiments, with Huber's objective function~\footnote{The results of the robust fits, using Tukey's 
(bisquare) objective function, were almost identical.} (along with the default value $1.345$ for the tuning constant), yielded the result $\delta z_+ (T)=-0.533429 \cdot 10^{-3} T + 0.068871$, where $T$ is expressed in MeV. The 
$\pi^+ p$ BERTIN76 and AULD79 datasets were assigned the normalisation uncertainty of $2 \, \delta z_+ (T_j)$, $T_j$ (in MeV) being the pion laboratory kinetic energy of the specific dataset~\footnote{The fitted $\delta z_+ (T)$ 
values are somewhat smaller than those used in the ZRH17 solution, which were based on a standard (non-robust) linear least-squares fit (to the same data). The consequence of the use of smaller assigned normalisation uncertainties 
for the BERTIN76 and AULD79 DCSs is a slight increase in the resulting $\chi^2_{\rm min}$ values of the fits to the $\pi^+ p$ measurements in ZRH19 and ZRH20.}. A robust fit (with the same objective function) to the normalisation 
uncertainties, reported in the $\pi^- p$ CX DCS experiments, yielded the nearly flat result $\delta z_0 (T)=-0.024238 \cdot 10^{-3} T + 0.062314$, where $T$ is again expressed in MeV. The $\pi^- p$ CX DUCLOS73 datasets were 
assigned the normalisation uncertainty of $2 \, \delta z_0 (T_j)$. In both cases, the use of generous uncertainties (i.e., double the fitted values at each $T=T_j$) was not meant as retribution for the lack of proper reporting, 
but as a precaution: it is unclear whether due attention was paid to the absolute normalisation of the datasets back in the 1970s and whether the normalisation effects were at those times recognised as potentially important 
sources of uncertainty. The remaining datasets with unknown normalisation uncertainties were assigned uncertainties as follows.
\begin{itemize}
\item All PTCSs and TCSs (FRIEDMAN90, FRIEDMAN99, CARTER71, PEDRONI78, BUGG71) were assigned a normalisation uncertainty of $6~\%$, double the reported uncertainty of the KRISS99 \cite{Kriss1999} TCSs.
\item The SALOMON84 measurements were assigned the normalisation uncertainty of $3.1~\%$, the normalisation uncertainty of the (similar, as well as temporally adjacent) BAGHERI88 \cite{Bagheri1988} experiment. It is likely 
that some normalisation effects are already contained in the SALOMON84 data.
\item The $\pi^- p$ CX BREITSCHOPF06 TCSs were assigned a normalisation uncertainty of $3~\%$; the experimental group had already combined statistical and systematic effects in quadrature and reported only the total uncertainty.
\end{itemize}

Just one reported normalisation uncertainty was replaced: the two SEVIOR89 \cite{Sevior1989} AP datasets at $98$ MeV, close to the upper $T$ limit, were assigned the normalisation uncertainty of $5~\%$, the maximal reported 
normalisation uncertainty in experiments pursuing the measurement of the AP. On p.~2785 of Ref.~\cite{Sevior1989}, one reads: ``The uncertainty in the magnitude of the target polarization was $1.6~\%$.'' However, it is unclear 
from the paper whether the quoted value represents the total normalisation uncertainty in that experiment, and whether or not the uncertainties of the reported AP values (see their Table I) already contain such effects. 
Importantly, the target polarisation, the main source of normalisation uncertainty in the measurements of the AP, has routinely been reported around $3~\%$ by all other experimental groups who measured that quantity, even two 
decades after Ref.~\cite{Sevior1989} appeared. In fact, there is no AP dataset in any of the three DBs with a normalisation uncertainty below $3~\%$. Using the reported normalisation uncertainty of $1.6~\%$ in the SEVIOR89 
datasets appeared to us to be unjust towards all other AP experiments at low energy.

As aforementioned, included now in the graphs of the ETH model are the $t$-channel contributions from the exchange of three $f_0$ resonances with masses between the $f_0(500)$-meson mass and $2$ GeV. In fact, four such mesons 
are found in the recent PDG compilation \cite{PDG2020}, but the branching fraction to $\pi \pi$ decay modes for one of them (namely, for the $f_0(1370)$) is unknown, see Ref.~\cite{Matsinos2020a} for details. The contributions 
of the other three mesons may easily be included using the relations of Section 3.1 of Ref.~\cite{Matsinos2014}. The Fermi-like coupling $G_{f_0}$ for each of these exchanges can be related to the model parameter $G_\sigma$, 
entering the $f_0(500)$ contributions:
\begin{equation} \label{eq:EQ016}
G_{f_0} = G_\sigma \frac{m_\sigma}{m_{f_0}} \sqrt{\frac{\Gamma_{f_0 \to \pi \pi}}{\Gamma_{\sigma \to \pi \pi}} \sqrt{\frac{m_\sigma^2-4 m_c^2}{m_{f_0}^2-4 m_c^2}}} \, \, \, ,
\end{equation}
which is obtained from the relation between the partial decay width $\Gamma_{f_0 \to \pi \pi}$ of each $f_0$ meson \cite{Narison2001} and the $\pi f_0$ coupling constant $g_{\pi \pi f_0}$:
\begin{equation*}
\Gamma_{f_0 \to \pi \pi}=\frac{g_{\pi \pi f_0}^2}{16 \pi} \left( \frac{m_c}{m_{f_0}} \right)^2 \sqrt{m_{f_0}^2-4 m_c^2} \, \, \, .
\end{equation*}
Equation (\ref{eq:EQ016}) is valid when all $f_0$ mesons couple to the nucleon with the same strength (with the same coupling constant $g_{f_0 N N}$). This is an approximation, but (in view of the smallness of the contributions 
of these states in comparison with those of the $f_0(500)$-exchange $t$-channel graph) its impact on the results is expected to be insignificant~\footnote{The ratios $G_{f_0}/G_\sigma$ decrease rapidly with increasing $f_0$ mass, 
and come out equal to about $0.109$, $0.052$, and $0.026$ for the $f_0(980)$, $f_0(1500)$, and $f_0(1710)$, respectively.}. In addition, all such states are assumed to share the $\kappa_\sigma$ value (ratio of the 
derivative-to-normal couplings to the pion) of the $f_0(500)$-exchange graph.

Regarding the $\rho(1700)$-exchange $t$-channel contributions, a relation between the $\rho(1700)$ Fermi-like coupling $G_{\rho^*}$ and the model parameter $G_\rho$ pertaining to the $\rho(770)$-meson graph, 
similar to the one given in Eq.~(\ref{eq:EQ016}), may be obtained:
\begin{equation} \label{eq:EQ016_2}
G_{\rho^*} = G_\rho \frac{m_\rho}{m_{\rho^*}} \sqrt{\frac{\Gamma_{\rho^* \to \pi^+ \pi^-}}{\Gamma_{\rho \to \pi^+ \pi^-}} \left( \frac{m^2_\rho - 4 m^2_c}{m^2_{\rho^*} - 4 m^2_c} \right)^{3/2}} \, \, \, ;
\end{equation}
the assumption again is that all $\rho$ mesons couple to the nucleon identically. The relation between the partial decay width $\Gamma_{\rho^*}$ of each $\rho^*$ meson (to the $\pi^+ \pi^-$ decay mode) and the $\pi \rho^*$ 
coupling constant $g_{\pi \pi \rho^*}$ may be found in Ref.~\cite{Hoehler1983}, p.~565 (Chapter 8.1.3, Eq.~(A.8.29)).

The last difference to the previous PSA relates to the treatment of the $f_0(500)$-meson mass (model parameter $m_\sigma$). To account for the $m_\sigma$ uncertainty, the PDG recommended in 2012 (and still do) the broad domain 
between $400$ and $550$ MeV. As a result, the fits involving the ETH model after 2014 were performed after fixing $m_\sigma$ to seven equidistant values in that domain. However, the PDG recommendation is marked as ``ESTIMATE,'' 
which they define as: ``Based on the observed range of the data. Not from a formal statistical procedure.'' It was recently discovered that the appropriate analysis of the entire dataset of the available $m_\sigma$ results may 
lead to a domain which is considerably narrower than the one recommended by the PDG; for instance, the estimate $m_\sigma=497^{+28}_{-33}$ MeV was obtained in Ref.~\cite{Matsinos2020a} on the basis of a robust procedure. To 
account for the variability of $m_\sigma$ from now on, the fits will be performed at a number of $m_\sigma$ values, randomly selected according to the result of Ref.~\cite{Matsinos2020a}: one hundred values have been generated 
for the purposes of this work~\footnote{The first version of this PSA was carried out using only ten simulated $m_\sigma$ values. To have confidence in the results, we examined their sensitivity to the number of simulated 
$m_\sigma$ values and thus increased that number by one order of magnitude. We found out that the average results (for the model parameters of Table \ref{tab:PSAPar}, for the phase shifts of Tables \ref{tab:PSAPhSh} and 
\ref{tab:PSAPhSha}, for the low-energy constants of Table \ref{tab:PSACnts}, and for the MC predictions of Section \ref{sec:PSAReproductionCHAOS}) were practically unaffected, but their associated uncertainties came out a few 
percent larger for the larger set of the $m_\sigma$ values. We thus decided to retain and report the results originating from the larger set.\label{ftn:FTN1}}.

\subsection{\label{sec:PSAKM}Fits of the phenomenological model}

The first fit to the DB$_+$ of $459$ datapoints resulted in $\chi^2_{\rm min} \approx 926.9$ (for $452$ DoFs, as the fit involves seven parameters). Following the procedure of eliminating one DoF per iteration step, the one with 
the largest contribution to $(\chi^2_j)_{\rm min}$ of the dataset with the lowest p-value (provided that that value did not exceed $\mathrm{p}_{\rm min}$), we obtained the tDB$_+$ with $\chi^2_{\rm min} \approx 546.8$ for $414$ 
DoFs (see Table \ref{tab:ProgressPIP}). The exclusion of entire datasets, on account of the number of outliers which they contain, has been detailed in Ref.~\cite{Matsinos2017}, p.~7. Three datasets, identified as problematic 
already in 1997 \cite{Fettes1997}, stick out of the DB$_+$ in a dramatic manner: the BRACK90 dataset at $66.80$ MeV (with eleven datapoints), the BERTIN76 dataset at $67.40$ MeV (with ten datapoints), and the JORAM95 dataset at 
$32.70$ MeV (with seven datapoints). Of the remaining ten outliers, two relate to the absolute normalisation: our procedure suggested that two of the four BRACK86 datasets, which are accompanied by suspiciously small normalisation 
uncertainties ($1.2$ and $1.4~\%$, respectively), be freely floated. In summary, the elimination of $38$ DoFs results in the reduction of the $\chi^2_{\rm min}$ by about $380.0$, i.e., by about $10.0$ per removed DoF. The tDB$_+$ 
is detailed in Table \ref{tab:DBPIP}. Apart from some experimental details, the table also contains the contribution $(\chi^2_j)_{\rm min}$ of each dataset to $\chi^2_{\rm min}$, the p-value associated with the quality of the 
description of each dataset in the final fit to the tDB$_+$, and the scale factor $z_j$ obtained from Eq.~(\ref{eq:EQ005}).

\begin{table}%[h!]
{\bf \caption{\label{tab:ProgressPIP}}}The results of the procedure of eliminating one DoF per iteration step when processing the DB$_+$. The quantities $T$ and $\theta$ denote the pion laboratory kinetic energy and the CM 
scattering angle, respectively. If the $\theta$ entry is missing, then the action applies to the entire dataset. The result `flagged' implies removal (of the datapoint, of the dataset, or of the absolute normalisation of the 
specific dataset) at the subsequent iteration step.
\vspace{0.2cm}
\begin{center}
\begin{tabular}{|c|c|c|c|c|}
\hline
$\chi^2_{\rm min}/{\rm NDF}$ & Identifier & $T$ (MeV) & $\theta$ (deg) & Result\\
\hline
\hline
$926.9/452$ & BERTIN76 & $67.40$ & $150.69$ & flagged\\
$890.5/451$ & BRACK90 & $66.80$ & $147.00$ & flagged\\
$866.8/450$ & BERTIN76 & $67.40$ & $133.31$ & flagged\\
$836.1/449$ & BRACK90 & $66.80$ & $47.60$ & flagged\\
$816.1/448$ & JORAM95 & $32.70$ & $131.28$ & flagged\\
$797.2/447$ & BRACK90 & $66.80$ & $59.00$ & flagged\\
 & BRACK90 & $66.80$ & & flagged\\
$741.2/438$ & JORAM95 & $44.60$ & $30.74$ & flagged\\
$727.5/437$ & BRACK86 & $66.80$ & & absolute normalisation flagged\\
$694.2/436$ & BERTIN76 & $67.40$ & $142.09$ & flagged\\
 & BERTIN76 & $67.40$ & & flagged\\
$656.3/428$ & JORAM95 & $32.70$ & $52.19$ & flagged\\
$646.7/427$ & JORAM95 & $44.60$ & $35.40$ & flagged\\
$630.9/426$ & JORAM95 & $32.20$ & $37.40$ & flagged\\
$613.4/425$ & JORAM95 & $32.70$ & $74.16$ & flagged\\
 & JORAM95 & $32.70$ & & flagged\\
$593.9/420$ & JORAM95 & $45.10$ & $124.42$ & flagged\\
$586.7/419$ & BERTIN76 & $39.50$ & $75.05$ & flagged\\
$582.1/418$ & BRACK86 & $86.80$ & & absolute normalisation flagged\\
$568.6/417$ & BERTIN76 & $39.50$ & $85.81$ & flagged\\
$563.0/416$ & BERTIN76 & $95.90$ & $65.67$ & flagged\\
$554.8/415$ & JORAM95 & $45.10$ & $131.69$ & flagged\\
$546.8/414$ & & & &\\
\hline
\end{tabular}
\end{center}
\vspace{0.5cm}
\end{table}

Details about the optimisation in the cases of the DB$_-$ and the DB$_0$ are given in Tables \ref{tab:DBPIMEL} and \ref{tab:DBPIMCX}, respectively. In the former case, the elimination of $8$ of the initial $334$ DoFs (see Table 
\ref{tab:ProgressPIMEL}) results in the reduction of the $\chi^2_{\rm min}$ by $157.0$ (from about $524.8$ to about $367.8$), i.e., by about $19.6$ per excluded DoF. As in our earlier PSAs since 2006, the five-point BRACK90 
dataset at $66.80$ MeV was marked for removal. In case of the BD$_0$, the removal of the absolute normalisation of three (out of seven) FITZGERALD86 datasets is noticeable (see Table \ref{tab:ProgressPIMCX}). Although this 
failure may provide arguments for calling into question the absolute normalisation of all FITZGERALD86 datasets, we decided not to deviate from our procedure: the remaining four datasets were retained, as the elimination of 
their absolute normalisation was not enforced when applying our rejection criteria. The elimination of $4$ DoFs of the $326$ initial DoFs of the fit to the DB$_0$ results in the reduction of the $\chi^2_{\rm min}$ by $72.9$ 
(from about $390.2$ to about $317.3$), i.e., by about $18.2$ per excluded DoF.

\begin{table}%[h!]
{\bf \caption{\label{tab:ProgressPIMEL}}}The equivalent of Table \ref{tab:ProgressPIP} when processing the DB$_-$.
\vspace{0.2cm}
\begin{center}
\begin{tabular}{|c|c|c|c|c|}
\hline
$\chi^2_{\rm min}/{\rm NDF}$ & Identifier & $T$ (MeV) & $\theta$ (deg) & Result\\
\hline
\hline
$523.0/334$ & BRACK90 & $66.80$ & $70.00$ & flagged\\
$489.9/333$ & BRACK95 & $98.10$ & $36.70$ & flagged\\
$436.3/332$ & WIEDNER89 & $54.30$ & & absolute normalisation flagged\\
$413.4/331$ & BRACK90 & $66.80$ & $80.80$ & flagged\\
$394.3/330$ & WIEDNER89 & $54.30$ & $15.55$ & flagged\\
$379.2/329$ & BRACK90 & $66.80$ & $111.00$ & flagged\\
 & BRACK90 & $66.80$ & & flagged\\
$367.8/326$ & & & &\\
\hline
\end{tabular}
\end{center}
\vspace{0.5cm}
\end{table}

\begin{table}%[h!]
{\bf \caption{\label{tab:ProgressPIMCX}}}The equivalent of Table \ref{tab:ProgressPIP} when processing the DB$_0$.
\vspace{0.2cm}
\begin{center}
\begin{tabular}{|c|c|c|c|c|}
\hline
$\chi^2_{\rm min}/{\rm NDF}$ & Identifier & $T$ (MeV) & $\theta$ (deg) & Result\\
\hline
\hline
$388.2/326$ & FITZGERALD86 & $40.26$ & & absolute normalisation flagged\\
$364.3/325$ & FITZGERALD86 & $36.11$ & & absolute normalisation flagged\\
$341.9/324$ & FITZGERALD86 & $32.48$ & & absolute normalisation flagged\\
$325.9/323$ & BREITSCHOPF06 & $75.10$ & & flagged\\
$317.3/322$ & & & &\\
\hline
\end{tabular}
\end{center}
\vspace{0.5cm}
\end{table}

Judged solely on the basis of the description of the measurements they contain, it appears that the tDB$_+$, the tDB$_-$, and the tDB$_0$ are not of the same quality. However, a formal statistical test does exist in order to 
support or refute such a thesis: in order to prove that the description of two tDBs $a$ and $b$ is different, the ratio
\begin{equation*}
F_{a/b}=\frac{\chi_a^2/{\rm NDF}_a}{\chi_b^2/{\rm NDF}_b}
\end{equation*}
must significantly differ from $1$. The quantity $F_{a/b}$ follows Fisher's ($F$) distribution with NDF$_a$ and NDF$_b$ DoFs. From the two final fits of the phenomenological model to the tDB$_+$ and to the tDB$_-$, one obtains 
$F_{+/-} \approx 1.17$ for $414$ and $326$ DoFs, which translates into a p-value of $6.73 \cdot 10^{-2}$. Therefore, the frequently expressed opinion that the tDB$_+$ and the tDB$_-$ are not of the same quality cannot be formally 
sustained. On the other hand, the two final fits to the tDB$_+$ and to the tDB$_0$ yield $F_{+/0} \approx 1.34$ for $414$ and $322$ DoFs, which translates into a p-value of $2.89 \cdot 10^{-3}$. Consequently, a significant (at 
our $\mathrm{p}_{\rm min}$ threshold) difference in quality between the tDB$_+$ and the tDB$_0$ can formally be sustained. This remark needs to be borne in mind when analysing these two reactions in a joint optimisation scheme.

Prior to submitting the data to further analysis, two additional fits were performed, using all fourteen parameters of the phenomenological model of Section \ref{sec:KM}. The fit to the tDB$_{+/-}$ yielded $\chi^2_{\rm min} \approx 903.7$ 
for $740$ DoFs. The fit to the tDB$_{+/0}$ yielded $\chi^2_{\rm min} \approx 848.2$ for $736$ DoFs. No additional DoFs were marked for removal in these fits. Therefore, the tDB$_{+/-}$ and the tDB$_{+/0}$, comprising the subsets 
detailed in Tables \ref{tab:DBPIP}-\ref{tab:DBPIMCX}, may be submitted to further analysis. In this work, just $50$ of the initial $1133$ DoFs had to be removed. This corresponds to about $4.4~\%$ of our initial low-energy DB.

\subsection{\label{sec:PSAETH}Fits of the ETH model}

The modelling of the $\pi N$ $s$- and $p$-wave scattering amplitudes with the phenomenological model of Section \ref{sec:KM} enables tests of the consistency of the DBs and serves as an unbiased method for the identification 
of the outliers. However, neither does it provide insight into the underlying physical processes nor can it easily incorporate the theoretical constraint of crossing symmetry. To this end, the ETH model is employed at the 
second stage of each PSA.

Each of our two PSAs (i.e., of the tDB$_{+/-}$ and the tDB$_{+/0}$) encompasses the results of one hundred fits to the corresponding data, each fit performed at one simulated $m_\sigma$ value, see Section \ref{sec:ETHModel}; 
the same set of $m_\sigma$ values is used in both PSAs. The results of each fit for the model parameters, as well as the corresponding Hessian matrices (all available upon request), enable the extraction of predictions: for 
the (average) model parameters, for the phase shifts and the scattering amplitudes, for the low-energy constants of the $\pi N$ system, and for the standard low-energy observables.

\subsubsection{\label{sec:PSAETHPar}Model parameters}

The optimal values of the seven parameters of the ETH model, obtained from the PSAs of the tDB$_{+/-}$ and the tDB$_{+/0}$, are given in Table \ref{tab:PSAPar}.

\begin{table}%[h!]
{\bf \caption{\label{tab:PSAPar}}}The values of the seven parameters of the ETH model, obtained from the PSAs of the tDB$_{+/-}$ and the tDB$_{+/0}$. To facilitate the comparison with other works, the estimate for the pseudoscalar 
coupling constant $g_{\pi N N}$ is converted into a value for the (square of the) pseudovector coupling ($f_{\pi N N}^2$) via Eq.~(\ref{eq:EQ017}).
\vspace{0.2cm}
\begin{center}
\begin{tabular}{|l|c|c|}
\hline
 & tDB$_{+/-}$ & tDB$_{+/0}$\\
\hline
\hline
$G_\sigma$ (GeV$^{-2}$) & $24.47 \pm 0.44$ & $22.4 \pm 1.6$\\
$\kappa_\sigma$ & $0.040 \pm 0.040$ & $0.033 \pm 0.062$\\
$G_\rho$ (GeV$^{-2}$) & $51.02 \pm 0.60$ & $55.55 \pm 0.32$\\ 
$\kappa_\rho$ & $1.04 \pm 0.42$ & $1.07 \pm 0.21$\\
$g_{\pi N N}$ & $13.09 \pm 0.13$ & $13.473 \pm 0.079$\\
$g_{\pi N \Delta}$ & $29.12 \pm 0.29$ & $28.82 \pm 0.20$\\
$Z$ & $-0.527 \pm 0.065$ & $-0.344 \pm 0.069$\\
\hline
$f_{\pi N N}^2$ & $0.0753 \pm 0.0015$ & $0.07980 \pm 0.00094$\\
\hline
\end{tabular}
\end{center}
\vspace{0.5cm}
\end{table}
The difference between the results of the fits to the two tDBs is significant for the model parameters $G_\rho$ (difference equivalent to about a $6.6 \sigma$ effect in the normal distribution) and, to a lesser extent (difference 
equivalent to about a $2.6 \sigma$ effect in the normal distribution), $g_{\pi N N}$. The $\kappa_\sigma$ values are small and compatible with $0$. The $\kappa_\rho$ values are also small, well below the corresponding results 
($6.1-6.6$) extracted at the $\rho(770)$-meson pole via dispersion relations (for details, see Ref.~\cite{Matsinos2018} and the works cited therein). The results for the $\pi N \Delta$ coupling constant are (and have always 
been) in good agreement with the value of $28.93(39)$, extracted directly from the decay width of the $\Delta(1232)$ resonance (see footnote 10 of Ref.~\cite{Matsinos2014}) using the values of the physical constants of this 
work. This agreement justifies the approach, put forward in 1994 \cite{Goudsmit1994a}, to determine the $\pi N R$ coupling constants from the partial decay width (to $\pi N$ decay modes) of each resonance $R$ in the 
contributions of the $s$ and $p$ HBRs to the model amplitudes. The $g_{\pi N \Delta}$ results of Table \ref{tab:PSAPar} strongly disagree with the typical values extracted at the $\Delta(1232)$-resonance pole via dispersion 
relations \cite{Hoehler1972}. The result for the model parameter $Z$ from the fit to the tDB$_{+/-}$ is compatible with $Z=-1/2$, which had been one of the popular theoretical preferences in the remote past; the $Z$ result from 
the fit to the tDB$_{+/0}$ is not incompatible with that preference.

The pseudoscalar and pseudovector couplings are linked via the equivalence relation~\footnote{Some authors define the two coupling constants differently, e.g., using the transformation $g_{\pi N N} \to g_{\pi N N} \sqrt{4 \pi}$ 
or $f_{\pi N N} \sqrt{4 \pi} \to f_{\pi N N}$ in Eq.~(\ref{eq:EQ017}).}:
\begin{equation} \label{eq:EQ017}
f_{\pi N N}^2 = \left( \frac{m_c}{m_1 + m_2} \right)^2 \frac{g_{\pi N N}^2}{4 \pi} \, \, \, ,
\end{equation}
where $m_1$ and $m_2$ stand for the masses of the two nucleons involved in the $\pi N N$ vertex: $m_1$ of the incoming (incident, initial-state) nucleon, $m_2$ of the outgoing (emitted, final-state) nucleon; the ETH model uses 
$m_1=m_2=m_p$. In case of the fit to the tDB$_{+/-}$, $f_{\pi N N}^2$ may be identified with the charged-pion coupling constant $f_c^2$: the extracted value, displayed in Table \ref{tab:PSAPar}, is in agreement with the $f_c^2$ 
estimate of Ref.~\cite{Matsinos2019a}.

\subsubsection{\label{sec:PSAPhases}$\pi N$ phase shifts}

\begin{figure}
\begin{center}
\includegraphics [width=15.5cm] {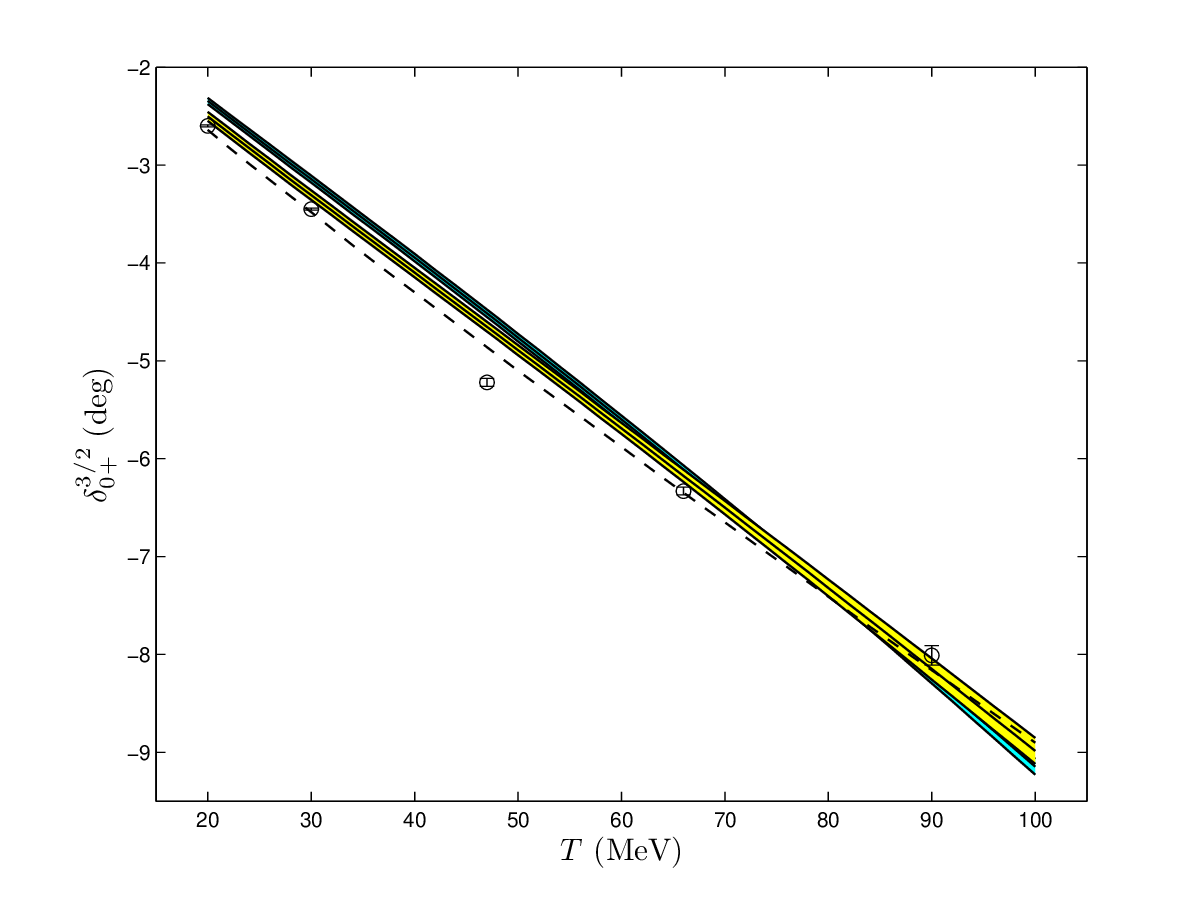}
\caption{\label{fig:s31}The phase shift $\delta_{0+}^{3/2}$ ($S_{31}$), obtained from the fits of the ETH model to the tDB$_{+/-}$ (blue band) and the tDB$_{+/0}$ (yellow band), as a function of the pion laboratory kinetic energy 
$T$. The bands represent $1 \sigma$ uncertainties around the average values. The XP15 solution \cite{XP15} is given by the dashed curve; the five points shown (at $T=20$, $30$, $47$, $66$, and $90$ MeV) are the XP15 single-energy 
values below $T=100$ MeV.}
\vspace{0.35cm}
\end{center}
\end{figure}

\begin{figure}
\begin{center}
\includegraphics [width=15.5cm] {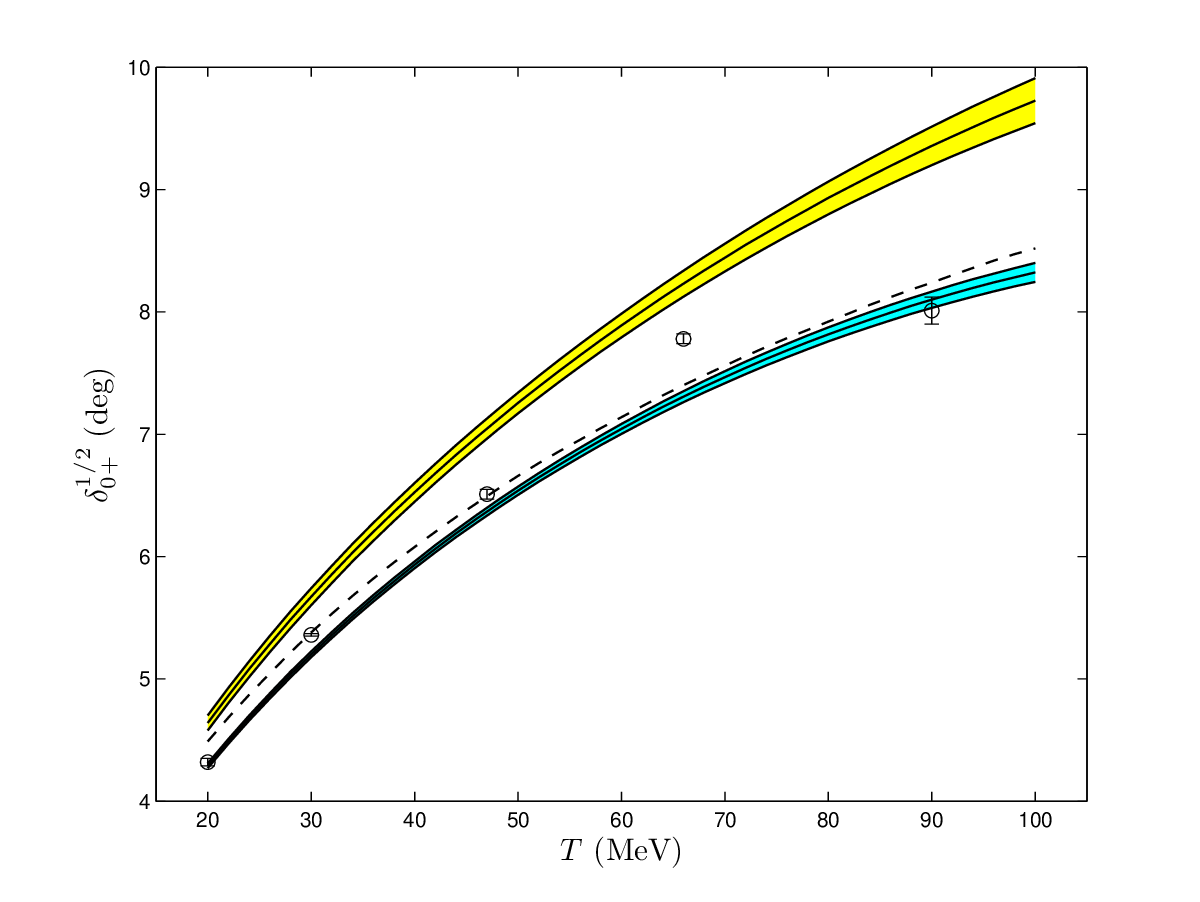}
\caption{\label{fig:s11}Same as Fig.~\ref{fig:s31} for the phase shift $\delta_{0+}^{1/2}$ ($S_{11}$).}
\vspace{0.35cm}
\end{center}
\end{figure}

\begin{figure}
\begin{center}
\includegraphics [width=15.5cm] {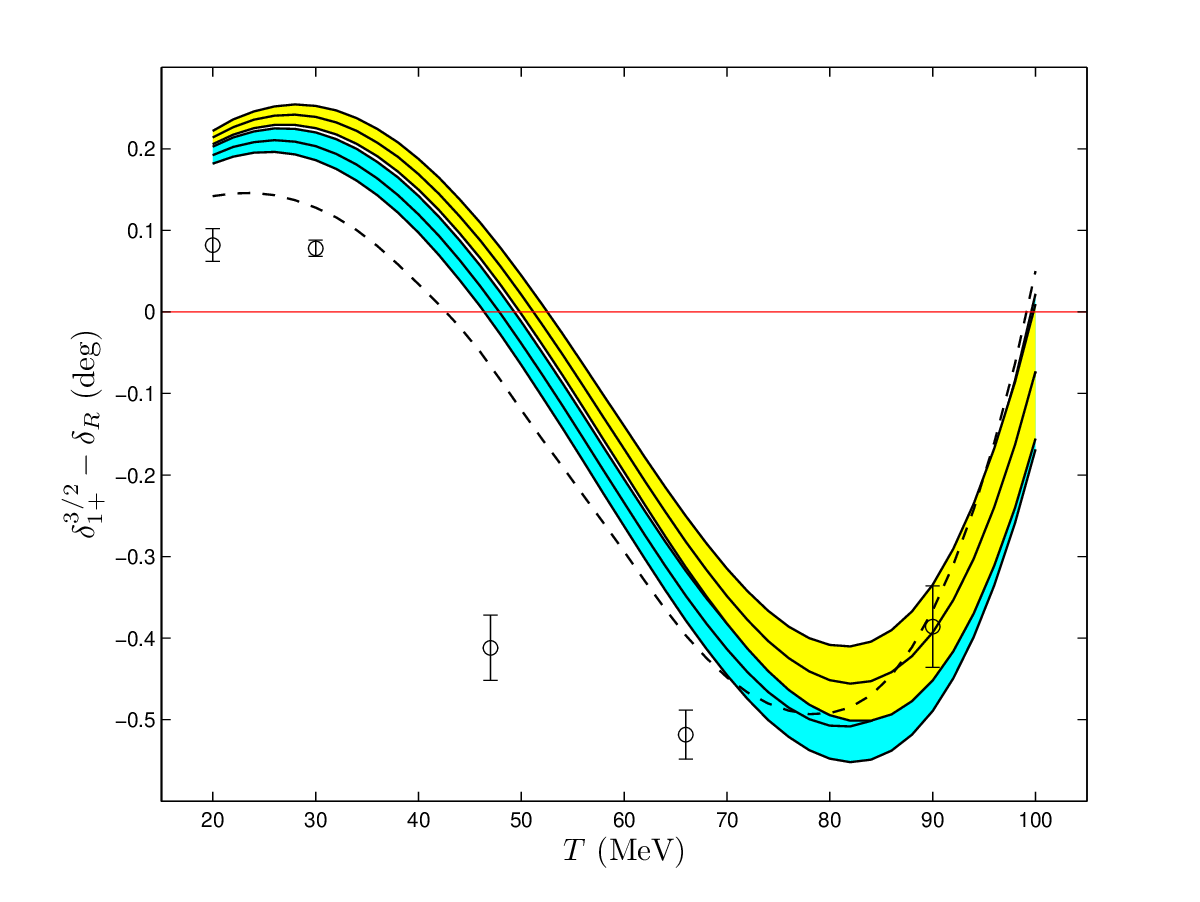}
\caption{\label{fig:p33}Same as Fig.~\ref{fig:s31} for the phase shift $\delta_{1+}^{3/2}$ ($P_{33}$). To facilitate the comparison of the values contained in this figure, the energy-dependent quantity $\delta_R$ 
($=(0.20 \cdot T+1.54) \cdot T \cdot 10^{-2}$, with $T$ in MeV and $\delta_R$ in degrees) has been subtracted from all data. The XP15 solution \cite{XP15} is given by the dashed curve; the five points shown (at $T=20$, $30$, 
$47$, $66$, and $90$ MeV) are the XP15 single-energy values below $T=100$ MeV.}
\vspace{0.35cm}
\end{center}
\end{figure}

\begin{figure}
\begin{center}
\includegraphics [width=15.5cm] {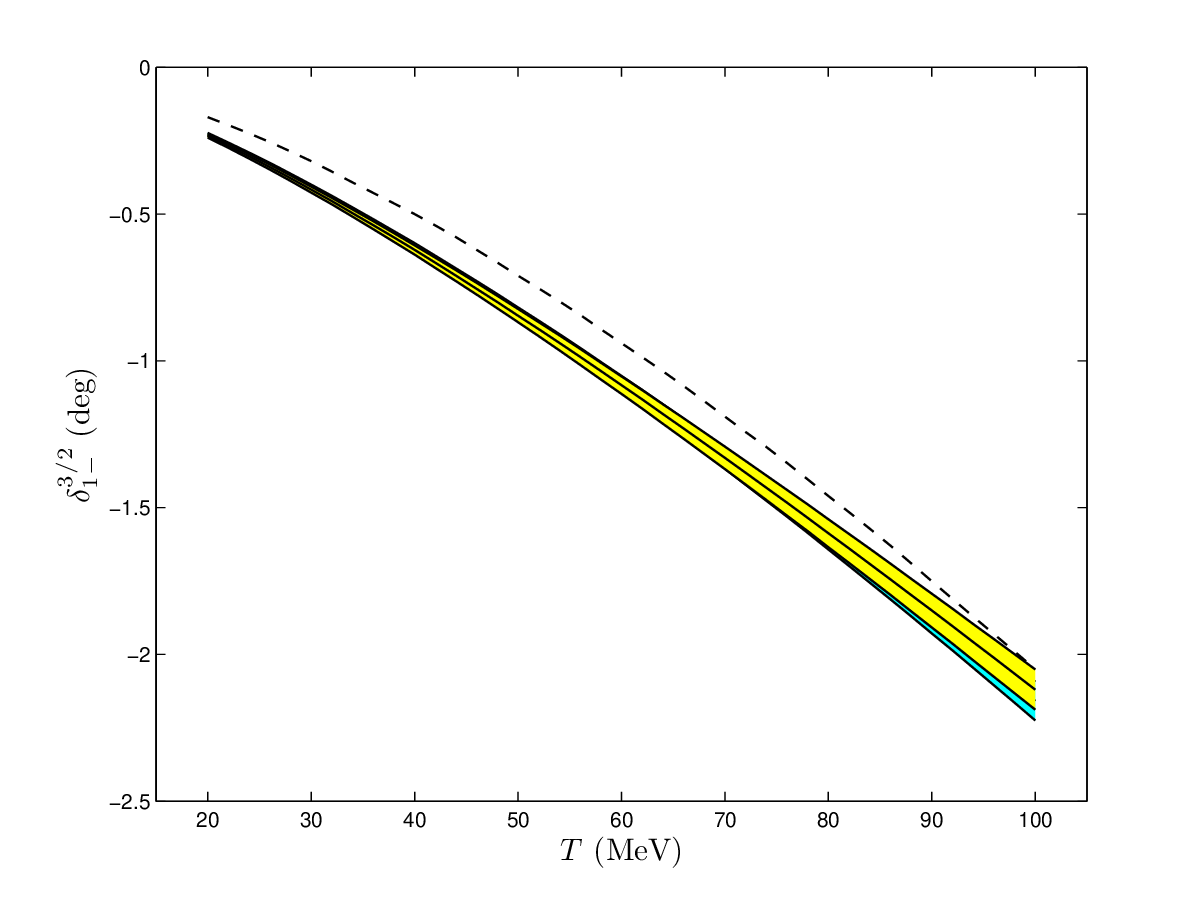}
\caption{\label{fig:p31}The phase shift $\delta_{1-}^{3/2}$ ($P_{31}$), obtained from the fits of the ETH model to the tDB$_{+/-}$ (blue band) and the tDB$_{+/0}$ (yellow band), as a function of the pion laboratory kinetic 
energy $T$. The bands represent $1 \sigma$ uncertainties around the average values. The XP15 solution \cite{XP15} is given by the dashed curve.}
\vspace{0.35cm}
\end{center}
\end{figure}

\begin{figure}
\begin{center}
\includegraphics [width=15.5cm] {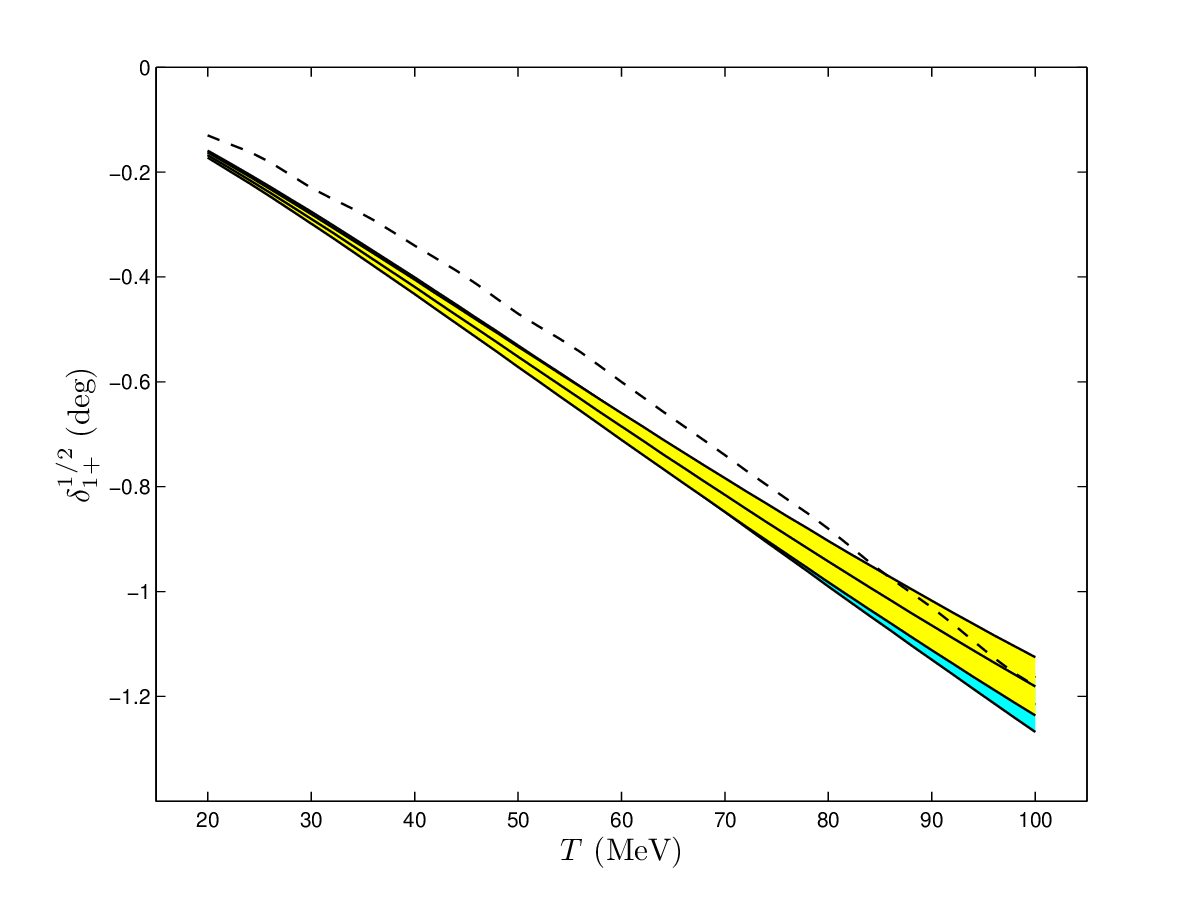}
\caption{\label{fig:p13}Same as Fig.~\ref{fig:p31} for the phase shift $\delta_{1+}^{1/2}$ ($P_{13}$).}
\vspace{0.35cm}
\end{center}
\end{figure}

\begin{figure}
\begin{center}
\includegraphics [width=15.5cm] {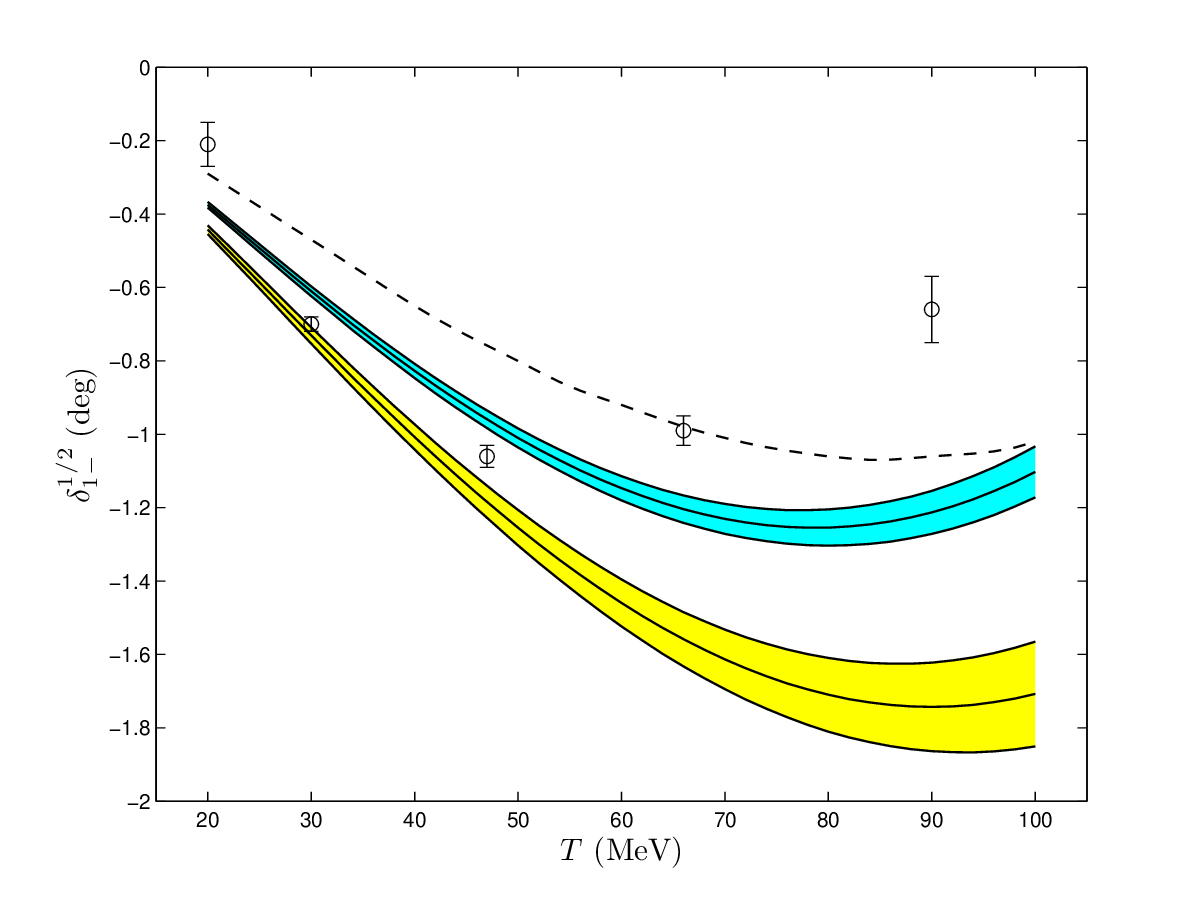}
\caption{\label{fig:p11}Same as Fig.~\ref{fig:s31} for the phase shift $\delta_{1-}^{1/2}$ ($P_{11}$).}
\vspace{0.35cm}
\end{center}
\end{figure}

The predictions for the $s$- and $p$-wave phase shifts from the fits of the ETH model to the tDB$_{+/-}$ and the tDB$_{+/0}$ are given in Tables \ref{tab:PSAPhSh} and \ref{tab:PSAPhSha}, respectively; they are also displayed 
in Figs.~\ref{fig:s31}-\ref{fig:p11}, along with the XP15 solution \cite{XP15}, including (wherever available) the five single-energy values of that solution for $T \leq 100$ MeV. The differences between our two predictions 
for the phase shifts $\delta_{0+}^{1/2}$ ($S_{11}$) and $\delta_{1-}^{1/2}$ ($P_{11}$) are noticeable throughout the energy range of this work. The two solutions also differ in the phase shift $\delta_{0+}^{3/2}$ ($S_{31}$) 
below about $40$ MeV. Although the XP15 phase shifts are not accompanied by uncertainties, there can be no doubt that they generally disagree with both our solutions.

\begin{table}%[h!]
{\bf \caption{\label{tab:PSAPhSh}}}The values of the six $s$- and $p$-wave phase shifts (in degrees), obtained from the fits of the ETH model to the tDB$_{+/-}$.
\vspace{0.2cm}
\begin{center}
\begin{tabular}{|c|c|c|c|c|c|c|}
\hline
$T$ (MeV) & $\delta_{0+}^{3/2}$ ($S_{31}$) & $\delta_{0+}^{1/2}$ ($S_{11}$) & $\delta_{1+}^{3/2}$ ($P_{33}$) & $\delta_{1-}^{3/2}$ ($P_{31}$) & $\delta_{1+}^{1/2}$ ($P_{13}$) & $\delta_{1-}^{1/2}$ ($P_{11}$)\\
\hline
\hline
$20$ & $-2.344(32)$ & $4.290(19)$ & $1.300(10)$ & $-0.2287(49)$ & $-0.1631(41)$ & $-0.3748(78)$\\
$25$ & $-2.745(34)$ & $4.773(21)$ & $1.845(14)$ & $-0.3151(68)$ & $-0.2215(57)$ & $-0.493(11)$\\
$30$ & $-3.144(36)$ & $5.202(23)$ & $2.465(17)$ & $-0.4087(91)$ & $-0.2831(75)$ & $-0.610(13)$\\
$35$ & $-3.543(37)$ & $5.586(26)$ & $3.161(20)$ & $-0.508(12)$ & $-0.3470(96)$ & $-0.722(16)$\\
$40$ & $-3.945(37)$ & $5.933(28)$ & $3.936(23)$ & $-0.614(14)$ & $-0.413(12)$ & $-0.827(19)$\\
$45$ & $-4.351(38)$ & $6.250(31)$ & $4.791(25)$ & $-0.724(17)$ & $-0.479(14)$ & $-0.924(23)$\\
$50$ & $-4.762(38)$ & $6.539(34)$ & $5.732(26)$ & $-0.838(20)$ & $-0.547(17)$ & $-1.010(26)$\\
$55$ & $-5.178(39)$ & $6.803(37)$ & $6.762(28)$ & $-0.956(24)$ & $-0.615(19)$ & $-1.085(29)$\\
$60$ & $-5.598(40)$ & $7.045(41)$ & $7.890(29)$ & $-1.078(27)$ & $-0.683(22)$ & $-1.147(33)$\\
$65$ & $-6.025(41)$ & $7.265(45)$ & $9.121(30)$ & $-1.204(31)$ & $-0.752(25)$ & $-1.196(37)$\\
$70$ & $-6.456(44)$ & $7.467(49)$ & $10.464(32)$ & $-1.332(35)$ & $-0.820(29)$ & $-1.230(41)$\\
$75$ & $-6.893(47)$ & $7.650(53)$ & $11.929(35)$ & $-1.463(40)$ & $-0.887(32)$ & $-1.250(45)$\\
$80$ & $-7.334(51)$ & $7.817(57)$ & $13.525(41)$ & $-1.597(45)$ & $-0.954(36)$ & $-1.254(49)$\\
$85$ & $-7.781(57)$ & $7.966(62)$ & $15.264(49)$ & $-1.734(49)$ & $-1.021(39)$ & $-1.242(54)$\\
$90$ & $-8.232(64)$ & $8.100(67)$ & $17.158(61)$ & $-1.873(55)$ & $-1.087(43)$ & $-1.213(59)$\\
$95$ & $-8.687(72)$ & $8.219(72)$ & $19.221(77)$ & $-2.015(60)$ & $-1.152(48)$ & $-1.167(64)$\\
$100$ & $-9.147(81)$ & $8.323(77)$ & $21.467(95)$ & $-2.158(66)$ & $-1.216(52)$ & $-1.102(70)$\\
\hline
\end{tabular}
\end{center}
\vspace{0.5cm}
\end{table}

\begin{table}%[h!]
{\bf \caption{\label{tab:PSAPhSha}}}Same as Table \ref{tab:PSAPhSh} for the phase shifts obtained from the fits of the ETH model to the tDB$_{+/0}$.
\vspace{0.2cm}
\begin{center}
\begin{tabular}{|c|c|c|c|c|c|c|}
\hline
$T$ (MeV) & $\delta_{0+}^{3/2}$ ($S_{31}$) & $\delta_{0+}^{1/2}$ ($S_{11}$) & $\delta_{1+}^{3/2}$ ($P_{33}$) & $\delta_{1-}^{3/2}$ ($P_{31}$) & $\delta_{1+}^{1/2}$ ($P_{13}$) & $\delta_{1-}^{1/2}$ ($P_{11}$)\\
\hline
\hline
$20$ & $-2.501(46)$ & $4.641(61)$ & $1.3219(82)$ & $-0.2341(55)$ & $-0.1675(49)$ & $-0.443(12)$\\
$25$ & $-2.908(48)$ & $5.185(65)$ & $1.874(11)$ & $-0.3219(78)$ & $-0.2268(69)$ & $-0.587(17)$\\
$30$ & $-3.307(49)$ & $5.673(69)$ & $2.501(14)$ & $-0.417(10)$ & $-0.2891(90)$ & $-0.731(22)$\\
$35$ & $-3.703(49)$ & $6.118(73)$ & $3.204(16)$ & $-0.517(13)$ & $-0.353(11)$ & $-0.872(28)$\\
$40$ & $-4.098(49)$ & $6.526(76)$ & $3.985(19)$ & $-0.622(16)$ & $-0.419(14)$ & $-1.008(34)$\\
$45$ & $-4.493(50)$ & $6.904(80)$ & $4.846(21)$ & $-0.732(19)$ & $-0.485(17)$ & $-1.136(41)$\\
$50$ & $-4.890(51)$ & $7.256(85)$ & $5.791(24)$ & $-0.846(23)$ & $-0.552(19)$ & $-1.255(48)$\\
$55$ & $-5.289(53)$ & $7.583(91)$ & $6.826(26)$ & $-0.963(26)$ & $-0.619(22)$ & $-1.363(56)$\\
$60$ & $-5.690(57)$ & $7.890(97)$ & $7.956(28)$ & $-1.083(30)$ & $-0.685(25)$ & $-1.459(64)$\\
$65$ & $-6.094(62)$ & $8.18(10)$ & $9.188(31)$ & $-1.205(34)$ & $-0.751(29)$ & $-1.543(72)$\\
$70$ & $-6.500(69)$ & $8.44(11)$ & $10.530(34)$ & $-1.330(38)$ & $-0.816(32)$ & $-1.614(81)$\\
$75$ & $-6.908(76)$ & $8.70(12)$ & $11.991(38)$ & $-1.457(43)$ & $-0.880(36)$ & $-1.669(91)$\\
$80$ & $-7.319(86)$ & $8.93(13)$ & $13.581(43)$ & $-1.587(48)$ & $-0.943(39)$ & $-1.71(10)$\\
$85$ & $-7.733(96)$ & $9.15(15)$ & $15.311(50)$ & $-1.718(52)$ & $-1.004(43)$ & $-1.73(11)$\\
$90$ & $-8.15(11)$ & $9.36(16)$ & $17.193(59)$ & $-1.850(58)$ & $-1.065(47)$ & $-1.74(12)$\\
$95$ & $-8.57(12)$ & $9.55(17)$ & $19.241(70)$ & $-1.984(63)$ & $-1.123(51)$ & $-1.73(13)$\\
$100$ & $-8.99(13)$ & $9.73(18)$ & $21.467(83)$ & $-2.120(68)$ & $-1.181(55)$ & $-1.71(14)$\\
\hline
\end{tabular}
\end{center}
\vspace{0.5cm}
\end{table}

\subsubsection{\label{sec:PSACnts}The low-energy constants of the $\pi N$ system}

The $\pi N$ scattering amplitude $\mathscr{F} (\vec{k}^{\, \prime}, \vec{k})$ at low energy may safely be confined to $s$- and $p$-wave contributions (e.g., see Ref.~\cite{Ericson1988}, pp.~17--18). Introducing the isospin of 
the pion as $\vec{t}$ and that of the nucleon as $\vec{\tau}/2$, one may write
\begin{equation} \label{eq:EQ018}
\mathscr{F} (\vec{k}^{\, \prime}, \vec{k}) = b_0 + b_1 \, \vec{\tau} \cdot \vec{t} + \left( c_0 + c_1 \, \vec{\tau} \cdot \vec{t} \, \right) \vec{k}^{\, \prime} \cdot \vec{k} + i \left( d_0 + d_1 \, \vec{\tau} \cdot \vec{t} \, \right) \vec\sigma \cdot 
(\vec{k}^{\, \prime} \times \vec{k}) \, ,
\end{equation}
where $\vec\sigma$ is (double) the spin of the nucleon and $\vec{k}^{\, \prime}$ is the CM $3$-momentum of the outgoing pion.

Equation (\ref{eq:EQ018}) defines the isoscalar and isovector $s$-wave scattering lengths ($b_0$ and $b_1$) and $p$-wave scattering volumes ($c_0$, $c_1$, $d_0$, and $d_1$), which may be projected onto the isospin-spin basis 
via standard transformations. Table \ref{tab:PSACnts} contains the predictions of the ETH model for these quantities, corresponding to the two PSAs of this work.

\begin{table}%[h!]
{\bf \caption{\label{tab:PSACnts}}}Upper part: The isoscalar and isovector $s$-wave scattering lengths (in $m_c^{-1}$) and $p$-wave scattering volumes (in $m_c^{-3}$) based on the fits of the ETH model to the tDB$_{+/-}$ and 
the tDB$_{+/0}$; these quantities are defined by Eq.~(\ref{eq:EQ018}). Middle part: the results for the corresponding spin-isospin quantities. Lower part: the predictions for the $\pi^- p$ ES and CX scattering lengths.
\vspace{0.2cm}
\begin{center}
\begin{tabular}{|l|c|c|}
\hline
Scattering length/volume & tDB$_{+/-}$ & tDB$_{+/0}$\\
\hline
\hline
$b_0 = \left( 2 a_{0+}^{3/2} + a_{0+}^{1/2} \right) / 3$ & $0.00675(79)$ & $0.0053(24)$\\
$b_1 = \left( a_{0+}^{3/2} - a_{0+}^{1/2} \right) / 3$ & $-0.07802(64)$ & $-0.08406(41)$\\
$c_0 = \left( 4 a_{1+}^{3/2} + 2 a_{1-}^{3/2} + 2 a_{1+}^{1/2} + a_{1-}^{1/2} \right) / 3$ & $0.2085(24)$ & $0.2086(27)$\\
$c_1 = \left( 2 a_{1+}^{3/2} + a_{1-}^{3/2} - 2 a_{1+}^{1/2} - a_{1-}^{1/2} \right) / 3$ & $0.1767(19)$ & $0.1842(13)$\\
$d_0 = \left( - 2 a_{1+}^{3/2} + 2 a_{1-}^{3/2} - a_{1+}^{1/2} + a_{1-}^{1/2} \right) / 3$ & $-0.1869(20)$ & $-0.1946(14)$\\
$d_1 = \left( -a_{1+}^{3/2} + a_{1-}^{3/2} + a_{1+}^{1/2} - a_{1-}^{1/2} \right) / 3$ & $-0.06870(85)$ & $-0.06683(97)$\\
\hline
$a_{0+}^{3/2}$ & $-0.0713(14)$ & $-0.0788(23)$\\
$a_{0+}^{1/2}$ & $0.16278(83)$ & $0.1734(28)$\\
$a_{1+}^{3/2}$ & $0.2136(22)$ & $0.2180(17)$\\
$a_{1-}^{3/2}$ & $-0.04200(83)$ & $-0.04336(94)$\\
$a_{1+}^{1/2}$ & $-0.03178(71)$ & $-0.03293(85)$\\
$a_{1-}^{1/2}$ & $-0.0813(16)$ & $-0.0939(21)$\\
\hline
$a_{cc} = \left( a_{0+}^{3/2} + 2 a_{0+}^{1/2} \right) / 3$ & $0.08477(49)$ & $0.0894(26)$\\
$a_{c0} = \sqrt{2} \left( a_{0+}^{3/2} - a_{0+}^{1/2} \right) / 3$ & $-0.11033(91)$ & $-0.11888(58)$\\
\hline
\end{tabular}
\end{center}
\vspace{0.5cm}
\end{table}

The long-standing discrepancy between the $a_{cc}$ value obtained from the PSA of the tDB$_{+/-}$ and those extracted from the pionic-hydrogen experiments at PSI \cite{Schroeder2001,Hennebach2014}, observed and discussed in 
Refs.~\cite{Matsinos2006,Matsinos2012,Matsinos2013b,Oades2007}, as well as in the first version of this preprint (ZRH17), is not an issue anymore. The prediction for $a_{c0}$, obtained from the results of the fits to the 
tDB$_{+/0}$, comes closer to the value extracted (via the second of the Deser formulae \cite{Deser1954,Trueman1961}) from $\Gamma_{1s}$ of pionic hydrogen \cite{Schroeder2001}, which (corrected for the EM effects according to 
Ref.~\cite{Oades2007}) reads as: $a_{c0} = -0.1284(30)(28)~m_c^{-1}$.

\subsubsection{\label{sec:PSAScFctrs}Scale factors}

Investigated in this section is the distribution of the scale factors $z_j$ obtained from the fits of the ETH model. When the Arndt-Roper formula of Eq.~(\ref{eq:EQ004_5}) is used in the optimisation, the expectation is that 
the datasets which are scaled `upwards' ($z_j<1$) balance (on average) those which are scaled `downwards' ($z_j>1$). Furthermore, the energy dependence of the scale factors must not be significant. If these prerequisites are 
not fulfilled, the description of the measurements cannot be considered satisfactory. As demonstrated in Ref.~\cite{Matsinos2017}, the fulfilment of these conditions should not only involve the entire set of the scale factors 
$z_j$ in each fit, but also those of arbitrary subsets of the DB, consistent with the basic principles of the Sampling Theory for representative sampling. One straightforward comparison is dictated by the compartmental structure 
of the DBs of this work: it must be verified that the scale factors, corresponding to the two distinct subsets of the two tDBs, i.e.,
\begin{itemize}
\item to the tDB$_+$ and the tDB$_-$ in the fits to the tDB$_{+/-}$, and 
\item to the tDB$_+$ and the tDB$_0$ in the fits to the tDB$_{+/0}$,
\end{itemize}
are centred on $1$ and exhibit no significant energy dependence.

\begin{figure}
\begin{center}
\includegraphics [width=15.5cm] {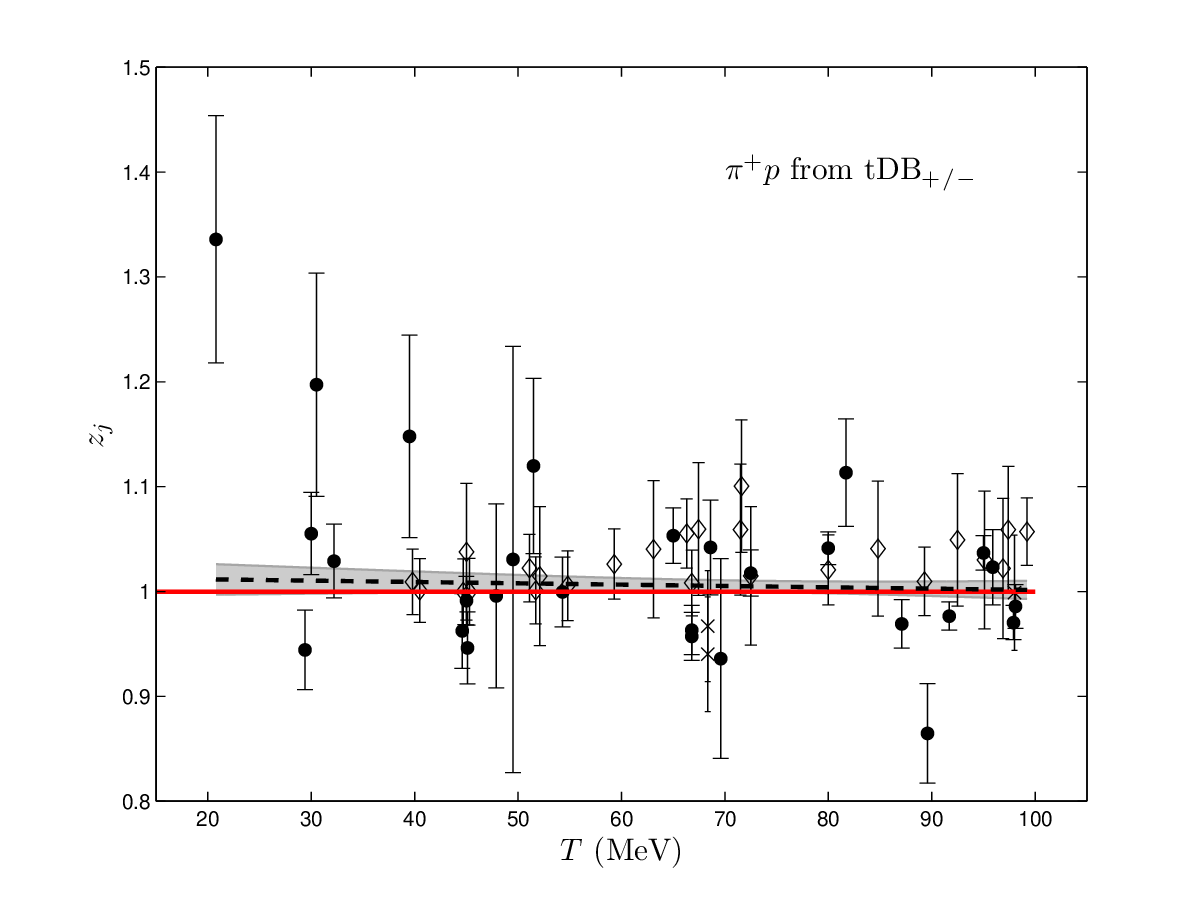}
\caption{\label{fig:sfPIPDB1}The scale factors $z_j$ of the datasets in the tDB$_+$, obtained from the fits of the ETH model to the tDB$_{+/-}$: solid points: DCS, diamonds: PTCS/TCS, crosses: AP. The values, corresponding to 
the datasets which were freely floated (see Table \ref{tab:DBPIP}), have not been included in this plot. Also not included are the entries for the three datasets of Ref.~\cite{Meier2004}; the pion laboratory kinetic energy $T$ 
was not kept constant within each of these datasets. The dashed straight line represents the result of the weighted linear least-squares fit to the data displayed and the shaded band $1 \sigma$ uncertainties around the fitted 
values. The red line represents the optimal, unbiased outcome of the optimisation.}
\vspace{0.35cm}
\end{center}
\end{figure}

\begin{figure}
\begin{center}
\includegraphics [width=15.5cm] {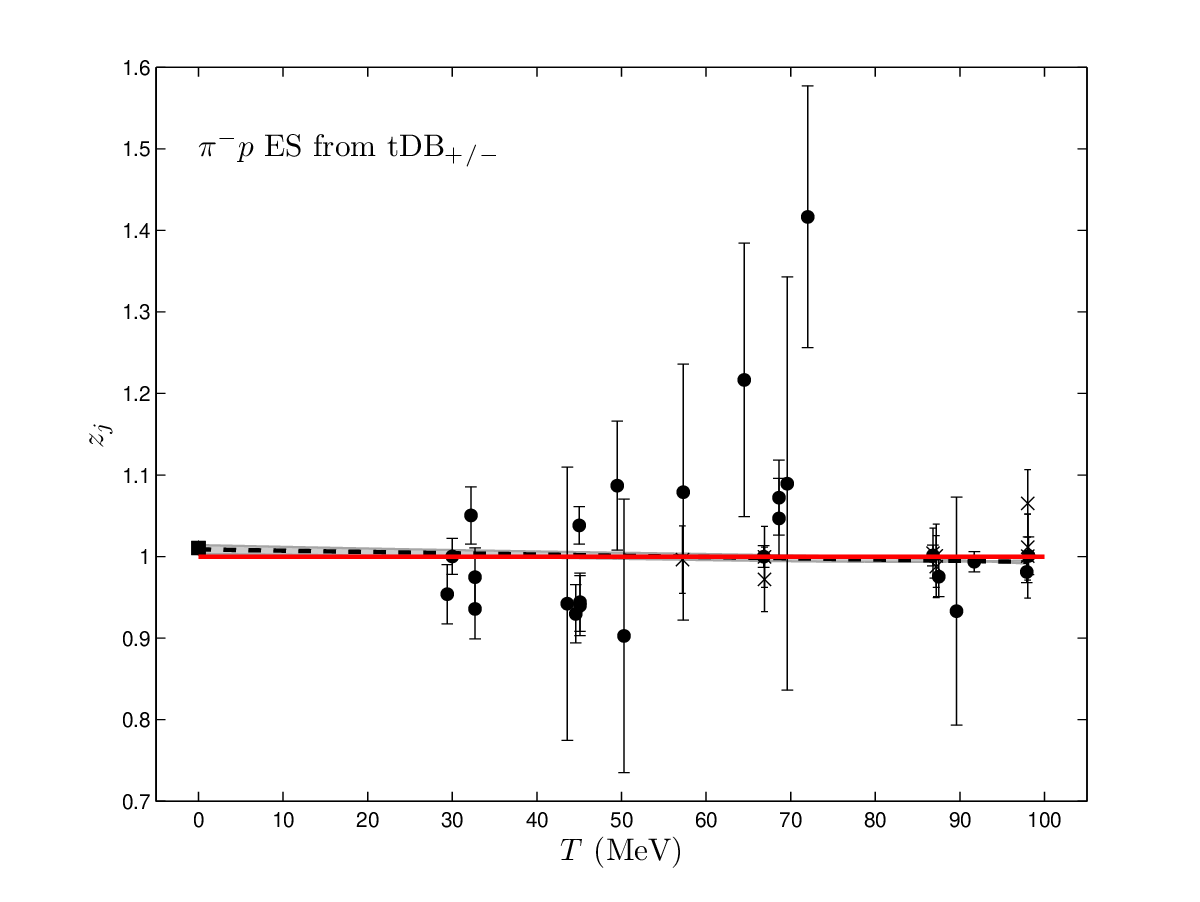}
\caption{\label{fig:sfPIMDB1}The scale factors $z_j$ of the datasets in the tDB$_-$, obtained from the fits of the ETH model to the tDB$_{+/-}$: solid points: DCS, crosses: AP. The squares at pion laboratory kinetic energy $T=0$ 
represent the $a_{cc}$ values, extracted from the measurements of the strong shift of the ground state in pionic hydrogen \cite{Schroeder2001,Hennebach2014}; the two datapoints overlap and their uncertainties are too small to 
be discernible. The value, corresponding to the dataset which was freely floated (see Table \ref{tab:DBPIMEL}), has not been included in this plot. Also not included is the entry for the dataset of Ref.~\cite{Meier2004}; not 
only did $T$ vary within the datasets in that experiment, the particular dataset contained AP measurements of both ES reactions as well. The dashed straight line represents the result of the weighted linear least-squares fit 
to the data displayed and the shaded band $1 \sigma$ uncertainties around the fitted values. The red line represents the optimal, unbiased outcome of the optimisation.}
\vspace{0.35cm}
\end{center}
\end{figure}

For both the $\pi^+ p$ (Fig.~\ref{fig:sfPIPDB1}) and $\pi^- p$ ES (Fig.~\ref{fig:sfPIMDB1}) datasets, the $z_j$ values above and below $1$ roughly balance, and their energy dependence is insignificant. The weighted linear 
least-squares fit ($T$ being the independent variable) to the scale factors of the $\pi^+ p$ reaction yields the intercept of $1.014(20)$ and the slope of $(-1.3 \pm 2.6) \cdot 10^{-4}$ MeV$^{-1}$. The fit to the scale factors 
of the $\pi^- p$ ES reaction yields the intercept of $1.0087(54)$ and the slope of $-1.55(94) \cdot 10^{-4}$ MeV$^{-1}$. (The uncertainties are substantially smaller in the $\pi^- p$ ES reaction because of the inclusion in the 
fit of the two $a_{cc}$ estimates from pionic hydrogen, see Section \ref{sec:MainDifference} for details.) In both cases, the departure from the expectation for an unbiased outcome of the optimisation 
(intercept $1$ and vanishing slope) is not significant.

\begin{figure}
\begin{center}
\includegraphics [width=15.5cm] {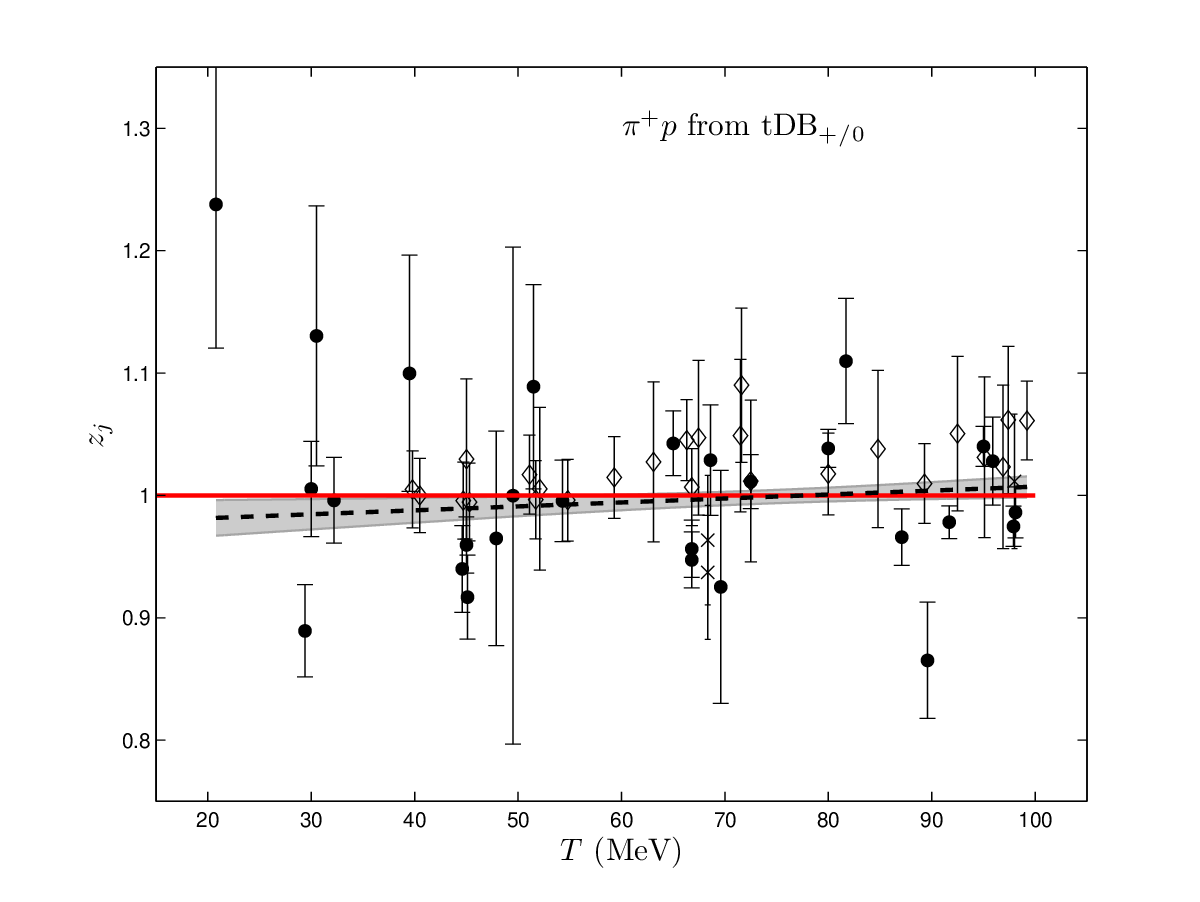}
\caption{\label{fig:sfPIPDB2}Same as Fig.~\ref{fig:sfPIPDB1} for the scale factors $z_j$ of the datasets in the tDB$_+$, obtained from the fits of the ETH model to the tDB$_{+/0}$.}
\vspace{0.35cm}
\end{center}
\end{figure}

\begin{figure}
\begin{center}
\includegraphics [width=15.5cm] {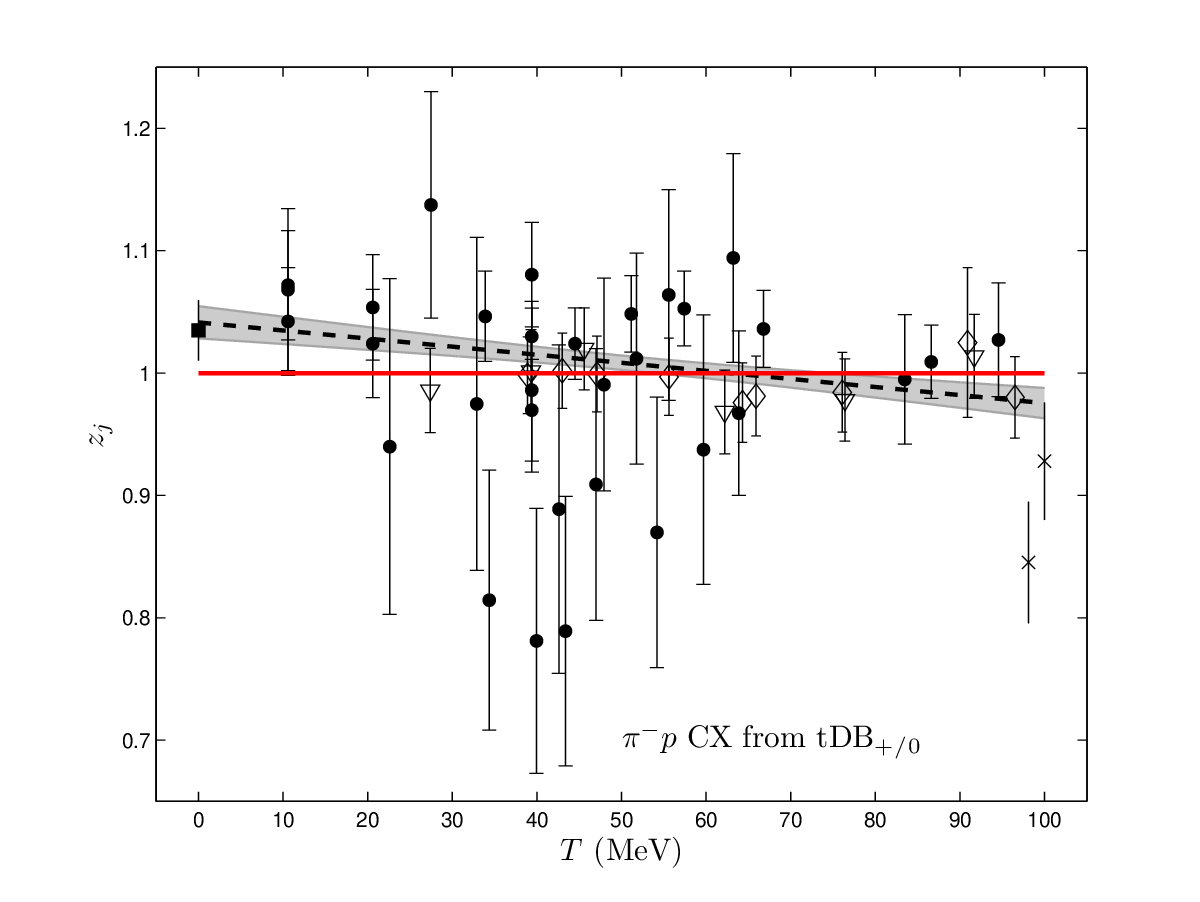}
\caption{\label{fig:sfPIMDB2}The scale factors $z_j$ of the datasets in the DB$_0$, obtained from the fits of the ETH model to the tDB$_{+/0}$: solid points: DCS, diamonds: PTCS/TCS, crosses: AP, triangles: coefficients in the 
Legendre expansion of the DCS. The square at pion laboratory kinetic energy $T=0$ represents the $a_{c0}$ value, extracted from the measurement of the total decay width of the ground state of pionic 
hydrogen \cite{Schroeder2001}. The values, corresponding to the three datasets which were freely floated (see Table \ref{tab:DBPIMCX}), have not been included. The dashed straight line represents the result of the weighted linear 
least-squares fit to the data displayed and the shaded band $1 \sigma$ uncertainties around the fitted values. The red line represents the optimal, unbiased outcome of the optimisation.}
\vspace{0.35cm}
\end{center}
\end{figure}

Two weighted linear least-squares fits to the $z_j$ values of Figs.~\ref{fig:sfPIPDB2} (for the $\pi^+ p$) and \ref{fig:sfPIMDB2} (for the $\pi^- p$ CX reaction) were also carried out in the PSA of the tDB$_{+/0}$. In this case 
however, the results do not match well the expectation for an unbiased outcome of the optimisation. The two values of the intercept are: $0.975(20)$ for the $\pi^+ p$ reaction and $1.042(13)$ for the $\pi^- p$ CX reaction. The 
slope was not found to be incompatible with $0$ in the former case: $(3.2 \pm 2.6) \cdot 10^{-4}$ MeV$^{-1}$. On the other hand, the slope in the $\pi^- p$ CX scale factors came out equal to $(-6.6 \pm 2.3) \cdot 10^{-4}$ MeV$^{-1}$. 
These results are indicative of a problematic situation: when forcing the data of these two reactions into a joint optimisation scheme with the ETH model, the modelling generates \emph{overestimated} fitted values for the $\pi^+ p$ 
reaction and \emph{underestimated} ones for the $\pi^- p$ CX reaction at low energy. Evidently, the optimisation of the description of the data is achieved at the expense of introducing a systematic bias in the description of 
\emph{both} subsets of the tDB$_{+/0}$. Equivalently, one might argue that the $I=3/2$ amplitudes, obtained with the model, have a difficulty to simultaneously account for the $\pi^+ p$ and $\pi^- p$ CX reactions. As we did not 
experience such difficulties in the PSA of the tDB$_{+/-}$, it is justifiable to raise the question whether the joint optimisation of the tDB$_0$ and another tDB is meaningful. It is also clear that the results, obtained from 
the PSA of the tDB$_{+/0}$ (given in Tables \ref{tab:PSAPar}, \ref{tab:PSAPhSha}, and \ref{tab:PSACnts}), need to be taken with caution.

Interestingly, the effects displayed in Figs.~\ref{fig:sfPIPDB2} and \ref{fig:sfPIMDB2} match well those observed in the analysis of the output of the SAID solution WI08, see Figs.~4 and 6 of Ref.~\cite{Matsinos2017}. The 
inevitable conclusion is that, when forcing the $\pi^- p$ CX reaction data into a joint optimisation scheme, the analysis becomes biased at low energy, in the sense of an unmistakable departure from the statistical expectation.

The results of this section demonstrate the importance of the steps which need to be taken in the data analysis, towards the understanding of the outcome of the optimisation. The quality of the data description of the input DB 
needs to be investigated. However, the same also applies to subsets of the DB, in this case autonomous ones, i.e., the measurement sets of the two reactions comprising the input data. Unless these steps are carried out (and yield 
acceptable results), there can be no confidence in the outcome of the optimisation in terms of the presence of bias. For the sake of example regarding the importance of this step, one should mention that the weighted linear 
least-squares fit to \emph{all} $z_j$ values for the PSA of the tDB$_{+/0}$ results in an intercept of $1.013(11)$ and a slope of $(-1.7 \pm 1.6) \cdot 10^{-4}$ MeV$^{-1}$, i.e., values which \emph{are not} incompatible with the 
statistical expectation (at our $\mathrm{p}_{\rm min}$ value). Unless one takes the step to carry out \emph{separate} tests for the two distinct subsets of the DB, the biased outcome of the optimisation is bound to remain as 
hidden as it is in the WI08 solution \cite{Matsinos2017}.

\subsubsection{\label{sec:PSAReproductionCHAOS}Reproduction of the DENZ04 DCSs}

The methodology of Section \ref{sec:Reproduction} enables the assessment of the quality of the reproduction of datasets on the basis of a BLS. This section investigates the reproduction of the ES DCSs of the CHAOS Collaboration 
\cite{Denz2006}, which amount to over two-thirds of the DB$_{+/-}$ comprising all other measurements at low energy. The results of two exclusive analyses of these data \cite{Matsinos2013b,Matsinos2015} had led us to the decision 
to refrain from using the CHAOS DCSs in our PSAs.

The reproduction of the DENZ04 data on the basis of the BLS obtained from the fits of the ETH model to the tDB$_{+/-}$ (henceforth, BLS$_{+/-}$) is shown in Figs.~\ref{fig:PIPPEDenz} and \ref{fig:PIMPEDenz}. The tests of the 
overall reproduction of the datasets, of the reproduction of their shape, and of the reproduction of their absolute normalisation left no doubt that none of the DENZ04 $\pi^+ p$ DCSs can be accounted for~\footnote{The best 
reproduced dataset is the one at $19.90$ MeV: the corresponding p-value is equal to $1.94 \cdot 10^{-6}$.}, and that, in all cases, the problems may be imputed to the \emph{shape} of the angular distribution of the DCS; the 
absolute normalisation of the $32.0$-MeV dataset also emerges as problematic. In addition, the angular distribution of three of the $\pi^- p$ ES DCSs (at $19.90$, $25.80$, and $43.30$(rot.) MeV) has problematic shape. Following 
a recommendation by two members of the CHAOS Collaboration \cite{Denz2011} and using their entire (unsegmented) datasets, we obtained (as a sum over the $(\chi^2_j)_{\rm min}$ contributions of Eq.~(\ref{eq:EQ012})) the reproduction 
$\chi^2$ value of $1305.1$ for $275$ $\pi^+ p$ datapoints and $554.9$ for $271$ $\pi^- p$ ES datapoints: an essential departure from the expectation, set by the bulk of the modern low-energy tDB$_+$, as that DB had been established 
before the measurements of the CHAOS Collaboration appeared, is evident. In comparison with the reproduction of these datasets with the ZRH19 solution, the situation somewhat deteriorated~\footnote{This deterioration was the 
main reason that we set out to investigate the sensitivity of our MC predictions to the number of simulated $m_\sigma$ values at which the first ZRH20 fits, using the ETH model, had been performed, see footnote \ref{ftn:FTN1}.} 
predominantly due to the more precise predictions of this work, which are to be traced to the narrower distribution (uniform distribution in $400-550$ MeV in ZRH19, Gaussian distribution centred on about $500$ MeV with a 
standard deviation of about $30$ MeV in this work) from which the parameter $m_\sigma$ is sampled in the fits of the ETH model.

In Refs.~\cite{Matsinos2013b,Matsinos2015}, it was argued that the segmentation of the datasets, into forward-, intermediate-, and backward-angle sets, is dictated by the procedure which yielded the DENZ04 DCSs; in fact, this 
is the form in which these data had been inserted into the SAID DB over sixteen years ago. Of course, in comparison with the unsegmented datasets, the segmented ones are bound to give better results, the reason being that the 
segmented pieces of one dataset are rescaled independently of each other. As a contribution from the DCSs to the $\chi^2_{\rm min}$ of their fit, the SAID group report a total of $1102.7$ for $545$ datapoints ($626.9/274$ for 
the $\pi^+ p$, $475.8/271$ for the $\pi^- p$ ES datasets). Therefore, even the segmented datasets are poorly described by the WI08 solution: when considering the description of the $29$ segmented datasets by that solution, $15$ 
p-values are below our $\mathrm{p}_{\rm min}$ threshold!

We have also analysed the segmented data of the CHAOS Collaboration, and have come to the conclusion that the following $\pi^+ p$ datasets cannot be reproduced by the BLS$_{+/-}$:
\begin{itemize}
\item the $19.90$-, $25.80$-, $32.00$-, $43.30$-, and $43.30$(rot.)-MeV datasets at forward angles (i.e., five out of the six datasets involving measurements in the Coulomb peak), and
\item the $19.90$-, $25.80$-, and $43.30$(rot.)-MeV datasets at intermediate angles;
\end{itemize}
in three cases, the problems are due to problematic shape; in four, to problematic normalisation; and in one, to both. From the $\pi^- p$ ES measurements, both $25.80$-MeV datasets, as well as the $43.30$(rot.)-MeV dataset at 
intermediate/backward angles, are poorly reproduced (all due to their problematic shape). Using the segmented datasets, reproduction $\chi^2$'s (sums over the $(\chi^2_j)_{\rm min}$ contributions of Eq.~(\ref{eq:EQ012})) of 
$517.1$ for $275$ $\pi^+ p$ datapoints and $433.0$ for $271$ $\pi^- p$ ES datapoints were obtained.

It would be instructive to enquire further into the description by the WI08 solution of the eight $\pi^+ p$ datasets (segmented data), which cannot be reproduced by the BLS$_{+/-}$. The sum of the relevant $\chi^2$ values, 
directly copied from the output of the SAID Analysis Program \cite{Arndt2006}, is equal to $337.8$; as just $121$ datapoints are contained within these sets, the reduced $\chi^2$ value, relating to the description of these 
datasets by the WI08 solution, is equal to about $2.8$, i.e., undoubtedly poor.

Further investigation might be revealing as to the identification of the source of the observed discrepancies. One of the possibilities would be to analyse the results obtained in the three angular intervals of the segmented 
data (i.e., forward, intermediate, and backward). The reproduction of the $\pi^+ p$ data at backward angles by the BLS$_{+/-}$ yields $\chi^2_{\rm b} \approx 66.2$ for $N_{\rm b} = 62$ datapoints; this result is satisfactory~\footnote{It 
thus follows that the hadronic part of the $\pi^+ p$ interaction, which is dominant at backward angles (the direct EM contributions vanish at $\theta=\pi$), is well accounted for by the BLS$_{+/-}$.} and may be used as reference 
in the $F$-tests, aiming at the assessment of the quality of the reproduction of the datasets at forward and intermediate angles. In the former case, $\chi^2_{\rm f} \approx 166.2$ for $N_{\rm f} = 48$ datapoints. Using the 
reduced $\chi^2$ value of the datasets at backward angles as reference, one obtains $F_{\rm f/b} = (\chi^2_{\rm f} / N_{\rm f}) / (\chi^2_{\rm b} / N_{\rm b}) \approx 3.24$ for $N_{\rm f}$ and $N_{\rm b}$ DoFs, resulting in the 
p-value of about $8.43 \cdot 10^{-6}$, therefore indicating a significant deterioration in the reproduction of the datasets at forward angles (in comparison with the reproduction of the measurements at backward angles); visual 
inspection of Fig.~\ref{fig:PIPPEDenz} confirms this result. The corresponding p-value result for the datasets at intermediate angles comes out equal to about $1.58 \cdot 10^{-2}$, yielding indication, but no significant evidence, 
for deterioration in the reproduction of the $\pi^+ p$ data as one shifts from backward to intermediate angles. It thus follows that (in our case) the problem predominantly lies with the $\pi^+ p$ datasets at \emph{forward} 
angles (Coulomb peak). It needs to be mentioned that the description of the $\pi^+ p$ data by the WI08 solution is equally poor in all three angular intervals (forward, intermediate, and backward): their reduced $\chi^2$ values 
for the $\pi^+ p$ datasets range between $2.22$ and $2.31$, thus pointing to a general underestimation of the uncertainties on the part of the CHAOS Collaboration. Therefore, the output of the SAID Analysis Program \cite{Arndt2006} 
provides no evidence that the description of the segmented data by the WI08 solution deteriorates in any of the three angular intervals. This is an interesting difference between the BLS, obtained in this work from the fits of 
the ETH model to the tDB$_{+/-}$, and the WI08 solution.

Provided that their number remains reasonably small, our methodology can handle the presence of outliers in the DBs both in terms of shape, as well as of absolute normalisation of the datasets. However, the breakdown point~\footnote{The 
breakdown point of an analysis is the fraction of discrepant data it can handle before yielding erroneous results.} of our approach is expected to be low, presumably not exceeding $20~\%$. Given the restriction of the DBs to 
low energy, the inclusion of the data in our DB cannot be considered as `small' by any means; the addition of $546$ datapoints to a DB$_{+/-}$, which contains $800$ measurements, makes sense only if the existing and the added 
parts are not incompatible. Figures \ref{fig:PIPPEDenz} and \ref{fig:PIMPEDenz} demonstrate beyond doubt that this is not the case, predominantly due to the severe discrepancies in the DB$_+$. On the other hand, the SAID group 
could afford to add the entirety of the CHAOS DCSs to their DBs because they (i.e., their 2004 DBs) already contained tens of thousands of datapoints ($\pi N$ measurements up to the few-GeV region); their sensitivity to the 
treatment of a few hundred of new measurements is surely lower than it would have been in our case.

The problematic nature of the DENZ04 DCSs may be revealed by examining the reports of the CHAOS Collaboration. For instance, upon inspection of the results of the single-energy phase-shift solution obtained from these data (see 
Table 6.1 of Denz's dissertation and Fig.~4 of the main publication of the CHAOS Collaboration \cite{Denz2006}), one cannot but feel uneasy about the quoted values. Of course, it is true that the single-energy phase-shift 
solutions cannot exhibit the smoothness of the results obtained when the energy dependence of the phase shifts is modelled by means of continuous functions. Having said that, the value of the phase shift $P_{31}$ ($+0.65^\circ$, 
no uncertainty has been quoted) in Table 6.1 of the dissertation, at $19.90$ MeV, is wrong by about $0.9^\circ$; the largest of the $p$-wave phase shifts ($P_{33}$) is itself about $1^\circ$ at that energy! At $20$ MeV, the 
WI08 result \cite{Arndt2006} for $P_{31}$ ($-0.22^\circ$) agrees well with the values we have obtained over time in this programme. This discrepancy alone should have sufficed in providing the motivation for the re-examination 
of the results of Ref.~\cite{Denz2006} by the CHAOS Collaboration.

Chiral-Perturbation Theory (ChPT) gloats over its capability of accounting for the low-energy $\pi N$ phenomena. As the beam energy in the CHAOS DCSs is low, an investigation of the description of these data within a ChPT 
framework (e.g., with the method of Ref.~\cite{Alarcon2013} or following Ref.~\cite{Hoferichter2016}) should be `within reach'; given the availability of the data for over one and a half decades, it is surprising that this 
subject has not been examined yet. Thus, the question arises: can ChPT accommodate the DENZ04 DCSs or can it not?

\subsubsection{\label{sec:PSAReproductionPTCS}Reproduction of the $\pi^- p$ PTCSs and TCSs}

In Section 3 of Ref.~\cite{Matsinos2006}, it was argued that the $\pi^- p$ PTCSs should not be used in this programme because they contain large contributions from the $\pi^- p$ CX reaction. Despite the fact that just nine 
such measurements are available for $T \leq 100$ MeV, their inclusion in any part of the analysis could provide ground for criticism and cast doubt on the interpretation of the results in terms of the isospin invariance in the 
$\pi N$ interaction. The results of the reproduction of the three available $\pi^- p$ PTCSs at low energy are given~\footnote{In Ref.~\cite{Friedman1990}, Friedman and collaborators reported $\pi^+ p$ and $\pi^- p$ ES PTCSs, 
originally identified as FRIEDMAN90. Corrections to the published values of the low-energy $\pi^+ p$ PTCSs appeared a few years later \cite{Friedman1999}, taking account of a revision in the energy calibration of the M11 pion 
channel at TRIUMF (which occurred in 1994); in Ref.~\cite{Friedman1999}, there is no mention of corrections to the low-energy $\pi^- p$ ES PTCS of Ref.~\cite{Friedman1990}. As a result, though both the $\pi^+ p$ and the $\pi^- p$ 
ES PTCSs originally appeared in the same paper, they are assigned different identifiers in the DB: the corrected low-energy $\pi^+ p$ PTCSs are identified as FRIEDMAN99, whereas the original low-energy $\pi^- p$ ES PTCS as 
FRIEDMAN90.} in Table \ref{tab:ReproductionPTCS}. The contributions from the $\pi^- p$ CX reaction are indeed large. Therefore, the line of argument of Ref.~\cite{Matsinos2006}, to avoid including these data in the DB$_-$, is 
justified.

The measurements of the $\pi^- p$ PTCS are compared with the predictions obtained after summing up the $\pi^- p$ ES PTCS and the entire $\pi^- p$ CX TCS. This operation is dictated by the experimental technique employed in the 
acquisition of the $\pi^- p$ PTCS measurements, namely by the detection only of the $\pi^-$'s (interacting or passing through) downstream of the target, within a cone of aperture $2 \theta_L$ with its apex at the geometrical 
centre of the target, where $\theta_L$ is the laboratory-angle cut associated with the PTCS measurement. Regarding the component of the $\pi^- p$ CX TCS, one may use the predictions obtained either from the fits of the ETH model 
to the tDB$_{+/-}$ or from those to the tDB$_{+/0}$. When comparing the values, obtained from the results of the two PSAs of this work on a point-to-point basis, only a slight preference for the tDB$_{+/0}$ predictions emerges.

In addition, there are six estimates for the so-called $\pi^- p$ total-nuclear cross section, obtained from measurements of the PTCS after the EM contributions of the Coulomb peak (as well as those of the Coulomb-nuclear 
interference) are removed. The proper treatment of these data is outlined in Ref.~\cite{Tromborg1977} (see Section `The total cross section' therein): the suggestion is to use the corrected (i.e., not the hadronic) phase shifts 
in the definitions of the partial-wave amplitudes and omit the Coulomb phase shift (and, of course, the direct Coulomb amplitude). The measurements and the predictions, based on the fits to both the tDB$_{+/-}$ and the tDB$_{+/0}$, 
are given in Table \ref{tab:ReproductionTNCS}. Although the uncertainties in the measurements are large to enable safe conclusions, the prediction based on the tDB$_{+/0}$ comes closer to the measurements.

An interesting conclusion may be drawn after performing an analysis of all values obtained in this section. Defining the ratio of corresponding cross sections as $r_i=\sigma^{\rm exp}_i/\sigma^{\rm th}_i$, where $\sigma^{\rm exp}_i$ 
stands for an experimentally obtained cross section (PTCS or TCS) and $\sigma^{\rm th}_i$ denotes the corresponding prediction from either tDB (i.e., from the tDB$_{+/-}$ or from the tDB$_{+/0}$), one may obtain average values 
$\avg{r_i}$ separately for the two options of selecting the BLS. The solution, obtained from the tDB$_{+/0}$ fits, accounts for the experimental data better:
\begin{itemize}
\item $\avg{r_i} = 0.955(21)$ for the results using the fits to the tDB$_{+/-}$ and
\item $\avg{r_i} = 1.001(22)$ for those obtained from the fits to the tDB$_{+/0}$.
\end{itemize}
The reason that the solution, obtained from the tDB$_{+/0}$, does a better job is because the largest component in the $\pi^- p$ PTCSs and TCSs relates to a $\pi^- p$ CX observable (i.e., to the $\pi^- p$ CX TCS), which (not 
surprisingly) is accounted for better by the solution obtained from the tDB$_{+/0}$ fits.

\subsubsection{\label{sec:PSAReproductionCX}Reproduction of the DB$_0$ by the results of the fits of the ETH model to the tDB$_{+/-}$}

The methodology developed in Ref.~\cite{Matsinos2015} was first applied to the reproduction of the DB$_0$ in the ZRH17 PSA. Investigated therein was the absolute normalisation of the DB$_0$ using the prediction based on the 
fits of the ETH model to the tDB$_{+/-}$ as BLS (BLS$_{+/-}$); the same approach will be followed herein. To this end, one must determine the amount at which the reference predictions (i.e., the $y_{ij}^{\rm th}$ set of values) 
need to be floated in order to optimally reproduce the datasets in the DB$_0$ (i.e., the $y_{ij}^{\rm exp}$ set of values). Therefore, relevant in this part of the analysis are the scale factors for free floating, obtained from 
Eq.~(\ref{eq:EQ013}).

The extracted values of the optimal scale factors $\hat{z}_j$ for the datasets, which are associated with measurements of the DCS or the TCS~\footnote{Being ratios of cross sections, the APs might not be suited for investigating 
the violation of the isospin invariance.}, and their total uncertainties, i.e., the purely statistical uncertainties
\begin{equation*}
\delta \hat{z}_j = \left( \sum_{i=1}^{N_j} w_{ij} \right) ^{-1/2}
\end{equation*}
(where the weights $w_{ij}$ were defined in Section \ref{sec:Reproduction}) combined (quadratically) with the normalisation uncertainty of the dataset $\delta z_j$, are given in Table \ref{tab:PSAReprCX} and, plotted separately 
for the DCS, for the TCS, and for the coefficients in the Legendre expansion of the DCS, in Fig.~\ref{fig:PSAReprCX}. Not included in the figure (but given in the table) are the entries for the three FITZGERALD86 datasets which 
had been freely floated in the optimisation, as well as the entry for the BREITSCHOPF06 one-point dataset which had been eliminated at the first stage of the analysis, see Table \ref{tab:ProgressPIMCX}. An interesting feature 
of Table \ref{tab:PSAReprCX} is that just one (out of a total of $52$) datasets is marked as having problematic shape; in contrast, $19$ datasets appear to have problematic absolute normalisation.

Inspection of Fig.~\ref{fig:PSAReprCX} leaves no doubt that, when using as BLS the prediction based on the fits of the ETH model to the tDB$_{+/-}$, the optimal scale factors $\hat{z}_j$ of the datasets in the DB$_0$ contain a 
large amount of fluctuation. As the tDB$_{+/-}$ prediction is smooth, this fluctuation reflects the variation of the absolute normalisation of the datasets in the DB$_0$. For the sake of example, the $\hat{z}_j$ value for the 
FRLE{\v Z}98 dataset comes out equal to $1.403(99)$. This dataset lies in-between three datasets with considerably smaller $\hat{z}_j$ values, i.e., between the two ISENHOWER99 $20.60$-MeV datasets and the MEKTEROVI{\'C}09 
$33.89$-MeV dataset. The values of the absolute normalisation of the two neighbouring datasets of DUCLOS73 ($22.60$ and $32.90$ MeV) and that of the JIA08 $34.37$-MeV dataset (both accompanied by large normalisation uncertainties), 
are compatible with the BLS$_{+/-}$.

The exponential function
\begin{equation*}
\hat{z} = \alpha \exp(-\beta T) + 1
\end{equation*}
has been fitted to the $\hat{z}_j$ values displayed in Fig.~\ref{fig:PSAReprCX}. The optimal values of the parameters $\alpha$ and $\beta$, corrected for the quality of the fit (which is evidently poor), are equal to $0.364(69)$ 
and $0.0263(52)$ MeV$^{-1}$, respectively. These values point to sizeable discrepancy in the low-energy region, increasing with decreasing beam energy.

Integrated between $0$ and $100$ MeV, this discrepancy between measured and predicted cross sections would be equivalent to an effect at the level of $12.8 \pm 1.3~\%$ or, naively converted into a relative difference between 
the two $\pi^- p$ CX scattering amplitudes, of $6.19(63)~\%$. Of course, given the pronounced energy dependence of the effect, these percentages should not to be taken too seriously.

\section{\label{sec:Causes}Possible reasons for the discrepancies in the low-energy $\pi N$ interaction}

Three discrepancies were established in this work.
\begin{itemize}
\item The results for the parameters of the ETH model, obtained from the PSAs of the tDB$_{+/-}$ and the tDB$_{+/0}$, are detailed in Table \ref{tab:PSAPar}. Significantly different results between these two PSAs are extracted 
for the model parameter $G_\rho$. Smaller differences (yet significant at our $\mathrm{p}_{\rm min}$ threshold) are observed in the values of the model parameter $g_{\pi N N}$. The model predictions, obtained from the two PSAs 
of this work, significantly differ in the phase shifts $\delta_{0+}^{1/2}$ ($S_{11}$) and $\delta_{1-}^{1/2}$ ($P_{11}$) throughout our energy domain, see Figs.~\ref{fig:s11} and \ref{fig:p11}. Significant effects are observed 
in $\delta_{0+}^{3/2}$ ($S_{31}$) in part of the energy range, see Fig.~\ref{fig:s31}.
\item The analysis of the scale factors from the fits of the ETH model to the tDB$_{+/-}$ and the tDB$_{+/0}$ is performed in Section \ref{sec:PSAScFctrs}. The scale factors in the former case (see Figs.~\ref{fig:sfPIPDB1} and 
\ref{fig:sfPIMDB1}) do not show a significant departure from the statistical expectation. Sizeable effects are observed in Figs.~\ref{fig:sfPIPDB2} and \ref{fig:sfPIMDB2}, when combining the tDB$_+$ with the tDB$_0$. A similar 
effect was found in Ref.~\cite{Matsinos2017} in the output of the SAID solution WI08 \cite{Arndt2006}. It appears that, when forcing the tDB$_0$ into a joint optimisation scheme, the modelling generates overestimated fitted DCS 
values for the $\pi^+ p$ reaction and underestimated ones for the $\pi^- p$ CX reaction at low energy. Evidently, the optimisation (of the description of the data) is achieved at the expense of introduction of a systematic bias 
in the description of both subsets of the input DB.
\item The reproduction of the $\pi^- p$ CX DCSs by the results of the fits of the ETH model to the tDB$_{+/-}$ is investigated in Section \ref{sec:PSAReproductionCX}. The scale factors, displayed in Fig.~\ref{fig:PSAReprCX}, 
represent the expected level of the absolute normalisation of the experimental data on the basis of the BLS$_{+/-}$. Assuming that the three assumptions, listed in the subsequent paragraph, are fulfilled, these quantities should 
come out centred on $1$ and should show no statistically significant energy dependence. On the contrary, significant effects are observed in the low-energy region, seemingly increasing with decreasing beam energy.
\end{itemize}

The aforementioned discrepancies suggest that at least one of the following assumptions is invalid.
\begin{itemize}
\item There are no significant systematic effects (e.g., a systematic underestimation) in the determination of the absolute normalisation of the datasets comprising the three low-energy DBs; the absolute normalisation of each 
dataset is subject only to statistical fluctuation, in accordance with the normalisation uncertainty as reported by each experimental group (or assigned by us).
\item The residual contributions in the EM corrections of the Aarhus-Canberra-Zurich Collaboration \cite{Gashi2001a,Gashi2001b,Oades2007} are negligible.
\item The isospin invariance holds in the hadronic part of the $\pi N$ interaction.
\end{itemize}
The three possibilities, arising from the non-fulfilment of these assumptions, will be examined further in the rest of this section.

\subsection{\label{sec:IntegrityDB}Experimental mismatches}

The first explanation for the discrepancies involves a trivial effect, namely the systematic inaccuracy of the absolute normalisation of the modern low-energy $\pi N$ data. For the sake of example, such a situation would arise 
if a part of the final-state pions evaded detection or if the flux of the incident beam were overestimated in the modern low-energy $\pi N$ experiments; in both cases, the DCS would be systematically underestimated.

Our first point concerns the absolute normalisation of some low-energy experiments: one frequently feels being at a loss to come up with an explanation for the sizeable effects in the absolute normalisation of some of the 
datasets. For instance, the absolute normalisation of the FITZGERALD86 datasets at the three lowest energies of that experiment appears to exceed the absolute normalisation of the bulk of the tDB$_0$ by nearly $70~\%$ on average 
(see Table \ref{tab:DBPIMCX}); the reported normalisation uncertainty in the experiment was $7.8~\%$. Such an effect may be due to one of the following reasons (or their combination): a) The energy of the incoming beam had not 
been what the experimental group expected. b) The effects of the contamination of the incoming beam were underestimated in the experiment. c) The normalisation uncertainty in the experiment had been grossly underestimated. d) 
The determination of the absolute normalisation in the experiment had been erroneous.

Our second point concerns the smallness of the normalisation uncertainty reported in several low-energy experiments. Arguments were presented in Ref.~\cite{Matsinos2006}, to substantiate the point of view that the reported 
uncertainties in the $\pi N$ experiments at low energy have been underestimated on average. In $23$ (out of $52$) DCS datasets (with known normalisation uncertainty) in the initial DB$_{+/-}$, normalisation uncertainties below 
$3~\%$ had been reported. The smallest normalisation uncertainty in the DB$_{+/-}$ is equal to a mere $1.2~\%$, which (based on the experience gained after three decades of relevant experimentation) borders on the impossible. 
While pondering over these issues, one cannot help thinking that it would make sense to disregard all claims of such exaggerated and unrealistic precision, and assign to all relevant datasets a reasonable normalisation 
uncertainty, e.g., $3~\%$, though this value might be optimistic too. However, such an approach would appear arbitrary and would surely provide ground for criticism. Such revisions must be instigated by the experimental groups 
responsible for the measurements, not by analysts.

\subsection{\label{sec:EMCorrections}Residual EM contributions}

The completeness of the EM corrections in the $\pi N$ interaction at low energy is an important issue which must properly be addressed. Underlying the work both of the NORDITA group \cite{Tromborg1976,Tromborg1977,Tromborg1978} 
as well as of the Aarhus-Canberra-Zurich Collaboration \cite{Gashi2001a,Gashi2001b,Oades2007} was the assumption that the hadronic masses of the various particles are equal to the corresponding physical ones. According to 
Refs.~\cite{Matsinos2006,Oades2007}, the residual EM contributions are predominantly related to the use of the physical masses for the proton, for the neutron, and for the charged and neutral pions in the determination of the 
EM corrections. On the other hand, one might expect that the inclusion of the residual EM contributions would result in an improved description of the input data. Provided that this is indeed the case, then the iterative 
procedure, which had been put forward in the determination of the EM corrections in Refs.~\cite{Gashi2001a,Gashi2001b}, might have captured part of these effects.

The EM corrections are identified as the changes (to the phase shifts and to the partial-wave amplitudes) when the EM interaction is switched from `on' (physical quantities in our world) to `off' (same quantities in a hypothetical 
world, devoid of the EM interaction), see also the discussion at the end of Appendix A in Ref.~\cite{Matsinos2019a}. Important issues, which need to be addressed (and resolved) in the re-assessment of the EM effects in the 
$\pi N$ interaction, include the following.
\begin{itemize}
\item Clear definitions of what is meant by `EM part' and by `hadronic part' of the $\pi N$ interaction.
\item Suitable methodology for dealing with the hadrons after they (or their constituents, i.e., the quarks and antiquarks) have been deprived of their EM features.
\item Suitable methodology for dealing with the altered kinematics, after the interacting hadrons have been deprived of the EM contributions to their rest masses.
\end{itemize}

It is unclear how knowledge of the hadronic part of the $\pi N$ interaction at low energy could be furthered by new experiments, while the subject of the EM corrections remains unresolved. Even for perfect data and no discrepancies 
in the low-energy DB, one would still need to address the extraction of the important (hadronic) information from those perfect measurements. It is therefore our opinion that new EM corrections, benefiting from the experience 
gained from the experimentation and the analysis of the $\pi N$ data during the last three decades, as well as from the facilitation of the dissemination of the information which the World Wide Web offers nowadays, is the first 
step forwards. These corrections need to be obtained in a wide energy range and need to be easy to implement and use, as straightforward as the corrections of the NORDITA group \cite{Tromborg1976,Tromborg1977,Tromborg1978} had 
been. It would be convenient to have the results in tabulated form (perhaps containing more energies than the NORDITA tables). Optimally, one could envisage that one dedicated group would undertake the responsibility not only 
of the development, but also of the implementation of the EM corrections: the application of these corrections could involve one software library with simple interfaces. In principle, the only input comprises the beam energy, 
the (model-dependent) phase shifts, and the reaction type, whereas the output should be the corrected phase shifts and (complex) partial-wave amplitudes, or directly the values of the standard observables (relating to the input 
reaction type). Such a scheme would disperse the current dubiousness regarding the comparison of results obtained in different PWAs of the $\pi N$ data: at present, one cannot be certain whether the differences, which the 
comparisons reveal, originate in the dissimilarity of the modelling schemes of the hadronic interaction, of the analysis techniques, or of the treatment of the EM effects~\footnote{To sidestep the issue of the EM corrections, 
several researchers in the $\pi N$ domain use as input the phase shifts (mostly those of the SAID group), rather than the experimental data, never pausing to reflect on how the EM effects have been taken into account by the 
SAID group in their various analyses of the experimental data.}.

\subsection{\label{sec:Isospin}The violation of the isospin invariance in the hadronic part of the $\pi N$ interaction}

The effects, described at the beginning of Section \ref{sec:Causes}, may be taken to suggest that the isospin invariance is broken in the $\pi N$ interaction at a level exceeding the theoretical expectation \cite{Hoferichter2010}. 
This is the last of the possibilities which may be put forward in an attempt to explain the discrepancies and, admittedly, the most compelling one in Physics terms.

Of course, the possibility of the violation of the isospin invariance in the $\pi N$ system is news to no one; such effects were investigated in a series of papers beginning in the late 1990s \cite{Matsinos1997a,Matsinos1997b,Matsinos2006,Matsinos2013a}, 
as well as - two years before Refs.~\cite{Matsinos1997a,Matsinos1997b} appeared - in the pioneering work of Gibbs and collaborators \cite{Gibbs1995}. Although the conclusions of Refs.~\cite{Matsinos1997a,Matsinos1997b,Gibbs1995} 
were not taken with enthusiasm, one is tempted to raise the question: `Why should the isospin invariance hold in the first place?' After all, the hadronic masses of the $u$ and $d$ quarks are different; similarly, the masses 
of the nucleons differ (beyond `trivial' EM effects), and so do those of the $\Delta(1232)$ isospin states. It appears, therefore, that the right question to ask is not whether the isospin invariance is broken in the $\pi N$ 
interaction, but at which level it is. At this point, it should be mentioned that systematic tests of the isospin invariance in the low-energy $\pi N$ system were recently proposed \cite{Matsinos2020}.

The subject of the isospin breaking in the $\pi N$ interaction may be put into context with what has long been known for the $N N$ system \cite{Miller2006}. The hadronic part of the low-energy $N N$ interaction is characterised 
by three scattering lengths, corresponding to the $^1S_0$ states $p p$, $n n$, and $n p$. If the charge independence (which is used in the $N N$ domain as a synonym for the isospin invariance) would hold, these three scattering 
lengths would be equal. In reality, after the removal of the EM effects, their values are \cite{Miller2006}:
\begin{equation}
a_{p p}=-17.3(4) \,\, {\rm fm}, \, \, \, a_{n n}=-18.8(3) \,\, {\rm fm}, \, \, \, a_{n p}=-23.77(9) \,\, {\rm fm} \, \, \, .
\end{equation}
(In Ref.~\cite{Miller2006}, these scattering lengths carry the superscript `N', indicating that they are nuclear ones, obtained after the EM corrections had been applied.) Evidently, these values violate charge independence and, 
to a lesser extent, charge symmetry, as
\begin{equation}
\Delta a_{CD}=(a_{p p}+a_{n n})/2-a_{n p}=5.7(3) {\rm fm}
\end{equation}
and
\begin{equation}
\Delta a_{CSD}=a_{p p}-a_{n n}=1.5(5) {\rm fm}
\end{equation}
are significantly non-zero. These values correspond to the violation of the charge independence in the low-energy $s$-wave part of the $N N$ scattering amplitude by about $27~\%$ and of charge symmetry by about $8~\%$. The level 
of the charge-independence breaking in the $N N$ interaction is to be compared with the $5$ to $10~\%$ effects which were reported in Refs.~\cite{Matsinos2017,Matsinos1997a,Matsinos1997b,Matsinos2006,Matsinos2013a,Gibbs1995} 
for the low-energy $\pi N$ system. If the $\pi N$ interaction is regarded as the basis for the description of the $N N$ interaction (as the case is in meson-exchange models of the $N N$ interaction), it is logical to expect 
that a part of the (large) isospin-breaking effects, observed in the $N N$ system, could originate from the $\pi N$ interaction.

There are two ways (or their combination) by which the isospin-breaking effects may be accounted for in a $\pi N$ interaction model.
\begin{itemize}
\item If the model is isospin-invariant by structure (as the ETH model is), then any isospin-breaking effects will manifest themselves as discrepancies in the data analysis, e.g., as those detailed in the beginning of Section 
\ref{sec:Causes} of this work.
\item If the model contains (the appropriate) isospin-breaking graphs, then - by fitting to the same data - one should come up with an \emph{improved} description and the \emph{removal} of any discrepancies established in 
analyses featuring the isospin-invariant version of that model. This possibility has never been pursued in this programme.
\end{itemize}

In 2017, two papers reported sizeable splitting effects in the $\pi N$ coupling constant. Under the assumption that the $N N$ force is modelled at low energy by means of the one-pion-exchange mechanism, Babenko and Petrov 
\cite{Babenko2017} obtained the significant result $f_c>f_0$, see Ref.~\cite{Matsinos2019a} for details. In the same year, Navarro P{\'e}rez and collaborators \cite{Navarro2017} came up with a (less significant) splitting effect 
in the $\pi N$ coupling constant, but appear to favour $f_c<f_0$. One should not forget that, aiming at providing an explanation for the surprising (at that time) result of Ref.~\cite{Gibbs1995}, Piekarewicz had (already in 
1995) attributed the isospin-breaking effects to changes in the coupling constant due to the mass difference between the $u$ and $d$ quarks \cite{Piekarewicz1995}. The splitting effects in the $\pi N$ coupling constant were 
also studied in Ref.~\cite{Meissner1997}: the authors reported that $g_{\pi^\pm p n}$ should be equal to the average of the two $g_{\pi^0 N N}$ values and provided an estimate for the splitting between $1.2$ and $3.7~\%$.

Table \ref{tab:PSAPar} demonstrates that the coupling constant $g_{\pi N N}$ is affected when replacing the tDB$_-$ with the tDB$_0$ in the PSA. The fits of the ETH model to the DB$_{+/-}$ essentially determine the 
charged-pion coupling constant $f_c$, whereas those involving the $\pi^- p$ CX reaction determine unusual combinations of these quantities: the $s$-channel graph involves the combination $f_- f_n$, whereas the 
$u$-channel graph involves $f_- f_p$, see Section 5 of Ref.~\cite{Matsinos2019a}. Under certain conditions, the analysis of the low-energy $\pi N$ data with the ETH model suggests that $f_0>f_c$, in agreement with the effect 
reported in Ref.~\cite{Navarro2017}.

Two mechanisms are known for a long time as potential sources of isospin-breaking effects in the hadronic interaction: the first one affects the ES processes ($\rho^0 - \omega$ mixing), whereas the second has an impact on the 
$\pi^- p$ CX reaction ($\pi^0 - \eta$ mixing). As both the $\omega(782)$ and the $\eta$ mesons are singlets, the coupling of the former to the $\rho^0(770)$ and of the latter to the $\pi^0$ explicitly violate the isospin 
invariance in the $\pi N$ interaction.

Regarding the possibility of the attribution of (some of) the discrepancies to isospin-breaking effects originating from the $\rho^0 - \omega$ mixing, there has been one development after the ZRH17 PSA appeared. The contributions 
from the graph of Fig.~\ref{fig:IsospinBreakingRhoOmega} were estimated in Ref.~\cite{Matsinos2018} and found to be small, below the $1~\%$ level in the low-energy region. Assuming the validity of the $t$ dependence of the 
effects of the $\rho^0 - \omega$ mixing of Ref.~\cite{Matsinos2018} (imported therein from other works), it does not appear likely that the $\rho^0 - \omega$ mixing could play an important role in the low-energy $\pi N$ interaction.

The $\pi^0 - \eta$ mixing was proposed as a source of isospin-breaking effects in the $\pi^- p$ CX reaction about four decades ago \cite{Cutkosky1979}. Given that just one graph (the $\rho(770)$-exchange $t$-channel graph) is 
affected in case of the ES, whereas all graphs are affected in case of the $\pi^- p$ CX reaction (see Fig.~\ref{fig:IsospinBreakingEtaPi0}), one might (perhaps, naively) expect that the isospin-breaking effects would be more 
significant in the latter case.

\section{\label{sec:Discussion}Summary and discussion}

The series of papers under this link serve two main purposes.
\begin{itemize}
\item They provide details about the development of the theoretical and analysis frameworks of this research programme, which aims at the study of the pion-nucleon ($\pi N$) interaction at low energy (pion laboratory kinetic 
energy $T \leq 100$ MeV).
\item They provide updated results, obtained from the low-energy $\pi N$ measurements via the use of the Arndt-Roper minimisation function \cite{Arndt1972} of Eq.~(\ref{eq:EQ004_5}).
\end{itemize}

To facilitate the discussion below, we remind the reader that our notation for the various databases of this work can be found in the introduction.

The essential difference of our last two phase-shift analyses (PSAs), i.e., of the ZRH19 PSA and of the PSA of this work (to be referred to as `ZRH20' in the future), to all our former PSAs relates to the inclusion in the DB$_-$ 
of the two estimates for the $\pi^- p$ $s$-wave elastic-scattering (ES) length $a_{cc}$, extracted from PSI measurements of the strong shift $\epsilon_{1 s}$ in pionic hydrogen \cite{Schroeder2001,Hennebach2014}. Regarding this 
modification, extensive comments may be found in Section \ref{sec:MainDifference}. As in the ZRH19 PSA, it has been demonstrated that the addition of these two datapoints to the input does not give rise to any pronounced 
differences to our earlier results: the $\chi^2$ minimum simply moves towards a solution characterised by enhanced isoscalar components in the low-energy $\pi N$ scattering amplitude. However, this modification has one important 
consequence: the long-standing discrepancy between the $a_{cc}$ result of our PSA of the tDB$_{+/-}$ and the estimates extracted from the results of Refs.~\cite{Schroeder2001,Hennebach2014}, observed and discussed in 
Refs.~\cite{Matsinos2006,Matsinos2012,Matsinos2013b,Oades2007}, as well as in the first version of this preprint (ZRH17), has been resolved.

In comparison with the ZRH19 PSA, the differences are detailed in Section \ref{sec:AdditionalDifferences}: the most important one relates to the variation of parameter $m_\sigma$ of the ETH model. Heretofore, the recommendation 
by the Particle-Data Group (PDG), to make use of a Breit-Wigner mass between $400$ and $550$ MeV, had been followed: to account for this broad domain, the approach in our recent PSAs was to perform the fits of the ETH model at 
seven (evenly spaced and equally weighted) $m_\sigma$ values between $400$ and $550$ MeV. In contrast, the fits of this work were performed at one hundred $m_\sigma$ values, randomly selected according to the recent result: 
$m_\sigma=497^{+28}_{-33}$ MeV \cite{Matsinos2020a}. All uncertainties herein (in the values of the model parameters, in the predictions for the low-energy constants of the $\pi N$ system, in the phase shifts, etc.) contain the 
effects of the $m_\sigma$ variation, as well as (if exceeding $1$) the Birge factor $\sqrt{\chi^2_{\rm min}/{\rm NDF}}$ which takes account of the quality of each fit. The remaining differences to the ZRH19 PSA do not have a 
significant impact on the results.

In the current analysis framework, the separate study of the ES DBs, as well as of the DB relating to the $\pi^- p$ charge-exchange (CX) reaction, is enabled by means of suitable low-energy parameterisations of the $s$- and 
$p$-wave $K$-matrix elements (representing the `phenomenological model' in this work). The analysis with the phenomenological model is free of theoretical constraints, other than the expected low-energy behaviour of the $K$-matrix 
elements. Deployed at the first stage of each PSA, the processing of the data with the phenomenological model enables an unbiased identification of the outliers in the DBs and the creation of consistent input for further analysis.

The hadronic model of this programme, the `ETH model', is mainly based on $f_0(500)$- and $\rho(770)$-exchange $t$-channel graphs, as well as on the $s$- and $u$-channel $N$ and $\Delta(1232)$ contributions (see Fig.~\ref{fig:FeynmanGraphsETHZ}). 
Additional (small) contributions from all well-established (four-star) $s$ and $p$ higher baryon resonances (HBRs) with masses below $2$ GeV had been part of the model since 1994. Henceforth, the $t$-channel contributions of 
all scalar-isoscalar and vector-isovector mesons with masses below $2$ GeV and known branching fractions to $\pi \pi$ decay modes will be included in the model amplitudes. At present, four such states are well-established 
\cite{PDG2020}: the $f_0(980)$, the $f_0(1500)$, the $f_0(1710)$, and the $\rho(1700)$. The physical properties of all these states have been fixed from two recent analyses of all available data \cite{Matsinos2020a,Matsinos2020b} 
as they appear listed in the recent compilation by the PDG \cite{PDG2020}.

The analysis of the measurements with the ETH model is performed at the second stage in each PSA; imposed at this stage are the theoretical constraints of crossing symmetry and isospin invariance, which the hadronic part of the 
$\pi N$ scattering amplitude of the ETH model obeys. To ensure that both isospin amplitudes (in each partial wave) are reliably determined from the measurements, joint analyses of the tDB$_+$ with the tDB$_-$ and with the tDB$_0$ 
are performed. The optimal values of the model parameters and the corresponding Hessian matrices are obtained from these fits, and are subsequently used in the generation of Monte-Carlo predictions for the low-energy constants 
of the $\pi N$ system, for the phase shifts, and for the standard low-energy $\pi N$ observables, i.e., for the differential cross section (DCS), for the analysing power (AP), for the partial-total cross section (PTCS), and for 
the total cross section (TCS). Predictions for these observables may be extracted for any of the three low-energy processes, at any value of the relevant kinematical variable(s), i.e., of the energy and the scattering angle for 
the DCS and the AP, of the energy and the laboratory-angle cut for the PTCS, and of the energy for the TCS. 

The analysis of the low-energy $\pi N$ DBs with the ETH model established a number of discrepancies, see beginning of Section \ref{sec:Causes}. In summary, the results for the model parameter $G_\rho$, obtained from the PSA of 
the DB$_{+/-}$ and of the tDB$_{+/0}$, significantly differ, see Table \ref{tab:PSAPar}; smaller effects are observed in the $\pi N$ coupling constant $g_{\pi N N}$. The model predictions, obtained from the two PSAs of this 
work, significantly differ in the phase shifts $\delta_{0+}^{1/2}$ ($S_{11}$) and $\delta_{1-}^{1/2}$ ($P_{11}$), throughout the energy range of this work, see Figs.~\ref{fig:s11} and \ref{fig:p11}. Significant effects are 
observed in $\delta_{0+}^{3/2}$ ($S_{31}$), in part of the energy range, see Fig.~\ref{fig:s31}. Sizeable effects are observed in the analysis of the scale factors obtained from the tDB$_{+/0}$ (see Figs.~\ref{fig:sfPIPDB2} 
and \ref{fig:sfPIMDB2}); similar effects were observed in Ref.~\cite{Matsinos2017} in the output of the SAID solution WI08 \cite{Arndt2006}. It appears that, when forcing the tDB$_0$ into a joint optimisation scheme, regardless 
of whether one uses the ETH model or dispersion relations \cite{Arndt2006} in the analysis, the modelling generates overestimated fitted DCS values for the $\pi^+ p$ reaction and underestimated ones for the $\pi^- p$ CX reaction 
at low energy. Finally, significant effects are observed in the reproduction of the $\pi^- p$ CX DCSs on the basis of the results of the fits of the ETH model to the tDB$_{+/-}$; these effects appear to increase with decreasing 
beam energy (see Fig.~\ref{fig:PSAReprCX}).

Our result for the $\pi N$ coupling constant $g_{\pi N N}$, obtained from the PSA of the tDB$_{+/-}$ is equal to $13.09 \pm 0.13$, which translates into (the square of) the pseudovector coupling $f_c^2=0.0753 \pm 0.0015$; this 
result is nearly identical to the value we had obtained in the ZRH19 PSA. The two values of the $\pi N \Delta$ coupling constant $g_{\pi N \Delta}$, obtained from the PSAs to the tDB$_{+/-}$ and the tDB$_{+/0}$, are in good 
agreement with the value of $28.93(39)$, extracted directly from the decay width of the $\Delta(1232)$ resonance.

Given that the electromagnetic (EM) corrections (which are applied to the phase shifts and to the partial-wave amplitudes on the way to the evaluation of the observables) of Refs.~\cite{Gashi2001a,Gashi2001b,Oades2007} were 
obtained with the physical (instead of the unknown hadronic) masses for the proton, for the neutron, and for the charged and neutral pions, a cautious attitude is assumed, by considering the physical quantities of the analysis 
(the fit parameters, the low-energy constants of the $\pi N$ system, the phase shifts, etc.) not purely hadronic, but `EM-modified', i.e., possibly containing residual EM contributions. Although part of these effects might have 
already been captured by the procedure put forward in the determination of the EM corrections in Refs.~\cite{Gashi2001a,Gashi2001b,Oades2007}, it remains unknown how important any residual contributions might be.

In Section \ref{sec:EMCorrections}, arguments are put forward to substantiate the recommendation to re-assess the EM corrections in the $\pi N$ interaction. New low-energy experiments cannot advance our knowledge of the hadronic 
part of the $\pi N$ interaction, while the subject of the EM corrections remains open. Some of the issues, which need to be addressed, are outlined in Section \ref{sec:EMCorrections}: it appears that the most important subject 
is the use of the physical masses in the determination of the EM corrections of Refs.~\cite{Gashi2001a,Gashi2001b,Oades2007} and \cite{Tromborg1976,Tromborg1977,Tromborg1978}.

Assuming the absence of significant systematic effects (e.g., a systematic underestimation) in the determination of the absolute normalisation of the datasets, comprising the low-energy DB, and the negligibility of the residual 
contributions in the EM corrections of Refs.~\cite{Gashi2001a,Gashi2001b,Oades2007}, the aforementioned discrepancies may be blamed on the violation of the isospin invariance in the $\pi N$ interaction at low energy. In this 
context, the findings of this study agree well with those reported since the mid 1990s, when the isospin invariance in the $\pi N$ interaction was first tested by using the then available experimental information. Considering 
the changes in the DBs (e.g., the DB$_0$ has been enhanced by a factor of seven in the meantime), in the analysis methods, and in the EM corrections applied to the data, this agreement is notable. These findings disagree with 
expectations by the Chiral-Perturbation Theory \cite{Hoferichter2010}, according to which the maximal isospin-breaking effects in the low-energy $\pi N$ interaction should be of order of $1~\%$.

During the last two decades, we have developed our methodology and brought it to a point at which further improvement is hard to envision. In spite of the changes made to our analysis framework, our results are stable and, only 
close to the $\pi N$ threshold, do they show dependence on the treatment (i.e., inclusion in the DB$_-$ or exclusion) of the two $a_{cc}$ estimates from pionic hydrogen. It is very unlikely that the inclusion in the ETH model 
of more $s$ and $p$ HBRs could have any sizeable impact on the description of the low-energy $\pi N$ data, and on the physical quantities extracted thereof; the same applies to the $t$-channel contributions from any missing 
scalar-isoscalar and vector-isovector mesons. Given their diminishing contributions with increasing mass, the inclusion in the ETH model of the states above $2$ GeV is also not expected to bring any surprises. As the differences 
between the various form-factor schemes are small at low $Q^2$ \cite{Matsinos2020c}, the simple dipole forms, used herein, more than suffice. Last but not least, experience has demonstrated that small changes in the values of 
the physical constants, which need to be imported into our PSAs, cannot have noticeable impact. In summary, the benefit-cost ratio is just not enticing enough to pursue a re-analysis of the low-energy $\pi N$ data on the basis 
of all aforementioned effects. At the present time, it appears to us that there are five reasons which might call for such a re-analysis.
\begin{itemize}
\item After considerable experimental activity has been carried out at low energy. At this time, there is no indication that new low-energy experiments are planned (excepting, of course, the upcoming data acquisition at PSI by 
the MUSE Collaboration).
\item After definite proof surfaces that the absolute normalisation of the low-energy $\pi N$ experiments had been underestimated, and corrections (applicable to those past experiments) have been established, see Section 
\ref{sec:IntegrityDB}.
\item After the publication of updates of the reported results of (some of) the low-energy $\pi N$ experiments. In this respect, our hope in 2013 was that our first paper on the analysis of the CHAOS DCSs \cite{Matsinos2013b} 
would instigate the reprocessing of their raw experimental data. Had this happened, our low-energy ES DB would have nearly doubled. Given that eight years passed since we first reported on the CHAOS DCSs, we believe that no 
such reprocessing (if possible at all) is likely to take place.
\item After a rigorous re-assessment of the EM effects at low energy, see Section \ref{sec:EMCorrections}.
\item After theoretical developments invalidate the use of the $\pi^- p$ scattering lengths $a_{cc}$ and $a_{c0}$, extracted from pionic hydrogen, in PSAs of the scattering data. For the sake of example, such a possibility 
could come to light if the application of Perturbative QED to bound systems emerged as questionable or problematic. (Such a development would also invalidate the use of the muonic-hydrogen results in the determination of the 
electric Sachs form factor of the proton.)
\end{itemize}
Regarding these five possibilities, near-future developments do not emerge as very likely. Consequently, we are bound to conclude at this time that the present work might be our final PSA of the low-energy $\pi N$ data.

We would like to remind the interested reader that predictions for the standard low-energy $\pi N$ observables (DCS, AP, PTCS, and TCS) for the three $\pi N$ reactions, extracted from the two PSAs of this work, are simple to 
obtain, free of charge, and available within a few days of a request. Unlike the results obtained from dispersion relations, \emph{our predictions are accompanied by uncertainties which reflect the statistical and systematic 
variation of the experimental data}. In Appendix \ref{App:AppA}, we publicise our predictions for the $\pi^\pm p$ ES DCS at three energies, corresponding to the forthcoming data acquisition at PSI by the MUSE Collaboration; our 
predictions are also available in an Excel file, which has been uploaded to arXiv\textregistered~as ancillary material.

\begin{ack}
This research programme has been shaped to its current form after the long-term interaction with B.L.~Birbrair, A.~Gashi, P.F.A.~Goudsmit, A.B.~Gridnev, H.J.~Leisi, G.C.~Oades (deceased), and W.S.~Woolcock (deceased). We are 
grateful to them for their contributions. As any further attempt to single out individual contributions is bound to yield a long list, the reader is referred to the acknowledgements in earlier papers.

We are grateful to I.I.~Strakovsky for having clarified a question regarding Refs.~\cite{XP15} and for having drawn our attention to the upcoming measurements of the $\pi^\pm p$ ES DCS at PSI by the MUSE Collaboration.

The Feynman graphs of this paper were drawn with the software package JaxoDraw \cite{Binosi2004,Binosi2009}, available from jaxodraw.sourceforge.net. The remaining figures have been created with MATLAB\textregistered~(The 
MathWorks, Inc., Natick, Massachusetts, United States).
\end{ack}

\newpage
\begin{table}[h!]
{\bf \caption{\label{tab:DBPIP}}}The datasets comprising the tDB$_+$. The columns correspond to details about each dataset as follows: the identifier of the dataset in the DB$_+$, the pion laboratory kinetic energy $T_j$ of 
the dataset (in MeV), the NDF after the removal of any outliers from the data, the scale factor $z_j$ of Eq.~(\ref{eq:EQ005}), the normalisation uncertainty $\delta z_j$ (reported or assigned), the value of $(\chi^2_j)_{\rm min}$ 
of Eq.~(\ref{eq:EQ006}), the corresponding p-value relating to the overall description of the dataset, and comments on any excluded DoFs. In case of free floating, $z_j$ is identified as the optimal scale factor $\hat{z}_j$ of 
Eq.~(\ref{eq:EQ013}). The entries of this table have been taken from the final fit of the phenomenological model of Section \ref{sec:KMPIP} at our $\mathrm{p}_{\rm min}$ threshold ($2.5 \sigma$ effects in the normal distribution).
\vspace{0.2cm}
\begin{center}
\begin{tabular}{|l|c|c|c|c|c|c|l|}
\hline
Identifier & $T_j$ & $N_j$ & $z_j$ & $\delta z_j$ & $(\chi^2_j)_{\rm min}$ & p-value & Comments\\
\hline
\hline
\multicolumn{8}{|c|}{DCSs}\\
\hline
BERTIN76 & $20.80$ & $10$ & $1.3814$ & $0.1156$ & $18.1415$ & $5.26 \cdot 10^{-2}$ &\\
BERTIN76 & $30.50$ & $10$ & $1.2325$ & $0.1052$ & $9.9974$ & $4.41 \cdot 10^{-1}$ &\\
BERTIN76 & $39.50$ & $8$ & $1.1670$ & $0.0956$ & $17.9613$ & $2.15 \cdot 10^{-2}$ & $75.05^\circ$, $85.81^\circ$ removed\\
BERTIN76 & $51.50$ & $10$ & $1.1199$ & $0.0828$ & $4.7106$ & $9.10 \cdot 10^{-1}$ &\\
BERTIN76 & $81.70$ & $10$ & $1.1009$ & $0.0506$ & $15.4857$ & $1.15 \cdot 10^{-1}$ &\\
BERTIN76 & $95.90$ & $9$ & $1.0199$ & $0.0354$ & $16.8587$ & $5.10 \cdot 10^{-2}$ & $65.67^\circ$ removed\\
AULD79 & $47.90$ & $11$ & $1.0010$ & $0.0866$ & $14.4613$ & $2.09 \cdot 10^{-1}$ &\\
RITCHIE83 & $65.00$ & $8$ & $1.0446$ & $0.0240$ & $17.1272$ & $2.88 \cdot 10^{-2}$ &\\
RITCHIE83 & $72.50$ & $10$ & $1.0077$ & $0.0200$ & $4.8002$ & $9.04 \cdot 10^{-1}$ &\\
RITCHIE83 & $80.00$ & $10$ & $1.0319$ & $0.0140$ & $20.0180$ & $2.91 \cdot 10^{-2}$ &\\
RITCHIE83 & $95.00$ & $10$ & $1.0329$ & $0.0150$ & $12.7192$ & $2.40 \cdot 10^{-1}$ &\\
FRANK83 & $29.40$ & $28$ & $0.9682$ & $0.0370$ & $20.0884$ & $8.61 \cdot 10^{-1}$ &\\
FRANK83 & $49.50$ & $28$ & $1.0352$ & $0.2030$ & $33.2761$ & $2.26 \cdot 10^{-1}$ &\\
FRANK83 & $69.60$ & $27$ & $0.9284$ & $0.0950$ & $22.9225$ & $6.89 \cdot 10^{-1}$ &\\
FRANK83 & $89.60$ & $27$ & $0.8586$ & $0.0470$ & $33.0339$ & $1.96 \cdot 10^{-1}$ &\\
BRACK86 & $66.80$ & $4$ & $0.8916$ & $0.0120$ & $2.8093$ & $5.90 \cdot 10^{-1}$ & freely floated\\
BRACK86 & $86.80$ & $8$ & $0.9400$ & $0.0140$ & $12.8296$ & $1.18 \cdot 10^{-1}$ & freely floated\\
BRACK86 & $91.70$ & $5$ & $0.9734$ & $0.0120$ & $11.3146$ & $4.55 \cdot 10^{-2}$ &\\
BRACK86 & $97.90$ & $5$ & $0.9700$ & $0.0150$ & $9.3286$ & $9.67 \cdot 10^{-2}$ &\\
BRACK88 & $66.80$ & $6$ & $0.9480$ & $0.0210$ & $10.9113$ & $9.12 \cdot 10^{-2}$ &\\
BRACK88 & $66.80$ & $6$ & $0.9569$ & $0.0210$ & $9.6389$ & $1.41 \cdot 10^{-1}$ &\\
WIEDNER89 & $54.30$ & $19$ & $0.9864$ & $0.0304$ & $15.2476$ & $7.07 \cdot 10^{-1}$ &\\
\hline
\end{tabular}
\end{center}
\end{table}

\newpage
\begin{table*}
{\bf Table \ref{tab:DBPIP} continued}
\vspace{0.2cm}
\begin{center}
\begin{tabular}{|l|c|c|c|c|c|c|l|}
\hline
Identifier & $T_j$ & $N_j$ & $z_j$ & $\delta z_j$ & $(\chi^2_j)_{\rm min}$ & p-value & Comments\\
\hline
\hline
BRACK90 & $30.00$ & $6$ & $1.0731$ & $0.0360$ & $14.4586$ & $2.49 \cdot 10^{-2}$ &\\
BRACK90 & $45.00$ & $8$ & $0.9984$ & $0.0220$ & $7.7323$ & $4.60 \cdot 10^{-1}$ &\\
BRACK95 & $87.10$ & $8$ & $0.9631$ & $0.0220$ & $13.3617$ & $1.00 \cdot 10^{-1}$ &\\
BRACK95 & $98.10$ & $8$ & $0.9729$ & $0.0200$ & $18.2003$ & $1.98 \cdot 10^{-2}$ &\\
JORAM95 & $45.10$ & $8$ & $0.9573$ & $0.0330$ & $14.7530$ & $6.41 \cdot 10^{-2}$ & $124.42^\circ$, $131.69^\circ$ removed\\
JORAM95 & $68.60$ & $9$ & $1.0445$ & $0.0440$ & $12.0324$ & $2.11 \cdot 10^{-1}$ &\\
JORAM95 & $32.20$ & $19$ & $0.9924$ & $0.0340$ & $27.7187$ & $8.90 \cdot 10^{-2}$ & $37.40^\circ$ removed\\
JORAM95 & $44.60$ & $18$ & $0.9432$ & $0.0340$ & $33.2700$ & $1.55 \cdot 10^{-2}$ & $30.74^\circ$, $35.40^\circ$ removed\\
\hline
\multicolumn{8}{|c|}{APs}\\
\hline
SEVIOR89 & $98.00$ & $6$ & $1.0184$ & $0.0500$ & $5.3257$ & $5.03 \cdot 10^{-1}$ &\\
WIESER96 & $68.34$ & $3$ & $0.9162$ & $0.0500$ & $3.9196$ & $2.70 \cdot 10^{-1}$ &\\
WIESER96 & $68.34$ & $4$ & $0.9387$ & $0.0500$ & $4.1846$ & $3.82 \cdot 10^{-1}$ &\\
MEIER04 & $57.20-87.20$ & $12$ & $0.9861$ & $0.0350$ & $14.0425$ & $2.98 \cdot 10^{-1}$ &\\
MEIER04 & $45.20$, $51.20$ & $6$ & $0.9668$ & $0.0350$ & $8.6506$ & $1.94 \cdot 10^{-1}$ &\\
MEIER04 & $57.30-87.20$ & $7$ & $1.0097$ & $0.0350$ & $11.7786$ & $1.08 \cdot 10^{-1}$ &\\
\hline
\multicolumn{8}{|c|}{PTCSs}\\
\hline
KRISS97 & $39.80$ & $1$ & $1.0106$ & $0.0300$ & $1.2918$ & $2.56 \cdot 10^{-1}$ &\\
KRISS97 & $40.50$ & $1$ & $1.0014$ & $0.0300$ & $0.0853$ & $7.70 \cdot 10^{-1}$ &\\
KRISS97 & $44.70$ & $1$ & $1.0008$ & $0.0300$ & $0.0067$ & $9.35 \cdot 10^{-1}$ &\\
KRISS97 & $45.30$ & $1$ & $1.0011$ & $0.0300$ & $0.0089$ & $9.25 \cdot 10^{-1}$ &\\
KRISS97 & $51.10$ & $1$ & $1.0228$ & $0.0300$ & $2.9805$ & $8.43 \cdot 10^{-2}$ &\\
KRISS97 & $51.70$ & $1$ & $1.0014$ & $0.0300$ & $0.0141$ & $9.05 \cdot 10^{-1}$ &\\
KRISS97 & $54.80$ & $1$ & $1.0052$ & $0.0300$ & $0.0802$ & $7.77 \cdot 10^{-1}$ &\\
KRISS97 & $59.30$ & $1$ & $1.0237$ & $0.0300$ & $1.1036$ & $2.93 \cdot 10^{-1}$ &\\
KRISS97 & $66.30$ & $2$ & $1.0494$ & $0.0300$ & $3.9635$ & $1.38 \cdot 10^{-1}$ &\\
KRISS97 & $66.80$ & $2$ & $1.0073$ & $0.0300$ & $0.5713$ & $7.52 \cdot 10^{-1}$ &\\
KRISS97 & $80.00$ & $1$ & $1.0138$ & $0.0300$ & $0.3493$ & $5.55 \cdot 10^{-1}$ &\\
KRISS97 & $89.30$ & $1$ & $1.0075$ & $0.0300$ & $0.2539$ & $6.14 \cdot 10^{-1}$ &\\
KRISS97 & $99.20$ & $1$ & $1.0527$ & $0.0300$ & $3.7620$ & $5.24 \cdot 10^{-2}$ &\\
\hline
\end{tabular}
\end{center}
\end{table*}

\newpage
\begin{table*}
{\bf Table \ref{tab:DBPIP} continued}
\vspace{0.2cm}
\begin{center}
\begin{tabular}{|l|c|c|c|c|c|c|l|}
\hline
Identifier & $T_j$ & $N_j$ & $z_j$ & $\delta z_j$ & $(\chi^2_j)_{\rm min}$ & p-value & Comments\\
\hline
\hline
FRIEDMAN99 & $45.00$ & $1$ & $1.0399$ & $0.0600$ & $1.8882$ & $1.69 \cdot 10^{-1}$ &\\
FRIEDMAN99 & $52.10$ & $1$ & $1.0153$ & $0.0600$ & $0.1933$ & $6.60 \cdot 10^{-1}$ &\\
FRIEDMAN99 & $63.10$ & $1$ & $1.0351$ & $0.0600$ & $0.4590$ & $4.98 \cdot 10^{-1}$ &\\
FRIEDMAN99 & $67.45$ & $2$ & $1.0515$ & $0.0600$ & $1.2268$ & $5.42 \cdot 10^{-1}$ &\\
FRIEDMAN99 & $71.50$ & $2$ & $1.0494$ & $0.0600$ & $0.8283$ & $6.61 \cdot 10^{-1}$ &\\
FRIEDMAN99 & $92.50$ & $2$ & $1.0411$ & $0.0600$ & $0.5393$ & $7.64 \cdot 10^{-1}$ &\\
\hline
\multicolumn{8}{|c|}{TCSs}\\
\hline
CARTER71 & $71.60$ & $1$ & $1.0924$ & $0.0600$ & $2.6905$ & $1.01 \cdot 10^{-1}$ &\\
CARTER71 & $97.40$ & $1$ & $1.0485$ & $0.0600$ & $0.6574$ & $4.17 \cdot 10^{-1}$ &\\
PEDRONI78 & $72.50$ & $1$ & $1.0122$ & $0.0600$ & $0.1355$ & $7.13 \cdot 10^{-1}$ &\\
PEDRONI78 & $84.80$ & $1$ & $1.0312$ & $0.0600$ & $0.3277$ & $5.67 \cdot 10^{-1}$ &\\
PEDRONI78 & $95.10$ & $1$ & $1.0222$ & $0.0600$ & $0.1893$ & $6.63 \cdot 10^{-1}$ &\\
PEDRONI78 & $96.90$ & $1$ & $1.0160$ & $0.0600$ & $0.1214$ & $7.28 \cdot 10^{-1}$ &\\
\hline
\end{tabular}
\end{center}
\end{table*}

\newpage
\begin{table}[h!]
{\bf \caption{\label{tab:DBPIMEL}}}The equivalent of Table \ref{tab:DBPIP} for the tDB$_-$. The entries of this table have been taken from the final fit of the phenomenological model of Section \ref{sec:KMPIM}.
\vspace{0.2cm}
\begin{center}
\begin{tabular}{|l|c|c|c|c|c|c|l|}
\hline
Identifier & $T_j$ & $N_j$ & $z_j$ & $\delta z_j$ & $(\chi^2_j)_{\rm min}$ & p-value & Comments\\
\hline
\hline
\multicolumn{8}{|c|}{DCSs}\\
\hline
FRANK83 & $29.40$ & $28$ & $0.9756$ & $0.0350$ & $31.4977$ & $2.95 \cdot 10^{-1}$ &\\
FRANK83 & $49.50$ & $28$ & $1.1025$ & $0.0780$ & $29.1661$ & $4.04 \cdot 10^{-1}$ &\\
FRANK83 & $69.60$ & $27$ & $1.0910$ & $0.2530$ & $24.3930$ & $6.08 \cdot 10^{-1}$ &\\
FRANK83 & $89.60$ & $27$ & $0.9437$ & $0.1390$ & $24.9176$ & $5.79 \cdot 10^{-1}$ &\\
BRACK86 & $66.80$ & $5$ & $0.9974$ & $0.0130$ & $13.8807$ & $1.64 \cdot 10^{-2}$ &\\
BRACK86 & $86.80$ & $5$ & $1.0032$ & $0.0120$ & $1.3746$ & $9.27 \cdot 10^{-1}$ &\\
BRACK86 & $91.70$ & $5$ & $0.9963$ & $0.0120$ & $2.8582$ & $7.22 \cdot 10^{-1}$ &\\
BRACK86 & $97.90$ & $5$ & $1.0003$ & $0.0120$ & $6.0855$ & $2.98 \cdot 10^{-1}$ &\\
WIEDNER89 & $54.30$ & $18$ & $1.1571$ & $0.0304$ & $23.5575$ & $1.70 \cdot 10^{-1}$ & $15.55^\circ$ removed,\\
 & & & & & & & freely floated\\
BRACK90 & $30.00$ & $5$ & $1.0151$ & $0.0200$ & $4.2504$ & $5.14 \cdot 10^{-1}$ &\\
BRACK90 & $45.00$ & $9$ & $1.0547$ & $0.0220$ & $13.0574$ & $1.60 \cdot 10^{-1}$ &\\
BRACK95 & $87.50$ & $6$ & $0.9792$ & $0.0220$ & $10.8205$ & $9.41 \cdot 10^{-2}$ &\\
BRACK95 & $98.10$ & $7$ & $1.0054$ & $0.0210$ & $7.4637$ & $3.82 \cdot 10^{-1}$ & $36.70^\circ$ removed\\
JORAM95 & $32.70$ & $4$ & $0.9905$ & $0.0330$ & $3.9748$ & $4.09 \cdot 10^{-1}$ &\\
JORAM95 & $32.70$ & $2$ & $0.9514$ & $0.0330$ & $6.0703$ & $4.81 \cdot 10^{-2}$ &\\
JORAM95 & $45.10$ & $4$ & $0.9564$ & $0.0330$ & $12.0238$ & $1.72 \cdot 10^{-2}$ &\\
JORAM95 & $45.10$ & $3$ & $0.9471$ & $0.0330$ & $9.0414$ & $2.87 \cdot 10^{-2}$ &\\
JORAM95 & $68.60$ & $7$ & $1.0820$ & $0.0440$ & $14.1650$ & $4.83 \cdot 10^{-2}$ &\\
JORAM95 & $68.60$ & $3$ & $1.0324$ & $0.0440$ & $2.2604$ & $5.20 \cdot 10^{-1}$ &\\
JORAM95 & $32.20$ & $20$ & $1.0585$ & $0.0340$ & $20.9047$ & $4.03 \cdot 10^{-1}$ &\\
JORAM95 & $44.60$ & $20$ & $0.9437$ & $0.0340$ & $30.2546$ & $6.58 \cdot 10^{-2}$ &\\
JANOUSCH97 & $43.60$ & $1$ & $1.0483$ & $0.1500$ & $0.2263$ & $6.34 \cdot 10^{-1}$ &\\
JANOUSCH97 & $50.30$ & $1$ & $1.0459$ & $0.1500$ & $0.2561$ & $6.13 \cdot 10^{-1}$ &\\
JANOUSCH97 & $57.30$ & $1$ & $1.0826$ & $0.1500$ & $4.9367$ & $2.63 \cdot 10^{-2}$ &\\
JANOUSCH97 & $64.50$ & $1$ & $1.0054$ & $0.1500$ & $0.0019$ & $9.65 \cdot 10^{-1}$ &\\
JANOUSCH97 & $72.00$ & $1$ & $1.2940$ & $0.1500$ & $4.4930$ & $3.40 \cdot 10^{-2}$ &\\
\hline
\end{tabular}
\end{center}
\end{table}

\newpage
\begin{table*}
{\bf Table \ref{tab:DBPIMEL} continued}
\vspace{0.2cm}
\begin{center}
\begin{tabular}{|l|c|c|c|c|c|c|l|}
\hline
Identifier & $T_j$ & $N_j$ & $z_j$ & $\delta z_j$ & $(\chi^2_j)_{\rm min}$ & p-value & Comments\\
\hline
\hline
\multicolumn{8}{|c|}{$a_{cc}$ from the strong shift $\epsilon_{1 s}$ in pionic hydrogen}\\
\hline
SCHROEDER01 & $0.00$ & $1$ & $1.0003$ & $0.0082$ & $0.0010$ & $9.75 \cdot 10^{-1}$ &\\
HENNEBACH14 & $0.00$ & $1$ & $1.0004$ & $0.0067$ & $0.0032$ & $9.55 \cdot 10^{-1}$ &\\
\hline
\multicolumn{8}{|c|}{APs}\\
\hline
ALDER83 & $98.00$ & $6$ & $1.0106$ & $0.0400$ & $5.4465$ & $4.88 \cdot 10^{-1}$ &\\
SEVIOR89 & $98.00$ & $5$ & $0.9947$ & $0.0500$ & $1.5600$ & $9.06 \cdot 10^{-1}$ &\\
HOFMAN98 & $86.80$ & $11$ & $1.0013$ & $0.0300$ & $6.2676$ & $8.55 \cdot 10^{-1}$ &\\
PATTERSON02 & $57.20$ & $10$ & $0.9459$ & $0.0370$ & $11.2906$ & $3.35 \cdot 10^{-1}$ &\\
PATTERSON02 & $66.90$ & $9$ & $0.9994$ & $0.0370$ & $5.2086$ & $8.16 \cdot 10^{-1}$ &\\
PATTERSON02 & $66.90$ & $10$ & $0.9598$ & $0.0370$ & $14.0940$ & $1.69 \cdot 10^{-1}$ &\\
PATTERSON02 & $87.20$ & $11$ & $0.9828$ & $0.0370$ & $8.0789$ & $7.06 \cdot 10^{-1}$ &\\
PATTERSON02 & $87.20$ & $11$ & $0.9932$ & $0.0370$ & $4.8343$ & $9.39 \cdot 10^{-1}$ &\\
PATTERSON02 & $98.00$ & $12$ & $1.0061$ & $0.0370$ & $6.0656$ & $9.13 \cdot 10^{-1}$ &\\
MEIER04 & $67.30$, $87.20$ & $3$ & $0.9935$ & $0.0350$ & $2.9768$ & $3.95 \cdot 10^{-1}$ &\\
\hline
\end{tabular}
\end{center}
\end{table*}

\newpage
\begin{table}[h!]
{\bf \caption{\label{tab:DBPIMCX}}}The equivalent of Table \ref{tab:DBPIP} for the tDB$_0$. The entries of this table have been taken from the final fit of the phenomenological model of Section \ref{sec:KMPIM}.
\vspace{0.2cm}
\begin{center}
\begin{tabular}{|l|c|c|c|c|c|c|l|}
\hline
Identifier & $T_j$ & $N_j$ & $z_j$ & $\delta z_j$ & $(\chi^2_j)_{\rm min}$ & p-value & Comments\\
\hline
\hline
\multicolumn{8}{|c|}{DCSs}\\
\hline
DUCLOS73 & $22.60$ & $1$ & $0.9099$ & $0.1235$ & $0.8229$ & $3.64 \cdot 10^{-1}$ &\\
DUCLOS73 & $32.90$ & $1$ & $0.9566$ & $0.1230$ & $0.1853$ & $6.67 \cdot 10^{-1}$ &\\
DUCLOS73 & $42.60$ & $1$ & $0.8763$ & $0.1226$ & $1.3972$ & $2.37 \cdot 10^{-1}$ &\\
FITZGERALD86 & $32.48$ & $2$ & $1.5043$ & $0.0780$ & $2.5186$ & $2.84 \cdot 10^{-1}$ & freely floated\\
FITZGERALD86 & $36.11$ & $2$ & $1.7233$ & $0.0780$ & $1.4481$ & $4.85 \cdot 10^{-1}$ & freely floated\\
FITZGERALD86 & $40.26$ & $2$ & $1.8199$ & $0.0780$ & $7.4565$ & $2.40 \cdot 10^{-2}$ & freely floated\\
FITZGERALD86 & $47.93$ & $3$ & $1.1554$ & $0.0780$ & $10.1351$ & $1.75 \cdot 10^{-2}$ &\\
FITZGERALD86 & $51.78$ & $3$ & $1.0812$ & $0.0780$ & $4.9665$ & $1.74 \cdot 10^{-1}$ &\\
FITZGERALD86 & $55.58$ & $3$ & $1.0593$ & $0.0780$ & $1.1684$ & $7.61 \cdot 10^{-1}$ &\\
FITZGERALD86 & $63.21$ & $3$ & $1.0336$ & $0.0780$ & $0.8127$ & $8.46 \cdot 10^{-1}$ &\\
FRLE{\v Z}98 & $27.50$ & $6$ & $1.0849$ & $0.0870$ & $10.5626$ & $1.03 \cdot 10^{-1}$ &\\
ISENHOWER99 & $10.60$ & $4$ & $1.0143$ & $0.0600$ & $2.0290$ & $7.30 \cdot 10^{-1}$ &\\
ISENHOWER99 & $10.60$ & $5$ & $1.0008$ & $0.0400$ & $1.4098$ & $9.23 \cdot 10^{-1}$ &\\
ISENHOWER99 & $10.60$ & $6$ & $1.0106$ & $0.0400$ & $7.8596$ & $2.49 \cdot 10^{-1}$ &\\
ISENHOWER99 & $20.60$ & $5$ & $0.9747$ & $0.0400$ & $1.8101$ & $8.75 \cdot 10^{-1}$ &\\
ISENHOWER99 & $20.60$ & $6$ & $1.0064$ & $0.0400$ & $8.1110$ & $2.30 \cdot 10^{-1}$ &\\
ISENHOWER99 & $39.40$ & $4$ & $1.0676$ & $0.0600$ & $7.2825$ & $1.22 \cdot 10^{-1}$ &\\
ISENHOWER99 & $39.40$ & $5$ & $1.0586$ & $0.0400$ & $8.5425$ & $1.29 \cdot 10^{-1}$ &\\
ISENHOWER99 & $39.40$ & $5$ & $0.9509$ & $0.0400$ & $5.1608$ & $3.97 \cdot 10^{-1}$ &\\
SADLER04 & $63.86$ & $20$ & $0.9574$ & $0.0650$ & $15.2518$ & $7.62 \cdot 10^{-1}$ &\\
SADLER04 & $83.49$ & $20$ & $0.9879$ & $0.0520$ & $12.0355$ & $9.15 \cdot 10^{-1}$ &\\
SADLER04 & $94.57$ & $20$ & $1.0277$ & $0.0450$ & $6.8628$ & $9.97 \cdot 10^{-1}$ &\\
JIA08 & $34.37$ & $4$ & $0.8414$ & $0.1000$ & $4.8722$ & $3.01 \cdot 10^{-1}$ &\\
JIA08 & $39.95$ & $4$ & $0.8599$ & $0.1000$ & $3.4418$ & $4.87 \cdot 10^{-1}$ &\\
JIA08 & $43.39$ & $4$ & $0.8648$ & $0.1000$ & $2.9961$ & $5.58 \cdot 10^{-1}$ &\\
JIA08 & $46.99$ & $4$ & $0.9616$ & $0.1000$ & $4.6958$ & $3.20 \cdot 10^{-1}$ &\\
\hline
\end{tabular}
\end{center}
\end{table}

\newpage
\begin{table*}
{\bf Table \ref{tab:DBPIMCX} continued}
\vspace{0.2cm}
\begin{center}
\begin{tabular}{|l|c|c|c|c|c|c|l|}
\hline
Identifier & $T_j$ & $N_j$ & $z_j$ & $\delta z_j$ & $(\chi^2_j)_{\rm min}$ & p-value & Comments\\
\hline
\hline
\multicolumn{8}{|c|}{DCSs}\\
\hline
JIA08 & $54.19$ & $4$ & $0.8768$ & $0.1000$ & $2.9406$ & $5.68 \cdot 10^{-1}$ &\\
JIA08 & $59.68$ & $4$ & $0.9061$ & $0.1000$ & $3.7182$ & $4.45 \cdot 10^{-1}$ &\\
MEKTEROVI{\'C}09 & $33.89$ & $20$ & $1.0215$ & $0.0340$ & $16.8203$ & $6.65 \cdot 10^{-1}$ &\\
MEKTEROVI{\'C}09 & $39.38$ & $20$ & $1.0135$ & $0.0260$ & $15.0774$ & $7.72 \cdot 10^{-1}$ &\\
MEKTEROVI{\'C}09 & $44.49$ & $20$ & $1.0101$ & $0.0270$ & $33.2469$ & $3.17 \cdot 10^{-2}$ &\\
MEKTEROVI{\'C}09 & $51.16$ & $20$ & $1.0369$ & $0.0290$ & $14.6495$ & $7.96 \cdot 10^{-1}$ &\\
MEKTEROVI{\'C}09 & $57.41$ & $20$ & $1.0413$ & $0.0290$ & $19.6849$ & $4.78 \cdot 10^{-1}$ &\\
MEKTEROVI{\'C}09 & $66.79$ & $20$ & $1.0251$ & $0.0300$ & $21.1509$ & $3.88 \cdot 10^{-1}$ &\\
MEKTEROVI{\'C}09 & $86.62$ & $20$ & $1.0012$ & $0.0290$ & $30.2117$ & $6.65 \cdot 10^{-2}$ &\\
\hline
\multicolumn{8}{|c|}{Coefficients in the Legendre expansion of the DCS}\\
\hline
SALOMON84 & $27.40$ & $3$ & $0.9705$ & $0.0310$ & $3.0071$ & $3.91 \cdot 10^{-1}$ &\\
SALOMON84 & $39.30$ & $3$ & $0.9937$ & $0.0310$ & $1.1235$ & $7.71 \cdot 10^{-1}$ &\\
BAGHERI88 & $45.60$ & $3$ & $1.0060$ & $0.0310$ & $0.1288$ & $9.88 \cdot 10^{-1}$ &\\
BAGHERI88 & $62.20$ & $3$ & $0.9607$ & $0.0310$ & $3.1952$ & $3.62 \cdot 10^{-1}$ &\\
BAGHERI88 & $76.40$ & $3$ & $0.9739$ & $0.0310$ & $3.2850$ & $3.50 \cdot 10^{-1}$ &\\
BAGHERI88 & $91.70$ & $3$ & $1.0129$ & $0.0310$ & $3.0538$ & $3.83 \cdot 10^{-1}$ &\\
\hline
\multicolumn{8}{|c|}{$b_1$ from the total decay width $\Gamma_{1s}$ of pionic hydrogen}\\
\hline
SCHROEDER01 & $0.00$ & $1$ & $0.9850$ & $0.0218$ & $0.9720$ & $3.24 \cdot 10^{-1}$ &\\
\hline
\multicolumn{8}{|c|}{APs}\\
\hline
STA{\v S}KO93 & $100.00$ & $4$ & $0.9926$ & $0.0440$ & $1.5254$ & $8.22 \cdot 10^{-1}$ &\\
GAULARD99 & $98.10$ & $6$ & $1.0145$ & $0.0450$ & $0.8793$ & $9.90 \cdot 10^{-1}$ &\\
\hline
\end{tabular}
\end{center}
\end{table*}

\newpage
\begin{table*}
{\bf Table \ref{tab:DBPIMCX} continued}
\vspace{0.2cm}
\begin{center}
\begin{tabular}{|l|c|c|c|c|c|c|l|}
\hline
Identifier & $T_j$ & $N_j$ & $z_j$ & $\delta z_j$ & $(\chi^2_j)_{\rm min}$ & p-value & Comments\\
\hline
\hline
\multicolumn{8}{|c|}{TCSs}\\
\hline
BUGG71 & $90.90$ & $1$ & $1.0206$ & $0.0600$ & $0.1228$ & $7.26 \cdot 10^{-1}$ &\\
BREITSCHOPF06 & $38.90$ & $1$ & $0.9958$ & $0.0300$ & $0.1774$ & $6.74 \cdot 10^{-1}$ &\\
BREITSCHOPF06 & $43.00$ & $1$ & $1.0011$ & $0.0300$ & $0.0249$ & $8.74 \cdot 10^{-1}$ &\\
BREITSCHOPF06 & $47.10$ & $1$ & $0.9981$ & $0.0300$ & $0.0553$ & $8.14 \cdot 10^{-1}$ &\\
BREITSCHOPF06 & $55.60$ & $1$ & $0.9954$ & $0.0300$ & $0.1912$ & $6.62 \cdot 10^{-1}$ &\\
BREITSCHOPF06 & $64.30$ & $1$ & $0.9730$ & $0.0300$ & $3.6665$ & $5.55 \cdot 10^{-2}$ &\\
BREITSCHOPF06 & $65.90$ & $1$ & $0.9783$ & $0.0300$ & $2.2598$ & $1.33 \cdot 10^{-1}$ &\\
BREITSCHOPF06 & $76.10$ & $1$ & $0.9816$ & $0.0300$ & $1.5823$ & $2.08 \cdot 10^{-1}$ &\\
BREITSCHOPF06 & $96.50$ & $1$ & $0.9798$ & $0.0300$ & $0.7390$ & $3.90 \cdot 10^{-1}$ &\\
\hline
\end{tabular}
\end{center}
\end{table*}

\newpage
\begin{table}%[h!]
{\bf \caption{\label{tab:ReproductionPTCS}}}Reproduction of the $\pi^- p$ PTCSs; the data of this observable have not been included in the DB$_-$. The first four columns correspond to the details about each datapoint as follows: 
the identifier of the datapoint, the pion laboratory kinetic energy $T_j$ of the datapoint (in MeV), the laboratory-angle cut $\theta_L$ (in degrees), and the experimental result; statistical and systematic effects have been 
combined in quadrature. The last three columns contain predictions obtained from the two PSAs of this work. The quantity $\sigma_{\rm ES}^{\rm th}$ is the $\pi^- p$ ES PTCS corresponding to the quoted $\theta_L$ value. The 
quantity $\sigma_{+/-}^{\rm th}$ represents the sum of $\sigma_{\rm ES}^{\rm th}$ (entry in the fifth column) and of the $\pi^- p$ CX TCS - i.e., of the DCS integrated over the entire solid angle - predicted from the fits of 
the ETH model to the tDB$_{+/-}$. Similarly, the quantity $\sigma_{+/0}^{\rm th}$ represents the sum of $\sigma_{\rm ES}^{\rm th}$ (entry in the fifth column) and of the $\pi^- p$ CX TCS predicted from the fits of the ETH model 
to the tDB$_{+/0}$. All cross sections are expressed in mb.
\vspace{0.2cm}
\begin{center}
\begin{tabular}{|l|c|c|c|c|c|c|}
\hline
Identifier & $T_j$ & $\theta_L$ & $\sigma^{\rm exp}$ & $\sigma_{\rm ES}^{\rm th}$ & $\sigma_{+/-}^{\rm th}$ & $\sigma_{+/0}^{\rm th}$\\
\hline
\hline
FRIEDMAN90 & $50.00$ & $30$ & $8.50 \pm 0.79$ & $2.27 \pm 0.15$ & $8.45 \pm 0.18$ & $9.07 \pm 0.16$\\
KRISS97 & $80.00$ & $30$ & $14.60 \pm 0.60$ & $2.966 \pm 0.098$ & $14.31 \pm 0.10$ & $14.93 \pm 0.12$\\
KRISS97 & $99.20$ & $30$ & $23.4 \pm 1.1$ & $4.509 \pm 0.049$ & $21.84 \pm 0.14$ & $22.36 \pm 0.13$\\
\hline
\end{tabular}
\end{center}
\vspace{0.5cm}
\end{table}

\newpage
\begin{table}%[h!]
{\bf \caption{\label{tab:ReproductionTNCS}}}Reproduction of the $\pi^- p$ total-nuclear cross sections; the data of this quantity have not been included in the DB$_-$. The first three columns correspond to the details about 
each datapoint as follows: the identifier of the datapoint, the pion laboratory kinetic energy $T_j$ of the datapoint (in MeV), and the measurement itself; statistical and systematic effects have been combined in quadrature. 
The last two columns contain the predictions obtained from the two PSAs of this work: the quantity $\sigma_{+/-}^{\rm th}$ represents the prediction based on the fits of the ETH model to the tDB$_{+/-}$, whereas the quantity 
$\sigma_{+/0}^{\rm th}$ has been obtained from the fits of the ETH model to the tDB$_{+/0}$. All cross sections are expressed in mb.
\vspace{0.2cm}
\begin{center}
\begin{tabular}{|l|c|c|c|c|c|}
\hline
Identifier & $T_j$ & $\sigma^{\rm exp}$ & $\sigma_{+/-}^{\rm th}$ & $\sigma_{+/0}^{\rm th}$\\
\hline
\hline
CARTER71 & $76.70$ & $15.80 \pm 0.97$ & $14.24 \pm 0.11$ & $15.33 \pm 0.16$\\
CARTER71 & $96.00$ & $23.1 \pm 1.4$ & $21.80 \pm 0.12$ & $22.84 \pm 0.13$\\
PEDRONI78 & $72.50$ & $13.6 \pm 2.1$ & $13.06 \pm 0.13$ & $14.15 \pm 0.17$\\
PEDRONI78 & $84.80$ & $17.7 \pm 1.4$ & $16.948 \pm 0.097$ & $18.04 \pm 0.14$\\
PEDRONI78 & $95.10$ & $21.2 \pm 1.6$ & $21.35 \pm 0.11$ & $22.41 \pm 0.13$\\
PEDRONI78 & $96.90$ & $22.2 \pm 2.4$ & $22.25 \pm 0.12$ & $23.29 \pm 0.13$\\
\hline
\end{tabular}
\end{center}
\vspace{0.5cm}
\end{table}

\newpage
\begin{table}[h!]
{\bf \caption{\label{tab:PSAReprCX}}}The optimal scale factors $\hat{z}_j$ of the DCS/TCS datasets in the DB$_0$ using the predictions based on the fits of the ETH model to the tDB$_{+/-}$ as baseline solution. The first three 
columns correspond to the details about each dataset as follows: the identifier of the dataset in the DB$_0$, the pion laboratory kinetic energy $T_j$ of the dataset (in MeV), and its number of datapoints $N_j$. The columns 
`Overall', `Shape', and `Abs.~norm.' contain the p-values corresponding: a) to the overall reproduction of the dataset, b) to the reproduction of its shape, and c) to the reproduction of its absolute normalisation, as explained 
in Section \ref{sec:Reproduction}. The quantity $\delta \hat{z}_j$ is the total uncertainty of the optimal scale factor $\hat{z}_j$ (see beginning of Section \ref{sec:PSAReproductionCX}).
\vspace{0.2cm}
\begin{center}
\begin{tabular}{|l|c|c|c|c|c|c|c|}
\hline
Identifier & $T_j$ & $N_j$ & Overall & Shape & Abs.~norm. & $\hat{z}_j$ & $\delta \hat{z}_j$\\
\hline
\hline
\multicolumn{8}{|c|}{DCSs}\\
\hline
DUCLOS73 & $22.60$ & $1$ & $9.36 \cdot 10^{-1}$ & $-$ & $9.36 \cdot 10^{-1}$ & $1.0132$ & $0.1643$\\
DUCLOS73 & $32.90$ & $1$ & $6.55 \cdot 10^{-1}$ & $-$ & $6.55 \cdot 10^{-1}$ & $1.0706$ & $0.1579$\\
DUCLOS73 & $42.60$ & $1$ & $6.63 \cdot 10^{-1}$ & $-$ & $6.63 \cdot 10^{-1}$ & $0.9351$ & $0.1490$\\
FITZGERALD86 & $32.48$ & $3$ & $1.82 \cdot 10^{-9}$ & $2.59 \cdot 10^{-1}$ & $1.59 \cdot 10^{-10}$ & $1.8970$ & $0.1402$\\
FITZGERALD86 & $36.11$ & $3$ & $5.58 \cdot 10^{-10}$ & $6.42 \cdot 10^{-1}$ & $1.83 \cdot 10^{-11}$ & $1.9797$ & $0.1458$\\
FITZGERALD86 & $40.26$ & $3$ & $2.07 \cdot 10^{-6}$ & $9.38 \cdot 10^{-1}$ & $7.09 \cdot 10^{-8}$ & $1.6654$ & $0.1235$\\
FITZGERALD86 & $47.93$ & $3$ & $3.87 \cdot 10^{-2}$ & $8.36 \cdot 10^{-2}$ & $6.43 \cdot 10^{-2}$ & $0.8224$ & $0.0960$\\
FITZGERALD86 & $51.78$ & $3$ & $2.05 \cdot 10^{-2}$ & $7.37 \cdot 10^{-2}$ & $3.27 \cdot 10^{-2}$ & $0.8073$ & $0.0902$\\
FITZGERALD86 & $55.58$ & $3$ & $3.92 \cdot 10^{-1}$ & $5.45 \cdot 10^{-1}$ & $1.82 \cdot 10^{-1}$ & $0.8797$ & $0.0901$\\
FITZGERALD86 & $63.21$ & $3$ & $8.52 \cdot 10^{-1}$ & $7.52 \cdot 10^{-1}$ & $6.40 \cdot 10^{-1}$ & $0.9585$ & $0.0886$\\
FRLE{\v Z}98 & $27.50$ & $6$ & $4.28 \cdot 10^{-5}$ & $2.26 \cdot 10^{-2}$ & $4.34 \cdot 10^{-5}$ & $1.4034$ & $0.0987$\\
ISENHOWER99 & $10.60$ & $4$ & $6.45 \cdot 10^{-3}$ & $6.09 \cdot 10^{-1}$ & $4.17 \cdot 10^{-4}$ & $1.4235$ & $0.1200$\\
ISENHOWER99 & $10.60$ & $5$ & $7.72 \cdot 10^{-3}$ & $8.43 \cdot 10^{-1}$ & $1.56 \cdot 10^{-4}$ & $1.3036$ & $0.0803$\\
ISENHOWER99 & $10.60$ & $6$ & $2.80 \cdot 10^{-5}$ & $1.79 \cdot 10^{-1}$ & $1.49 \cdot 10^{-6}$ & $1.2747$ & $0.0571$\\
ISENHOWER99 & $20.60$ & $5$ & $2.84 \cdot 10^{-2}$ & $9.65 \cdot 10^{-1}$ & $5.53 \cdot 10^{-4}$ & $1.1803$ & $0.0522$\\
ISENHOWER99 & $20.60$ & $6$ & $2.58 \cdot 10^{-4}$ & $1.95 \cdot 10^{-1}$ & $1.90 \cdot 10^{-5}$ & $1.1999$ & $0.0467$\\
ISENHOWER99 & $39.40$ & $4$ & $2.08 \cdot 10^{-1}$ & $6.64 \cdot 10^{-1}$ & $3.80 \cdot 10^{-2}$ & $1.2084$ & $0.1004$\\
ISENHOWER99 & $39.40$ & $5$ & $1.45 \cdot 10^{-5}$ & $3.01 \cdot 10^{-1}$ & $5.26 \cdot 10^{-7}$ & $1.2262$ & $0.0451$\\
ISENHOWER99 & $39.40$ & $5$ & $3.22 \cdot 10^{-1}$ & $5.59 \cdot 10^{-1}$ & $9.15 \cdot 10^{-2}$ & $1.0718$ & $0.0425$\\
SADLER04 & $63.86$ & $20$ & $8.90 \cdot 10^{-1}$ & $8.74 \cdot 10^{-1}$ & $5.12 \cdot 10^{-1}$ & $1.0447$ & $0.0680$\\
SADLER04 & $83.49$ & $20$ & $2.76 \cdot 10^{-1}$ & $2.52 \cdot 10^{-1}$ & $4.40 \cdot 10^{-1}$ & $1.0410$ & $0.0532$\\
SADLER04 & $94.57$ & $20$ & $8.39 \cdot 10^{-1}$ & $8.92 \cdot 10^{-1}$ & $1.60 \cdot 10^{-1}$ & $1.0660$ & $0.0469$\\
\hline
\end{tabular}
\end{center}
\end{table}

\newpage
\begin{table*}
{\bf Table \ref{tab:PSAReprCX} continued}
\vspace{0.2cm}
\begin{center}
\begin{tabular}{|l|c|c|c|c|c|c|c|}
\hline
Identifier & $T_j$ & $N_j$ & Overall & Shape & Abs.~norm. & $\hat{z}_j$ & $\delta \hat{z}_j$\\
\hline
\hline
JIA08 & $34.37$ & $4$ & $7.57 \cdot 10^{-1}$ & $6.03 \cdot 10^{-1}$ & $8.65 \cdot 10^{-1}$ & $0.9807$ & $0.1141$\\
JIA08 & $39.95$ & $4$ & $5.31 \cdot 10^{-1}$ & $6.08 \cdot 10^{-1}$ & $2.48 \cdot 10^{-1}$ & $0.8624$ & $0.1192$\\
JIA08 & $43.39$ & $4$ & $1.84 \cdot 10^{-1}$ & $4.38 \cdot 10^{-1}$ & $6.15 \cdot 10^{-2}$ & $0.7662$ & $0.1250$\\
JIA08 & $46.99$ & $4$ & $8.44 \cdot 10^{-1}$ & $9.77 \cdot 10^{-1}$ & $2.74 \cdot 10^{-1}$ & $0.8613$ & $0.1269$\\
JIA08 & $54.19$ & $4$ & $1.51 \cdot 10^{-1}$ & $9.71 \cdot 10^{-1}$ & $1.09 \cdot 10^{-2}$ & $0.6990$ & $0.1182$\\
JIA08 & $59.68$ & $4$ & $1.50 \cdot 10^{-1}$ & $3.22 \cdot 10^{-1}$ & $7.10 \cdot 10^{-2}$ & $0.7907$ & $0.1159$\\
MEKTEROVI{\'C}09 & $33.89$ & $20$ & $4.85 \cdot 10^{-3}$ & $5.89 \cdot 10^{-1}$ & $1.55 \cdot 10^{-6}$ & $1.1913$ & $0.0398$\\
MEKTEROVI{\'C}09 & $39.38$ & $20$ & $4.09 \cdot 10^{-3}$ & $7.50 \cdot 10^{-1}$ & $3.20 \cdot 10^{-7}$ & $1.1624$ & $0.0318$\\
MEKTEROVI{\'C}09 & $44.49$ & $20$ & $8.29 \cdot 10^{-5}$ & $2.83 \cdot 10^{-2}$ & $5.78 \cdot 10^{-6}$ & $1.1424$ & $0.0314$\\
MEKTEROVI{\'C}09 & $51.16$ & $20$ & $1.33 \cdot 10^{-2}$ & $8.81 \cdot 10^{-1}$ & $7.74 \cdot 10^{-7}$ & $1.1655$ & $0.0335$\\
MEKTEROVI{\'C}09 & $57.41$ & $20$ & $7.44 \cdot 10^{-4}$ & $2.47 \cdot 10^{-1}$ & $1.27 \cdot 10^{-6}$ & $1.1528$ & $0.0315$\\
MEKTEROVI{\'C}09 & $66.79$ & $20$ & $1.33 \cdot 10^{-3}$ & $3.18 \cdot 10^{-2}$ & $4.13 \cdot 10^{-4}$ & $1.1141$ & $0.0323$\\
MEKTEROVI{\'C}09 & $86.62$ & $20$ & $2.83 \cdot 10^{-3}$ & $3.90 \cdot 10^{-3}$ & $1.13 \cdot 10^{-1}$ & $1.0480$ & $0.0303$\\
\hline
\multicolumn{8}{|c|}{Coefficients in the Legendre expansion of the DCS}\\
\hline
SALOMON84 & $27.40$ & $3$ & $3.83 \cdot 10^{-1}$ & $5.45 \cdot 10^{-1}$ & $1.75 \cdot 10^{-1}$ & $1.0759$ & $0.0560$\\
SALOMON84 & $39.30$ & $3$ & $2.04 \cdot 10^{-1}$ & $6.65 \cdot 10^{-1}$ & $5.18 \cdot 10^{-2}$ & $1.1148$ & $0.0590$\\
BAGHERI88 & $45.60$ & $3$ & $3.85 \cdot 10^{-3}$ & $9.48 \cdot 10^{-1}$ & $2.66 \cdot 10^{-4}$ & $1.1337$ & $0.0367$\\
BAGHERI88 & $62.20$ & $3$ & $7.24 \cdot 10^{-1}$ & $8.66 \cdot 10^{-1}$ & $3.09 \cdot 10^{-1}$ & $1.0395$ & $0.0389$\\
BAGHERI88 & $76.40$ & $3$ & $3.81 \cdot 10^{-1}$ & $4.95 \cdot 10^{-1}$ & $1.97 \cdot 10^{-1}$ & $1.0464$ & $0.0360$\\
BAGHERI88 & $91.70$ & $3$ & $2.40 \cdot 10^{-1}$ & $5.16 \cdot 10^{-1}$ & $8.94 \cdot 10^{-2}$ & $1.0674$ & $0.0397$\\
\hline
\multicolumn{8}{|c|}{$b_1$ from the total decay width $\Gamma_{1s}$ of pionic hydrogen}\\
\hline
SCHROEDER01 & $0.00$ & $1$ & $4.93 \cdot 10^{-6}$ & $-$ & $4.93 \cdot 10^{-6}$ & $1.1639$ & $0.0359$\\
\hline
\end{tabular}
\end{center}
\end{table*}

\newpage
\begin{table*}
{\bf Table \ref{tab:PSAReprCX} continued}
\vspace{0.2cm}
\begin{center}
\begin{tabular}{|l|c|c|c|c|c|c|c|}
\hline
Identifier & $T_j$ & $N_j$ & Overall & Shape & Abs.~norm. & $\hat{z}_j$ & $\delta \hat{z}_j$\\
\hline
\hline
\multicolumn{8}{|c|}{TCSs}\\
\hline
BUGG71 & $90.90$ & $1$ & $2.86 \cdot 10^{-1}$ & $-$ & $2.86 \cdot 10^{-1}$ & $1.0661$ & $0.0620$\\
BREITSCHOPF06 & $38.90$ & $1$ & $3.42 \cdot 10^{-1}$ & $-$ & $3.42 \cdot 10^{-1}$ & $1.0980$ & $0.1031$\\
BREITSCHOPF06 & $43.00$ & $1$ & $3.04 \cdot 10^{-1}$ & $-$ & $3.04 \cdot 10^{-1}$ & $1.1543$ & $0.1500$\\
BREITSCHOPF06 & $47.10$ & $1$ & $4.58 \cdot 10^{-1}$ & $-$ & $4.58 \cdot 10^{-1}$ & $1.0917$ & $0.1235$\\
BREITSCHOPF06 & $55.60$ & $1$ & $4.93 \cdot 10^{-1}$ & $-$ & $4.93 \cdot 10^{-1}$ & $1.0636$ & $0.0928$\\
BREITSCHOPF06 & $64.30$ & $1$ & $5.29 \cdot 10^{-1}$ & $-$ & $5.29 \cdot 10^{-1}$ & $0.9568$ & $0.0686$\\
BREITSCHOPF06 & $65.90$ & $1$ & $8.28 \cdot 10^{-1}$ & $-$ & $8.28 \cdot 10^{-1}$ & $0.9855$ & $0.0668$\\
BREITSCHOPF06 & $75.10$ & $1$ & $1.11 \cdot 10^{-1}$ & $-$ & $1.11 \cdot 10^{-1}$ & $0.9207$ & $0.0497$\\
BREITSCHOPF06 & $76.10$ & $1$ & $8.48 \cdot 10^{-1}$ & $-$ & $8.48 \cdot 10^{-1}$ & $0.9875$ & $0.0652$\\
BREITSCHOPF06 & $96.50$ & $1$ & $9.79 \cdot 10^{-1}$ & $-$ & $9.79 \cdot 10^{-1}$ & $0.9990$ & $0.0396$\\
\hline
\end{tabular}
\end{center}
\end{table*}

\clearpage
\begin{figure}
\begin{center}
\includegraphics [width=15.5cm] {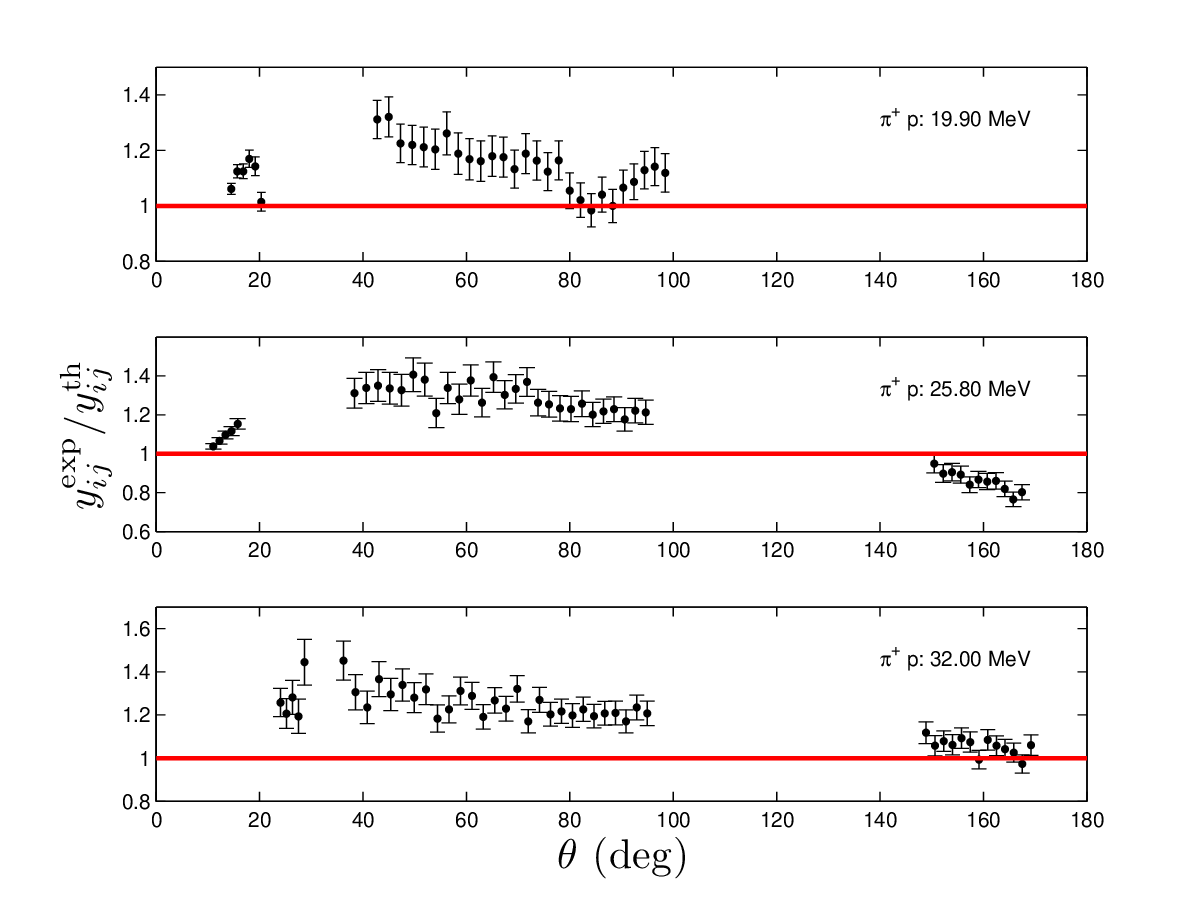}
\end{center}
\end{figure}

\clearpage
\begin{figure}
\begin{center}
\includegraphics [width=15.5cm] {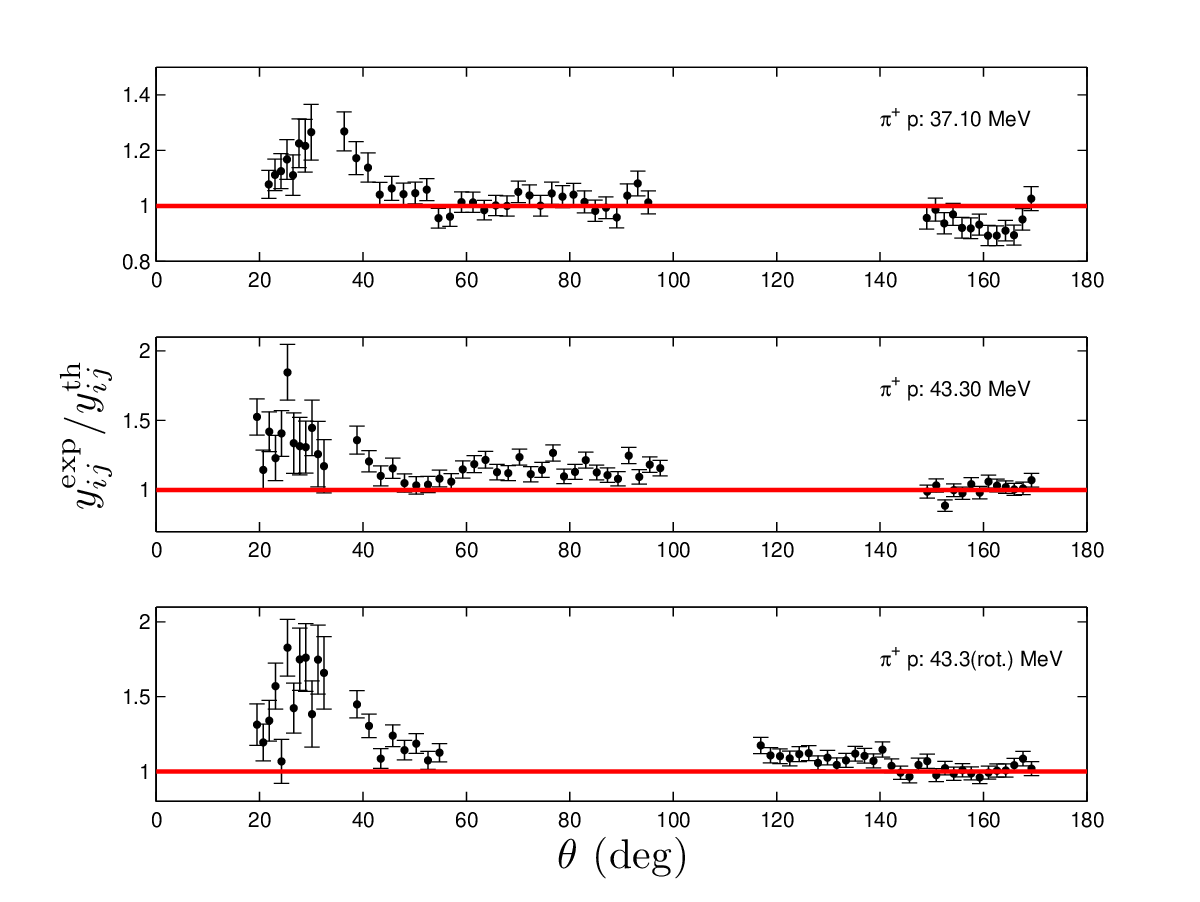}
\caption{\label{fig:PIPPEDenz}The DENZ04 $\pi^+ p$ DCSs \cite{Denz2006} ($y_{ij}^{\rm exp}$), normalised to the corresponding predictions ($y_{ij}^{\rm th}$) obtained from the results of the fits of the ETH model to the tDB$_{+/-}$. 
The normalisation uncertainties of the experimental datasets (see Refs.~\cite{Matsinos2012,Denz2006} for details) are not contained in the uncertainties shown.}
\vspace{0.35cm}
\end{center}
\end{figure}

\clearpage
\begin{figure}
\begin{center}
\includegraphics [width=15.5cm] {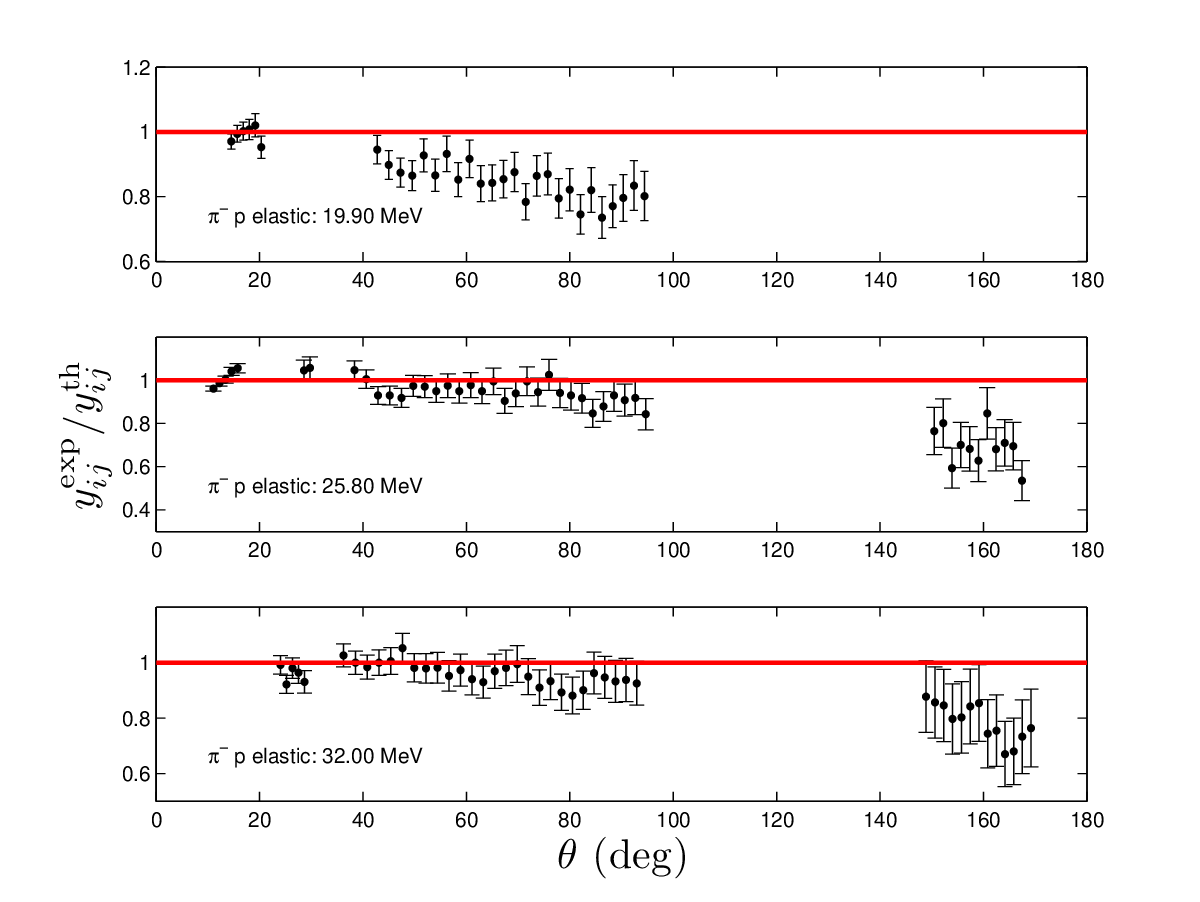}
\end{center}
\end{figure}

\clearpage
\begin{figure}
\begin{center}
\includegraphics [width=15.5cm] {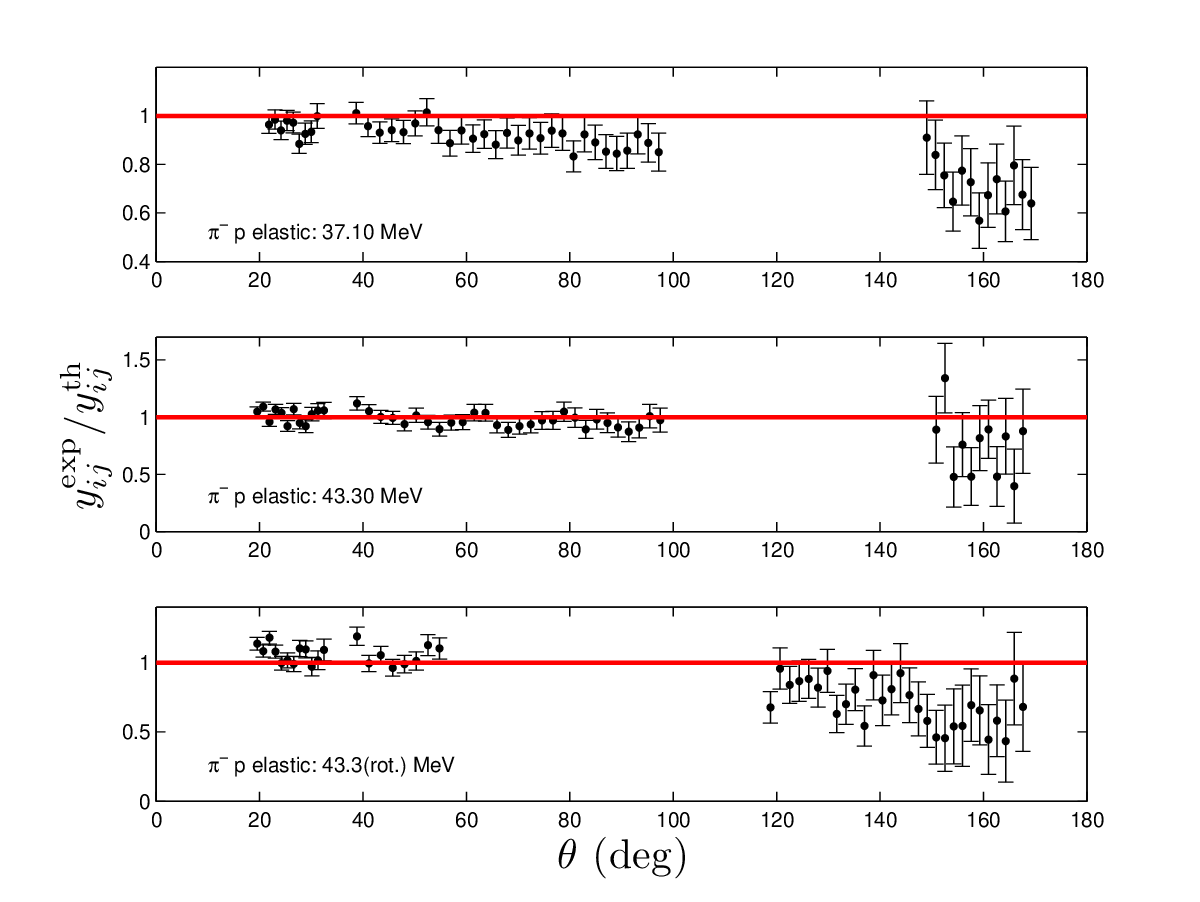}
\caption{\label{fig:PIMPEDenz}The equivalent of Fig.~\ref{fig:PIPPEDenz} for the DENZ04 $\pi^- p$ ES DCSs \cite{Denz2006}.}
\vspace{0.35cm}
\end{center}
\end{figure}

\clearpage
\begin{figure}
\begin{center}
\includegraphics [width=15.5cm] {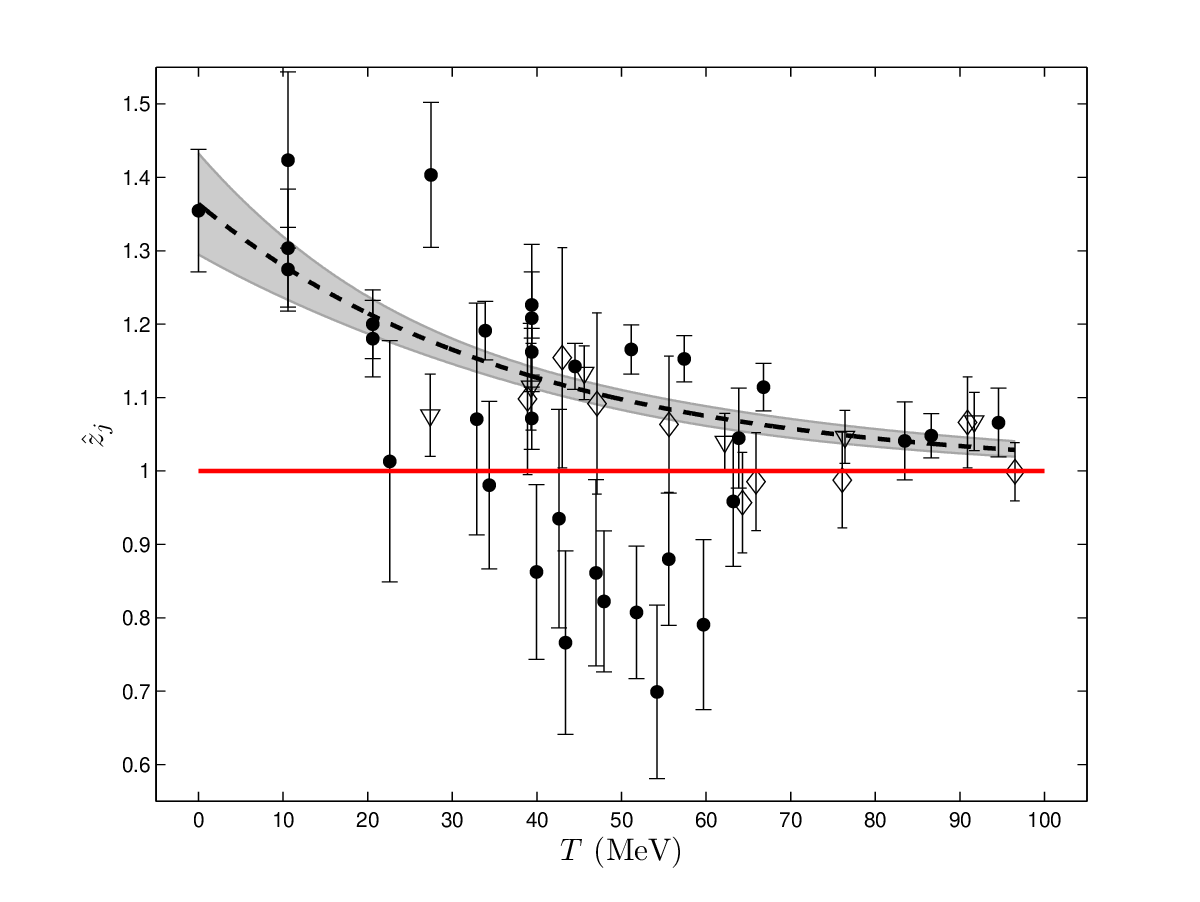}
\caption{\label{fig:PSAReprCX}The optimal scale factors $\hat{z}_j$ of Eq.~(\ref{eq:EQ013}), for those of the datasets in the DB$_0$ which are associated with measurements of the $\pi^- p$ CX DCS. The baseline solution has been 
obtained from the fits of the ETH model to the tDB$_{+/-}$; solid points: DCS, diamonds: TCS, inverse triangles: coefficients in the Legendre expansion of the DCS. The square of the $b_1$ result of Ref.~\cite{Schroeder2001} was 
assigned to the DCS set. Not included in the plot are the three FITZGERALD86 datasets which were freely floated, as well as the (excluded from the optimisation) BREITSCHOPF06 $75.10$ MeV entry, see Table \ref{tab:ProgressPIMCX}. 
The shaded band represents $1 \sigma$ uncertainties around the fitted values of a simple exponential form (see Section \ref{sec:PSAReproductionCX}).}
\vspace{0.35cm}
\end{center}
\end{figure}

\clearpage
\begin{figure}
\begin{center}
\includegraphics [width=7.75cm] {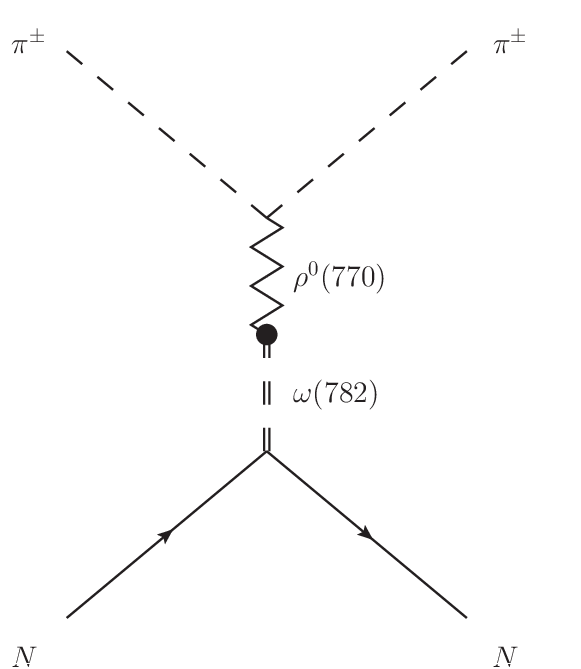}
\caption{\label{fig:IsospinBreakingRhoOmega}Feynman graphs involving the $\rho^0 - \omega$ mixing, a potential mechanism for the violation of the isospin invariance in the hadronic part of the $\pi N$ interaction in case of the 
$\pi^+ p$ and $\pi^- p$ ES reactions.}
\vspace{0.35cm}
\end{center}
\end{figure}

\begin{figure}
\begin{center}
\includegraphics [width=15.5cm] {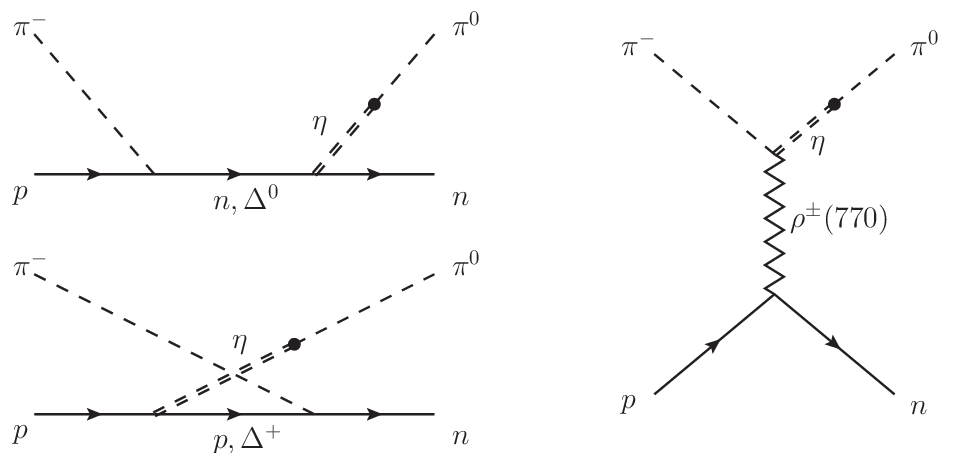}
\caption{\label{fig:IsospinBreakingEtaPi0}Feynman graphs involving the $\pi^0 - \eta$ mixing, a potential mechanism for the violation of the isospin invariance in the hadronic part of the $\pi N$ interaction in case of the 
$\pi^- p$ CX reaction.}
\vspace{0.35cm}
\end{center}
\end{figure}

\clearpage
\newpage
\appendix
\section{\label{App:AppA}Extraction of predictions for the $\pi^\pm p$ DCS for the MUSE experiment}

The MUon proton Scattering Experiment (MUSE) aims at a determination of the rms electric-charge radius of the proton from low-$Q^2$ measurements of the $\mu p$ and $e p$ DCS at projectile laboratory momenta of $115$, $153$ (or 
$160$), and $210$ MeV \cite{Roy2020}. The data acquisition will take place at PSI, though (given the current situation) it is not clear exactly when. According to the most recent plans \cite{Gilman2020}, measurements of the 
$\pi^\pm p$ ES DCS at the same energies are also foreseen. We take this opportunity to generate (and publicise) our predictions for the $\pi^\pm p$ ES DCS at pion laboratory momenta of $115$, $153$, and $160$ MeV, which lie 
within the energy domain of our PSAs. As they plan to carefully examine the systematic effects in their data, our predictions might be useful to the MUSE Collaboration. We will refrain from giving predictions at $210$ MeV, as 
the corresponding pion laboratory kinetic energy $T$ exceeds the upper limit of $100$ MeV in our PSAs of the $\pi N$ data: at this time, to provide predictions at $210$ MeV would have necessitated the determination of the EM 
corrections of the basis of extrapolations, the validity of which has not been investigated.

According to Ref.~\cite{Roy2020}, the MUSE detector covers an angular domain between $20$ and $100^\circ$, which (translated into ranges in the CM) corresponds to about $23.65-110.17^\circ$, $24.10-111.21^\circ$, and 
$24.19-111.42^\circ$ for the $115$-, $153$-, and $160$-MeV beams, respectively. To allow for small changes in the experimental apparatus, we will generate predictions for CM scattering angles $\theta$ between $20$ and $120^\circ$ 
for all three energies: these predictions will be accompanied by meaningful uncertainties, which reflect all important sources of statistical and systematic variation of the input data. (Not included in the uncertainties are 
the effects of the variation of the physical constants, of the EM form factors, etc. However, throughout the years, we have conducted various analyses with different sets of input values for these quantities, and have found 
out that their impact on our results is small in comparison with the uncertainties of the input $\pi N$ data.)

\begin{figure}
\begin{center}
\includegraphics [width=15.5cm] {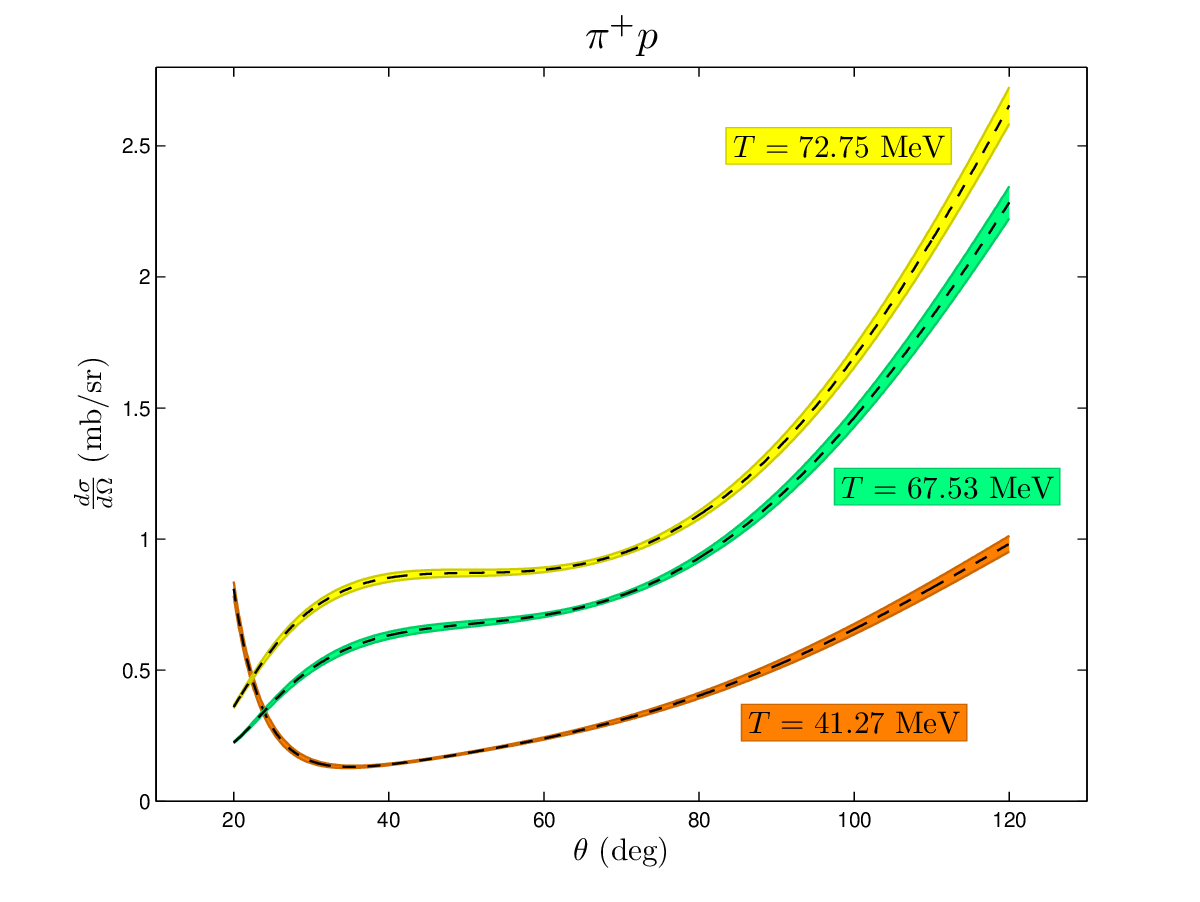}
\caption{\label{fig:PIPPEMUSE}Our predictions for the $\pi^+ p$ DCS at pion laboratory momenta of $115$, $153$, and $160$ MeV, at which the MUSE Collaboration plan to acquire experimental data at PSI. These predictions are 
based on the results of our PSA of the tDB$_{+/-}$. The bands represent $1 \sigma$ uncertainties around the average values (dashed curves); the uncertainties reflect all important sources of statistical and systematic variation 
of the input data. We have refrained from giving predictions at their last momentum value ($210$ MeV), as (in that case) the corresponding pion laboratory kinetic energy exceeds the upper limit of the energy domain of our PSAs.}
\vspace{0.35cm}
\end{center}
\end{figure}

\begin{figure}
\begin{center}
\includegraphics [width=15.5cm] {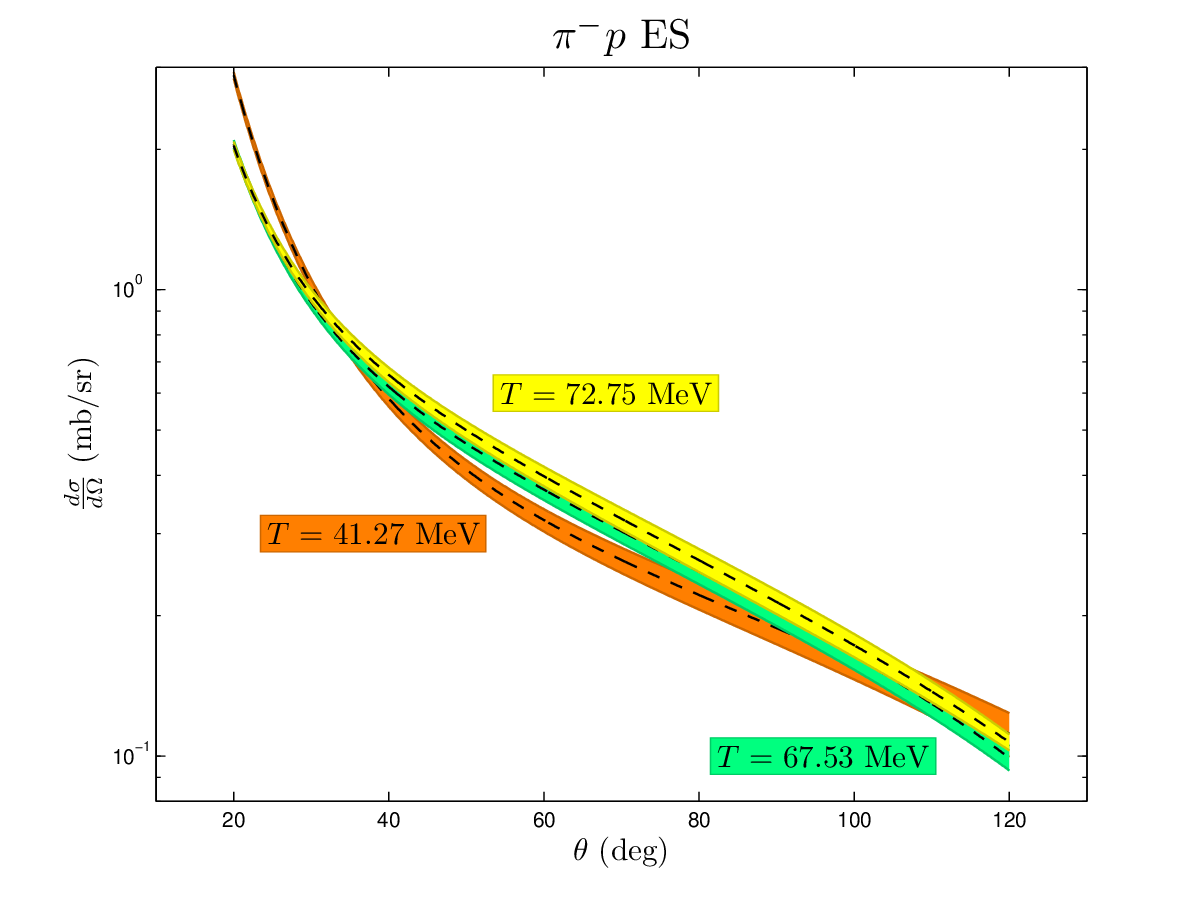}
\caption{\label{fig:PIMPEMUSE}Same as Fig.~\ref{fig:PIPPEMUSE} for the $\pi^- p$ ES DCS. In contrast to Fig.~\ref{fig:PIPPEMUSE}, a logarithmic scale is plotted on the ordinate axis.}
\vspace{0.35cm}
\end{center}
\end{figure}

The plots of the CM DCS as functions of $\theta$ are given in Fig.~\ref{fig:PIPPEMUSE} for the $\pi^+ p$ reaction and in Fig.~\ref{fig:PIMPEMUSE} for the $\pi^- p$ ES reaction. The transformation of these quantities from the 
CM to the laboratory frame of reference has been detailed in Section 2.2.1 of Ref.~\cite{Matsinos2014}. The absolute normalisation of our predictions reflects that of the bulk of the data on which our current PSA relies. For 
convenience, we also make our predictions available in an Excel file, uploaded to arXiv\textregistered~as ancillary material.

\clearpage

\clearpage


\begin{thebibliography}{99}
\bibitem{Arndt1972} R.A.~Arndt, L.D.~Roper, `The use of partial-wave representations in the planning of scattering measurements. Application to $330$ MeV $n p$ scattering', Nucl.~Phys.~B 50, 285 (1972). DOI: 10.1016/S0550-3213(72)80019-1
\bibitem{Deser1954} S.~Deser, M.L.~Goldberger, K.~Baumann, W.~Thirring, `Energy level displacements in pi-mesonic atoms', Phys.~Rev.~96, 774 (1954). DOI: 10.1103/PhysRev.96.774
\bibitem{Trueman1961} T.L.~Trueman, `Energy level shifts in atomic states of strongly-interacting particles', Nucl.~Phys.~26, 57 (1961). DOI: 10.1016/0029-5582(61)90115-8
\bibitem{Schroeder2001} H.-Ch.~Schr{\"o}der \etal, `The pion-nucleon scattering lengths from pionic hydrogen and deuterium', Eur.~Phys.~J.~C 21, 473 (2001). DOI: 10.1007/s100520100754
\bibitem{Hennebach2014} M.~Hennebach \etal, `Hadronic shift in pionic hydrogen', Eur.~Phys.~J.~A 50, 190 (2014). DOI: 10.1140/epja/i2014-14190-x
\bibitem{Arndt2006} R.A.~Arndt, W.J.~Briscoe, I.I.~Strakovsky, R.L.~Workman, `Extended partial-wave analysis of $\pi N$ scattering data', Phys.~Rev.~C 74, 045205 (2006). DOI: 10.1103/PhysRevC.74.045205; SAID Analysis Program: gwdac.phys.gwu.edu
\bibitem{Matsinos2017} E.~Matsinos, G.~Rasche, `Systematic effects in the low-energy behavior of the current SAID solution for the pion-nucleon system', Int.~J.~Mod.~Phys.~E 26, 1750002 (2017). DOI: 10.1142/S0218301317500021
\bibitem{Alarcon2012} J.M.~Alarc{\'o}n, J.~Martin Camalich, J.A.~Oller, `Chiral representation of the $\pi N$ scattering amplitude and the pion-nucleon sigma term', Phys.~Rev.~D 85, 051503 (2012). DOI: 10.1103/PhysRevD.85.051503
\bibitem{Gashi2001a} A.~Gashi, E.~Matsinos, G.C.~Oades, G.~Rasche, W.S.~Woolcock, `Electromagnetic corrections to the phase shifts in low energy $\pi^+ p$ elastic scattering', Nucl.~Phys.~A 686, 447 (2001). DOI: 10.1016/S0375-9474(00)00603-5
\bibitem{Gashi2001b} A.~Gashi, E.~Matsinos, G.C.~Oades, G.~Rasche, W.S.~Woolcock, `Electromagnetic corrections for the analysis of low energy $\pi^- p$ scattering data', Nucl.~Phys.~A 686, 463 (2001). DOI: 10.1016/S0375-9474(00)00604-7
\bibitem{Goudsmit1991} P.F.A.~Goudsmit, H.J.~Leisi, E.~Matsinos, `Pionic atoms, the relativistic mean-field theory and the pion-nucleon scattering lengths', Phys.~Lett.~B 271, 290 (1991). DOI: 10.1016/0370-2693(91)90089-9
\bibitem{Goudsmit1992a} P.F.A.~Goudsmit, H.J.~Leisi, E.~Matsinos, `A pion-nucleon interaction model at the tree level', $\pi N$ Newsl.~6, 60 (1992), available from gwdac.phys.gwu.edu
\bibitem{Goudsmit1992b} P.F.A.~Goudsmit, H.J.~Leisi, E.~Matsinos, `Pionic atoms, the relativistic mean-field theory and the pion-nucleon scattering lengths', Few-Body Syst., Suppl.~5, 416 (1992). DOI: 10.1007/978-3-7091-7617-7\_53
\bibitem{Goudsmit1992c} P.F.A.~Goudsmit, H.J.~Leisi, E.~Matsinos, `A new pion-nucleon interaction model', Helv.~Phys.~Acta 65, 878 (1992).
\bibitem{Goudsmit1993a} P.F.A.~Goudsmit, H.J.~Leisi, E.~Matsinos, `A dynamical model for the pion-nucleon interaction', $\pi N$ Newsl.~8, 98 (1993), available from gwdac.phys.gwu.edu
\bibitem{Goudsmit1993b} P.F.A.~Goudsmit, H.J.~Leisi, E.~Matsinos, `A pion-nucleon interaction model', Phys.~Lett.~B 299, 6 (1993). DOI: 10.1016/0370-2693(93)90875-I
\bibitem{Goudsmit1994a} P.F.A.~Goudsmit, H.J.~Leisi, E.~Matsinos, B.L.~Birbrair, A.B.~Gridnev, `The extended tree-level model of the pion-nucleon interaction', Nucl.~Phys.~A 575, 673 (1994). DOI: 10.1016/0375-9474(94)90162-7
\bibitem{Goudsmit1994b} P.F.A.~Goudsmit, H.J.~Leisi, E.~Matsinos, `The low-energy pion-nucleon interaction', Helv.~Phys.~Acta 67, 369 (1994).
\bibitem{Fettes1997} N.~Fettes, E.~Matsinos, `Analysis of recent $\pi^+ p$ low-energy differential cross-section measurements', Phys.~Rev.~C 55, 464 (1997). DOI: 10.1103/PhysRevC.55.464
\bibitem{Matsinos1997a} E.~Matsinos, `$\pi N$ scattering below 100 MeV', $\pi N$ Newsl.~13, 132 (1997), available from gwdac.phys.gwu.edu
\bibitem{Matsinos1997b} E.~Matsinos, `Isospin violation in the $\pi N$ system at low energy', Phys.~Rev.~C 56, 3014 (1997). DOI: 10.1103/PhysRevC.56.3014
\bibitem{Tromborg1976} B.~Tromborg, S.~Waldestr{\o}m, I.~{\O}verb{\o}, `Electromagnetic corrections to $\pi^+ p$ scattering', Ann.~Phys.~100, 1 (1976). DOI: 10.1016/0003-4916(76)90055-5
\bibitem{Tromborg1977} B.~Tromborg, S.~Waldestr{\o}m, I.~{\O}verb{\o}, `Electromagnetic corrections to $\pi N$ scattering', Phys.~Rev.~D 15, 725 (1977). DOI: 10.1103/PhysRevD.15.725
\bibitem{Tromborg1978} B.~Tromborg, S.~Waldestr{\o}m, I.~{\O}verb{\o}, `Electromagnetic corrections in hadron scattering, with application to $\pi N \to \pi N$', Helv.~Phys.~Acta 51, 584 (1978).
\bibitem{Matsinos2006} E.~Matsinos, W.S.~Woolcock, G.C.~Oades, G.~Rasche, A.~Gashi, `Phase-shift analysis of low-energy $\pi^\pm p$ elastic-scattering data', Nucl.~Phys.~A 778, 95 (2006). DOI: 10.1016/j.nuclphysa.2006.07.040
\bibitem{Matsinos2012} E.~Matsinos, G.~Rasche, `Analysis of the low-energy $\pi^\pm p$ elastic-scattering data', J.~Mod.~Phys.~3, 1369 (2012). DOI: 10.4236/jmp.2012.310174
\bibitem{Matsinos2013a} E.~Matsinos, G.~Rasche, `Analysis of the low-energy $\pi^- p$ charge-exchange data', Int.~J.~Mod.~Phys.~A 28, 1350039 (2013). DOI: 10.1142/S0217751X13500395
\bibitem{Matsinos2013b} E.~Matsinos, G.~Rasche, `Analysis of the low-energy $\pi^\pm p$ differential cross sections of the CHAOS Collaboration', Nucl.~Phys.~A 903, 65 (2013). DOI: 10.1016/j.nuclphysa.2012.12.129
\bibitem{Matsinos2015} E.~Matsinos, G.~Rasche, `New analysis of the low-energy $\pi^\pm p$ differential cross sections of the CHAOS Collaboration', Int.~J.~Mod.~Phys.~E 24, 1550050 (2015). DOI: 10.1142/S0218301315500500
\bibitem{Denz2006} H.~Denz \etal, `$\pi^\pm p$ differential cross sections at low energy', Phys.~Lett.~B 633, 209 (2006); H.~Denz, Ph.D.~dissertation, T{\"u}bingen University, 2004; available from: publikationen.uni-tuebingen.de/xmlui/handle/10900/48622.\\DOI: 10.1016/j.physletb.2005.12.017
\bibitem{Matsinos2014} E.~Matsinos, G.~Rasche, `Aspects of the ETH model of the pion-nucleon interaction', Nucl.~Phys.~A 927, 147 (2014).\\DOI: 10.1016/j.nuclphysa.2014.04.021
\bibitem{PDG2020} P.A.~Zyla \etal~(Particle Data Group), `The Review of Particle Physics (2020)', Prog.~Theor.~Exp.~Phys.~2020, 083C01 (2020).
\bibitem{Matsinos2020a} E.~Matsinos, `Determination of the masses and decay widths of the scalar-isoscalar and vector-isovector mesons below $2$ GeV', {\tt arXiv:2007.13130 [hep-ph]}.
\bibitem{Birge1932} R.T.~Birge, `The calculation of errors by the method of least squares', Phys.~Rev.~40, 207 (1932). DOI: 10.1103/PhysRev.40.207
\bibitem{Matsinos2020b} E.~Matsinos, `Determination of the masses and decay widths of the well-established $s$ and $p$ baryon resonances below $2$ GeV', {\tt arXiv:2008.06919 [hep-ph]}.
\bibitem{XP15} R.L.~Workman, R.A.~Arndt, W.J.~Briscoe, M.W.~Paris, I.I.~Strakovsky, `Parameterization dependence of $T$-matrix poles and eigenphases from a fit to $\pi N$ elastic scattering data', Phys.~Rev.~C 86, 035202 (2012). DOI: 10.1103/PhysRevC.86.035202; 
I.G.~Alekseev \etal~(EPECUR Collaboration and GW INS Data Analysis Center), `High-precision measurements of $\pi p$ elastic differential cross sections in the second resonance region', Phys.~Rev.~C 91, 025205 (2015). DOI: 10.1103/PhysRevC.91.025205; 
A.~Gridnev \etal, `Search for narrow resonances in $\pi p$ elastic scattering from the EPECUR experiment', Phys.~Rev.~C 93, 062201(R) (2016). DOI: 10.1103/PhysRevC.93.062201
\bibitem{James} F.~James, `MINUIT - Function Minimization and Error Analysis', CERN Program Library Long Writeup D506.
\bibitem{Nelder1965} J.A.~Nelder, R.~Mead, `A simplex method for function minimization', Comput.~J.~7, 308 (1965). DOI: 10.1093/comjnl/7.4.308
\bibitem{Johnson2013} V.E.~Johnson, `Revised standards for statistical evidence', P.~Natl.~Acad.~Sci.~USA 110, 19313 (2013).\\DOI: 10.1073/pnas.1313476110
\bibitem{Abramowitz1972} M.~Abramowitz, I.A.~Stegun, `Handbook of Mathematical Functions with Formulas, Graphs, and Mathematical Tables', United States Department of Commerce, National Bureau of Standards (1972). ISBN: 9780318117300
\bibitem{Press2007} W.H.~Press, S.A.~Teukolsky, W.T.~Vetterling, B.P.~Flannery, `Numerical Recipes: The Art of Scientific Computing' (3rd edn.), Cambridge University Press (2007). ISBN: 9780521880688
\bibitem{Oades2007} G.C.~Oades, G.~Rasche, W.S.~Woolcock, E.~Matsinos, A.~Gashi, `Determination of the $s$-wave pion-nucleon threshold scattering parameters from the results of experiments on pionic hydrogen', Nucl.~Phys.~A 794, 73 (2007). DOI: 10.1016/j.nuclphysa.2007.07.007
\bibitem{Frank1983} J.S.~Frank \etal, `Measurement of low-energy elastic $\pi^\pm p$ differential cross sections', Phys.~Rev.~D 28, 1569 (1983). DOI: 10.1103/PhysRevD.28.1569
\bibitem{Brack1990} J.T.~Brack \etal, `Absolute differential cross section for the $\pi^\pm p$ elastic scattering at $30 \leq T_\pi \leq 67$ MeV', Phys.~Rev.~C 41, 2202 (1990). DOI: 10.1103/PhysRevC.41.2202
\bibitem{Joram1995a} Ch.~Joram \etal, `Low-energy differential cross section of pion-proton ($\pi^\pm p$) scattering. I. The isospin-even forward scattering amplitude at $T_\pi=32.2$ and $44.6$ MeV.', Phys.~Rev.~C 51, 2144 (1995). DOI: 10.1103/PhysRevC.51.2144
\bibitem{Joram1995b} Ch.~Joram \etal, `Low-energy differential cross section of pion-proton ($\pi^\pm p$) scattering. II. Phase shifts at $T_\pi=32.7$, $45.1$, and $68.6$ MeV.', Phys.~Rev.~C 51, 2159 (1995). DOI: 10.1103/PhysRevC.51.2159
\bibitem{Schlesser2011} S.~Schlesser, E.-O.~Le Bigot, P.~Indelicato, K.~Pachucki, `Quantum-electrodynamics corrections in pionic hydrogen', Phys.~Rev.~C 84, 015211 (2011). DOI: 10.1103/PhysRevC.84.015211
\bibitem{Matsinos2019a} E.~Matsinos, `A brief history of the pion-nucleon coupling constant', {\tt arXiv:1901.01204 [nucl-th]}.
\bibitem{Matsinos2020c} E.~Matsinos, `Comparison of results for the electromagnetic form factors of the proton at low $Q^2$', {\tt arXiv:2009.04156 [hep-ph]}.
\bibitem{Venkat2011} S.~Venkat, J.~Arrington, G.A.~Miller, X.~Zhan, `Realistic transverse images of the proton charge and magnetization densities', Phys.~Rev.~C 83, 015203 (2011). DOI: 10.1103/PhysRevC.83.015203
\bibitem{Pohl2010} R.~Pohl \etal, `The size of the proton', Nature 466, 213 (2010). DOI: 10.1038/nature09250
\bibitem{Antognini2013} A.~Antognini \etal, `Proton structure from the measurement of $2S - 2P$ transition frequencies of muonic hydrogen', Science 339, 417 (2013). DOI: 10.1126/science.1230016
\bibitem{Belushkin2007} M.A.~Belushkin, H.-W.~Hammer, Ulf-G.~Mei{\ss}ner, `Dispersion analysis of the nucleon form factors including meson continua', Phys.~Rev.~C 75, 035202 (2007). DOI: 10.1103/PhysRevC.75.035202
\bibitem{Borisyuk2010} D.~Borisyuk, `Proton charge and magnetic rms radii from the elastic $e p$ scattering data', Nucl.~Phys.~A 843, 59 (2010).\\DOI: 10.1016/j.nuclphysa.2010.05.054
\bibitem{Bernauer2010} J.C.~Bernauer \etal~(A1 Collaboration), `High-precision determination of the electric and magnetic form factors of the proton', Phys.~Rev.~Lett.~105, 242001 (2010). DOI: 10.1103/PhysRevLett.105.242001
\bibitem{Zhan2011} X.~Zhan \etal, `High-precision measurement of the proton elastic form factor ratio $\mu_p G_E / G_M$ at low $Q^2$', Phys.~Lett.~B 705, 59 (2011). DOI: 10.1016/j.physletb.2011.10.002
\bibitem{Lorenz2012} I.T.~Lorenz, H.-W.~Hammer, Ulf-G.~Mei{\ss}ner, `The size of the proton: Closing in on the radius puzzle', Eur.~Phys.~J.~A 48, 151 (2012). DOI: 10.1140/epja/i2012-12151-1
\bibitem{Epstein2014} Z.~Epstein, G.~Paz, J.~Roy, `Model independent extraction of the proton magnetic radius from electron scattering', Phys.~Rev.~D 90, 074027 (2014). DOI: 10.1103/PhysRevD.90.074027
\bibitem{Lorenz2015} I.T.~Lorenz, Ulf-G.~Mei{\ss}ner, H.-W.~Hammer, Y.-B.~Dong, `Theoretical constraints and systematic effects in the determination of the proton form factors', Phys.~Rev.~D 91, 014023 (2015). DOI: 10.1103/PhysRevD.91.014023
\bibitem{Lee2015} G.~Lee, J.R.~Arrington, R.J.~Hill, `Extraction of the proton radius from electron-proton scattering data', Phys.~Rev.~D 92, 013013 (2015). DOI: 10.1103/PhysRevD.92.013013
\bibitem{Bertin1976} P.Y.~Bertin \etal, `$\pi^+ p$ scattering below $100$ MeV', Nucl.~Phys.~B 106, 341 (1976). DOI: 10.1016/0550-3213(76)90383-7
\bibitem{Auld1979} E.G.~Auld \etal, `$\pi^+-p$ scattering at $47.9$ MeV', Can.~J.~Phys.~57, 73 (1979). DOI: 10.1139/p79-008
\bibitem{Friedman1990} E.~Friedman \etal, `Integral cross sections for $\pi^+ p$ interactions at low energies', Nucl.~Phys.~A 514, 601 (1990). DOI: 10.1016/0375-9474(90)90012-B
\bibitem{Friedman1999} E.~Friedman, `Partial-total $\pi N$ cross sections', $\pi N$ Newsl.~15, 37 (1999), available from gwdac.phys.gwu.edu
\bibitem{Carter1971} A.A.~Carter, J.R.~Williams, D.V.~Bugg, P.J.~Bussey, D.R.~Dance, `The total cross sections for pion-proton scattering between $70$ MeV and $290$ MeV', Nucl.~Phys.~B 26, 445 (1971). DOI: 10.1016/0550-3213(71)90188-X
\bibitem{Pedroni1978} E.~Pedroni \etal, `A study of charge independence and symmetry from $\pi^+$ and $\pi^-$ total cross sections on hydrogen and deuterium near the $3,3$ resonance', Nucl.~Phys.~A 300, 321 (1978). DOI: 10.1016/0375-9474(78)96136-5
\bibitem{Duclos1973} J.~Duclos \etal, `A measurement of the pion-nucleon charge-exchange reaction $\pi^- + p \to \pi^0 + n$ at $22.6$, $32.9$ and $42.6$ MeV', Phys.~Lett.~B 43, 245 (1973). DOI: 10.1016/0370-2693(73)90280-3
\bibitem{Salomon1984} M.~Salomon, D.F.~Measday, J-M.~Poutissou, B.C.~Robertson, `Radiative capture and charge exchange of negative pions on protons at $26.4$ and $39.3$ MeV', Nucl.~Phys.~A 414, 493 (1984). DOI: 10.1016/0375-9474(84)90615-8
\bibitem{Bugg1971} D.V.~Bugg \etal, `The $\pi^- p \to \pi^0 n$ charge-exchange cross sections between $90$ MeV and $290$ MeV', Nucl.~Phys.~B 26, 588 (1971). DOI: 10.1016/0550-3213(71)90197-0
\bibitem{Breitschopf2006} J.~Breitschopf \etal, `Pionic charge exchange on the proton from $40$ to $250$ MeV', Phys.~Lett.~B 639, 424 (2006). DOI: 10.1016/j.physletb.2006.07.009
\bibitem{Kriss1999} B.J.~Kriss \etal, `Pion-proton integral cross section measurements', $\pi N$ Newsl.~12, 20 (1997), available from gwdac.phys.gwu.edu; B.J.~Kriss \etal, `Pion-proton integral cross sections at $T_\pi=40$ to 
$284$ MeV', Phys.~Rev.~C 59, 1480 (1999). DOI: 10.1103/PhysRevC.59.1480
\bibitem{Bagheri1988} A.~Bagheri \etal, `Reaction $\pi^- p \to \pi^0 n$ below the $\Delta$ resonance', Phys.~Rev.~C 38, 885 (1988). DOI: 10.1103/PhysRevC.38.885
\bibitem{Sevior1989} M.E.~Sevior \etal, `Analyzing powers in $\pi^\pm \vec{p}$ elastic scattering from $T_\pi=98$ to $263$ MeV', Phys.~Rev.~C 40, 2780 (1989). DOI: 10.1103/PhysRevC.40.2780
\bibitem{Narison2001} S.~Narison, `Scalar Mesons in QCD', Nucl.~Phys.~B - Proc.~Sup.~96, 244 (2001). DOI: 10.1016/S0920-5632(01)01137-9
\bibitem{Hoehler1983} G.~H{\"o}hler, `Pion Nucleon Scattering. Part 2: Methods and Results of Phenomenological Analyses', Landolt-B{\"o}rnstein, Vol.~9b2, ed.~H.~Schopper, Springer, Berlin (1983). ISBN: 9783540112822
\bibitem{Matsinos2018} E.~Matsinos, `The contribution of the $\rho^0 - \omega$ interference to the violation of the isospin invariance in the $\pi N$ system', {\tt arXiv:1804.09344 [nucl-th]}.
\bibitem{Hoehler1972} G.~H{\"o}hler, H.P.~Jakob, R.~Strauss, `A critical test of models for the low energy $\pi N$ amplitude', Nucl.~Phys.~B 39, 237 (1972). DOI: 10.1016/0550-3213(72)90372-0
\bibitem{Ericson1988} T.E.O.~Ericson, W.~Weise, `Pions and Nuclei', Clarendon Press (1988). ISBN: 9780198520085
\bibitem{Meier2004} R.~Meier \etal, `Low energy analyzing powers in pion-proton elastic scattering', Phys.~Lett.~B 588, 155 (2004). DOI: 10.1016/j.physletb.2004.02.071
\bibitem{Denz2011} H.~Denz, G.J.~Wagner, private communication.
\bibitem{Alarcon2013} J.M.~Alarc{\'o}n, J.~Martin Camalich, J.A.~Oller, `Improved description of the $\pi N$-scattering phenomenology at low energies in covariant baryon chiral perturbation theory', Ann.~Phys.~336, 413 (2013).\\DOI: 10.1016/j.aop.2013.06.001
\bibitem{Hoferichter2016} M.~Hoferichter, J.~Ruiz de Elvira, B.~Kubis, Ulf-G.~Mei{\ss}ner, `Roy-Steiner-equation analysis of pion-nucleon scattering', Phys.~Rep.~625, 1 (2016). DOI: 10.1016/j.physrep.2016.02.002
\bibitem{Hoferichter2010} M.~Hoferichter, B.~Kubis, Ulf-G.~Mei{\ss}ner, `Isospin violation in low-energy pion-nucleon scattering revisited', Nucl.~Phys.~A 833, 18 (2010). DOI: 10.1016/j.nuclphysa.2009.11.012
\bibitem{Gibbs1995} W.R.~Gibbs, Li Ai, W.B.~Kaufmann, `Isospin breaking in low-energy pion-nucleon scattering', Phys.~Rev.~Lett.~74, 3740 (1995). DOI: 10.1103/PhysRevLett.74.3740
\bibitem{Matsinos2020} E.~Matsinos, G.~Rasche, `Proposals for the test of the isospin invariance in the pion-nucleon interaction at low energy', {\tt arXiv:2001.01229 [nucl-th]}.
\bibitem{Miller2006} G.A.~Miller, A.K.~Opper, E.J.~Stephenson, `Charge symmetry breaking and QCD', Annu.~Rev.~Nucl.~Part.~S.~56, 253 (2006). DOI: 10.1146/annurev.nucl.56.080805.140446
\bibitem{Babenko2017} V.A.~Babenko, N.M.~Petrov, `Relation between the charged and neutral pion-nucleon coupling constants in the Yukawa model', Phys.~Part~Nuclei Lett.~14, 58 (2017). DOI: 10.1134/S1547477117010083
\bibitem{Navarro2017} R.~Navarro P{\'e}rez, J.E.~Amaro, E.~Ruiz Arriola, `Precise determination of charge dependent pion-nucleon-nucleon coupling constants', Phys.~Rev.~C 95, 064001 (2017). DOI: 10.1103/PhysRevC.95.064001
\bibitem{Piekarewicz1995} J.~Piekarewicz, `Isospin violations in the pion-nucleon system', Phys.~Lett.~B 358, 27 (1995). DOI: 10.1016/0370-2693(95)01026-M
\bibitem{Meissner1997} T.~Meissner, E.M.~Henley, `Isospin breaking in the pion-nucleon coupling from QCD sum rules', Phys.~Rev.~C 55, 3093 (1997).\\DOI: 10.1103/PhysRevC.55.3093
\bibitem{Cutkosky1979} R.E.~Cutkosky, `Isospin violation in $\pi N$ scattering from $\pi^0-\eta$ mixing', Phys.~Lett.~B 88, 339 (1979). DOI: 10.1016/0370-2693(79)90482-9
\bibitem{Binosi2004} D.~Binosi, L.~Theu{\ss}l, `JaxoDraw: A graphical user interface for drawing Feynman diagrams', Comput.~Phys.~Commun.~161, 76 (2004). DOI: 10.1016/j.cpc.2004.05.001
\bibitem{Binosi2009} D.~Binosi, J.~Collins, C.~Kaufhold, L.~Theu{\ss}l, `JaxoDraw: A graphical user interface for drawing Feynman diagrams. Version 2.0 release notes', Comput.~Phys.~Commun.~180, 1709 (2009).\\DOI: 10.1016/j.cpc.2009.02.020
\bibitem{Roy2020} P.~Roy \etal, `A liquid hydrogen target for the MUSE experiment at PSI', Nucl.~Instrum.~Methods Phys.~Res.~A 949, 162874 (2020). DOI: 10.1016/j.nima.2019.162874
\bibitem{Gilman2020} R.~Gilman, private communication.
\end{thebibliography}
\end{document}